\documentclass[aps,pra,superscriptaddress,twocolumn,floatfix,longbibliography]{revtex4-1}
\usepackage{dcolumn}
\usepackage{graphicx,color}
\usepackage[utf8]{inputenc}
\usepackage{amsmath,amssymb,bm,amsfonts,dsfont,mathrsfs,amsthm}
\usepackage{braket}
\usepackage{txfonts,comment}
\usepackage[colorlinks=true,linkcolor=blue,citecolor=blue,urlcolor=blue]{hyperref}
\usepackage{soul}

\begin{document}

\title{Near-Ultimate Quantum-Enhanced Sensitivity in Dissipative Critical Sensing with Partial Access}

\author{Dingwei Zhao}
\email{202311210340@std.uestc.edu.cn}
\affiliation{Institute of Fundamental and Frontier Sciences, University of Electronic Science and Technology of China, Chengdu 611731, China}

\author{Abolfazl Bayat}
\email{abolfazl.bayat@uestc.edu.cn}
\affiliation{Institute of Fundamental and Frontier Sciences, University of Electronic Science and Technology of China, Chengdu 611731, China}
\affiliation{Key Laboratory of Quantum Physics and Photonic Quantum Information, Ministry of Education, University of Electronic Science and
Technology of China, Chengdu 611731, China}
\affiliation{Shimmer Center, Tianfu Jiangxi Laboratory, Chengdu 641419, China}

\author{Victor Montenegro}
\email{victor.montenegro@ku.ac.ae}
\affiliation{College of Computing and Mathematical Sciences, Department of Applied Mathematics and Sciences, Khalifa University of Science and Technology, 127788 Abu Dhabi, United Arab Emirates}
\affiliation{Institute of Fundamental and Frontier Sciences, University of Electronic Science and Technology of China, Chengdu 611731, China}
\affiliation{Key Laboratory of Quantum Physics and Photonic Quantum Information, Ministry of Education, University of Electronic Science and
Technology of China, Chengdu 611731, China}

\date{\today}

\begin{abstract}
Quantum sensors are powerful devices that exploit quantum effects to detect minute quantities with extremely high precision. Two obstacles to harnessing the full capacity of quantum probes are the resource-intensive preparation of the probe and the need for sophisticated measurements that typically require full access to the entire probe. Here, we address these challenges by investigating the driven Jaynes-Cummings system undergoing a dissipative quantum phase transition as a quantum sensor. We show that detuning the system off resonance significantly improves sensing performance by adequately selecting a preferred bistable state in phase space. Our dissipative sensor, independent of the initial probe preparation, exhibits a super-linear enhancement in sensitivity with respect to a specific sensing resource---the strong-coupling regime ratio---which manifests in both the full system and partial subsystem. Hence, quantum-enhanced sensitivity persists even when only partial system accessibility is available. Remarkably, we show that a homodyne detection of the field state, combined with Bayesian estimation, nearly saturates the ultimate sensitivity limit of the entire system.
\end{abstract}

\maketitle 

\textit{Introduction---} By exploiting distinctive quantum phenomena, quantum sensors have been shown to surpass the standard quantum limit of precision~\cite{degen2017quantum,giovannetti2006quantummetrology, giovannetti2011advances, giovannetti2004quantum, montenegro2025review,toth2014quantum,polino2020photonic}, enabling a wide range of high-precision field estimations~\cite{mishra2021driving,mishra2022integrable,montenegro2021global,marciniak2022optimal,he2023stark,mukhopadhyay2025current,qvarfort2018gravimetry,montenegro2025heisenberg,kwon2020magnetic} as well as the estimation of both Hamiltonian~\cite{campos2007quantum,zanardi2008quantum,shabani2011estimation,dooley2021robust,sarkar2022free,montenegro2022probing,mukhopadhyay2024modular,sahoo2024localization,ye2024essay,zhang2025quantum,burgarth2017evolution} and non-Hamiltonian parameters~\cite{Mehboudi_2019,Razavian2019,mihailescu2023thermometry,montenegro2020mechanical,srivastava2023topological,Sekatski_2022,glatthard2022optimal,mok2021optimal,rubio2021global,campbell2017global}. Among the diverse platforms and sensing strategies~\cite{hou2019control,liu2017quantum,hentschel2011efficient,cimini2019calibration,yuan2015optimal,lang2015dynamical,montenegro2022sequential,burgarth2015quantum,yang2023extractable}, critical quantum sensors~\cite{lu2022critical,salvatori2014quantum,Zhu2023criticality,garbe2020critical,salvia2023critical,hotter2024combining,invernizzi2008optimal, zanardi2008quantum,zanardi2007mixed,zanardi2006ground, frerot2018quantum,mukhopadhyay2023modular,mihailescu2025uncertain,gietka2022understanding} stand out as promising candidates for achieving quantum-enhanced sensing by efficiently using available resources~\cite{zanardi2006ground,zanardi2008quantum}. Intuitively, a system near a phase transition can act as a quantum sensor, since infinitesimal changes in the control parameter induce divergent responses~\cite{montenegro2025review}. In particular, the ground state of a quantum many-body system at a quantum phase transition has been extensively explored for this purpose~\cite{alushi2025collective,zhao2009singularities,zanardi2006ground,venuti2007quantum,schwandt2009quantum,albuquerque2010quantum,gritsev2009universal,
gu2008fidelity,greschner2013fidelity,invernizzi2008optimal, zanardi2008quantum,zanardi2007mixed,frerot2018quantum}. However, their applicability faces several challenges: (i) preparing the ground state of a many-body system is demanding; (ii) exploiting the enhanced sensitivity often requires unfeasible measurements across the entire system; and (iii) accessing subsystems severely diminish the sensing capabilities. Failure to meet any of these conditions usually nullifies any quantum advantage. It is therefore desirable to identify a system that achieves quantum-enhanced sensitivity with local, feasible measurements and without complex state preparation.

To overcome the accessibility challenge, certain probes have been shown to extract either part~\cite{mishra2021driving,mishra2021integrable,montenegro2022probing} or all~\cite{qvarfort2018gravimetry,montenegro2025heisenberg} of the relevant information from only a reduced portion of the system. However, there is no universal sensing method that can reach or approach the full sensitivity of the entire system by measuring only part of it. To address the state-preparation challenge, probes undergoing dissipative quantum phase transitions~\cite{dicandia2023critical,minganti2018spectral,kessler2012dissipative,dalla2012dynamics,marino2016driven}, which occurs due to the interplay between Hamiltonian dynamics and Lindbladian processes~\cite{kessler2012dissipative}, have been explored as an alternative. In this approach, instead of preparing a specific initial state, the system evolves towards a steady state that exhibits critical properties suitable for sensing~\cite{fernandez2017quantum}. Indeed, dissipative quantum phase transitions, which have been extensively studied theoretically~\cite{lledo2020dissipative, bartolo2016exact, rota2017critical, letscher2016bistability, overbeck2017multicritical, iemini2018boundary, capriotti2005dissipation, heugel2019quantum, sieberer2014nonequilibrium, garbe2020critical, carmichael2015breakdown, benito2016degenerate, arenas2016beyond, casteels2017critical, casteels2017quantum, savona2017spontaneous, sieberer2013dynamical, jin2016cluster, lee2011antiferromagnetic, chan2015limit, maghrebi2016nonequilibrium} and realized experimentally~\cite{cai2022probing, ferioli2022observation, benary2022experimental, brennecke2013real, ferri2021emerging, baumann2010dicke, fink2018signatures, ohadi2015spontaneous, fitzpatrick2017observation, collodo2019observation, ding2016phase, beaulieu2023observationfirstsecondorderdissipative, letscher2016bistability}, have also been proposed as a resource for quantum sensing~\cite{xie2020dissipative, raghunandan2018highdensity, ivanov2020enhanced, arandes2025quantum, montenegro2023quantum, gribben2025boundary, pavlov2023quantum, fernandez2017quantum}. Even though dissipative probes solve the state-preparation challenge, it is still unknown whether they can reach quantum-enhanced sensitivity when only practical, partial-access measurements are allowed. In this work, we show that they can.

In this Letter, we study the driven-dissipative Jaynes-Cummings system as a quantum sensor undergoing a dissipative quantum phase transition~\cite{carmichael2015breakdown}. In the thermodynamic limit, this system exhibits quantum-critical behavior characterized by polarization of the two-level (qubit) system and bistability of the field mode. Remarkably, at the critical point, the steady state becomes weakly entangled in the thermodynamic limit, effectively encoding the unknown parameter in the field subsystem and suggesting that local probing can attain the system's ultimate sensitivity. We present three main results. First, we demonstrate that detuning the system off resonance favors one of the two bistable states, leading to a significant enhancement in sensing precision. Second, our sensor achieves quantum-enhanced sensitivity off resonance, under both full and partial probing. Third, a feasible homodyne field measurement combined with a Bayesian estimator nearly saturates the ultimate sensing precision of the whole quantum state. Even with this practical local measurement, the quantum-enhanced sensitivity is clearly demonstrated.

\textit{Quantum estimation background---} For a given measurement, the uncertainty in estimating an unknown parameter $\theta$ parameterized in a quantum state $\rho(\theta)$ is lower bounded by $\mathrm{Var}[\tilde{\theta}]{\geq}[M\mathcal{F}(\theta)]^{-1}$~\cite{cramer1999mathematical,LeCam-1986,rao1992information}. Here, $\mathrm{Var}[\tilde{\theta}]$ is the variance of the unbiased local estimator $\tilde{\theta}$, $M$ is the number of measurements trials, $\mathcal{F}(\theta)=\sum_k[p_k(\theta)]^{-1}(\partial p_k(\theta)/\partial\theta)^2$ is the classical Fisher information (CFI)~\cite{paris2004quantum} and $p_k{=}\mathrm{Tr}[\Pi_k\rho(\theta)]$ is the probability distribution for a given positive operator-valued measure (POVM) $\Pi_k$ with measurement outcome $k$. By maximizing over all POVMs, a more fundamental lower bound known as quantum Cram\'{e}r-Rao theorem is derived~\cite{helstrom1976quantum, vantrees2004detection,holevo2011probabilistic,braunstein1994statistical,paris2004quantum}:
\begin{equation}
\mathrm{Var}[\tilde{\theta}] {\geq} \frac{1}{M \mathcal{F}\left( \theta \right)}{\geq} \frac{1}{M Q\left( \theta \right)}, \label{eq_QCRB} 
\end{equation}
where $Q(\theta)=\max_{\{\Pi_k\}}\mathcal{F}(\theta)$ is the quantum Fisher information (QFI), given by~\cite{paris2004quantum}:
\begin{equation}
 Q\left( \theta \right) {=}2\sum_{n,m}{\frac{\left| \langle \psi _m|\partial\rho\right( {\theta})/\partial\theta\left|\psi _n\rangle \right|^2}{\lambda _m{+}\lambda _n}}.~\label{eq_qfi}
 \end{equation}
In the above, $\rho(\theta){=}\sum_i{\lambda_i|\psi_i\rangle \langle \psi_i|}$ is expressed in spectral decomposition with eigenvalue $\lambda_i$ and corresponding eigenvector $|\psi_i\rangle$. While the CFI determines the best possible precision for estimating an unknown parameter given a specific measurement, the QFI sets the ultimate precision limit for estimating a parameter encoded in the quantum state $\rho(\theta)$. A higher QFI corresponds to a lower estimation uncertainty.

\textit{The model---} We consider the dissipative quantum dynamics of a driven qubit-field system described by the master equation~\cite{carmichael2015breakdown}:
\begin{equation}
\frac{d \rho}{dt} = -\frac{i}{\hbar} \left[ H_{\text{int}}, \rho \right] + \kappa \left( 2a \rho a^\dagger - \rho a^\dagger a - a^\dagger a \rho \right), \label{eq_master_equation}
\end{equation}
where $a$ and $a^\dagger$ are the annihilation and creation operators of the field mode, $\kappa$ is the decay rate, and $H_{\text{int}}{=}T H_{\mathrm{DJC}} T^\dagger$ is the Hamiltonian in the interaction picture with the unitary transformation $T{=}\exp\left[i \Omega t \left( a^\dagger a{+}\sigma^+ \sigma^- \right) \right]$ switching the system to a frame rotating at the drive frequency $\Omega$. The lab-frame Hamiltonian $H_\mathrm{DJC}$ describes a driven Jaynes–Cummings interaction between a qubit and a field mode given by $\frac{H_\mathrm{DJC}}{\hbar}{=}\omega_f a^\dagger a{+}\omega_q \sigma^+ \sigma^-{+}g \left( \sigma^+ a + \sigma^- a^\dagger \right){+}\mathcal{E} \left( a e^{i\Omega t}{+}a^\dagger e^{-i\Omega t} \right)$, where $\omega_f$ is the field resonance frequency, $\omega_q$ is the qubit transition frequency, $g$ is the qubit-field coupling strength, $\mathcal{E}$ is the amplitude of the external drive, and $\sigma^+$, $\sigma^-$ are the qubit raising and lowering operators. Assuming $\omega_f = \omega_q = \omega$, the Hamiltonian simplifies to a time-independent form:
\begin{equation}
H_{\text{int}}{=}{-}\hbar\Delta(a^{\dagger}a{+}\sigma^+ \sigma^-){+}\hbar g\left( \sigma ^{\dagger}a{+}\sigma ^-a^{\dagger} \right) {+}\hbar \mathcal{E} \left( a{+}a^{\dagger} \right), \label{eq_Hint}
\end{equation}
where $\Delta{=}\Omega{-}\omega$ is the detuning between the drive frequency $\Omega$ and the subsystem frequency $\omega$. In this work, we focus on estimating $\mathcal{E}$ from the steady state $\rho_\mathrm{SS}$ of Eq.~\eqref{eq_master_equation}. Following Ref.~\cite{carmichael2015breakdown}, we take the ratio
\begin{equation}
N = \left(\frac{g}{2\kappa}\right)^2
\end{equation}
as a bona-fide sensing resource, which quantifies the competition between coherent interactions and incoherent losses (for a detailed discussion of the thermodynamic limit, see Ref.~\cite{carmichael2015breakdown}). The steady state $\rho_\mathrm{SS}$ is computed numerically by finding the zero eigenvalue of the Liouvillian using standard methods~\cite{lambert2024qutip5quantumtoolbox}.

\begin{figure}[t]
    \centering
    \includegraphics[width=\linewidth]{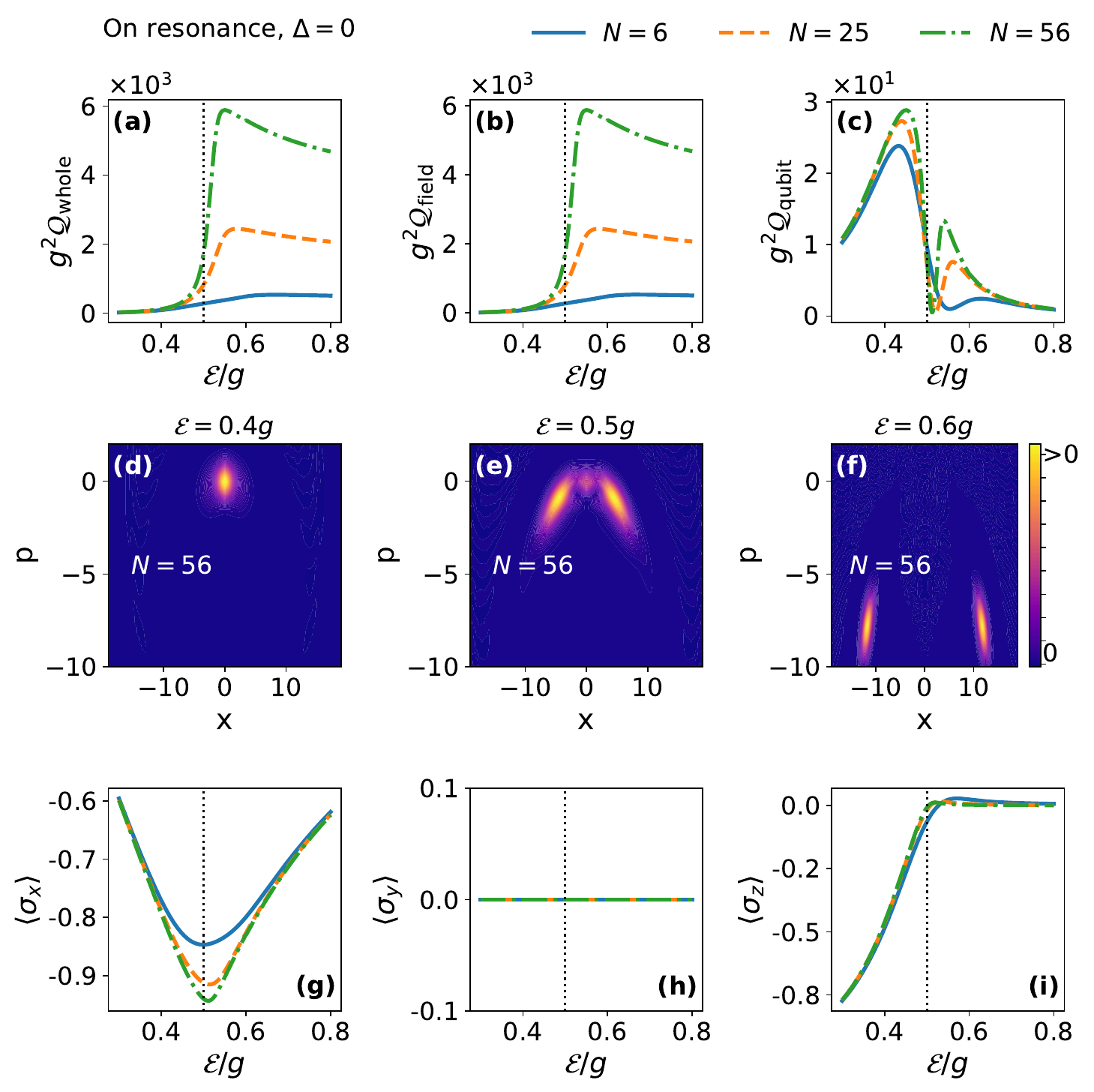}
    \caption{On resonance case, $\Delta=0$: (a)-(c) Quantum Fisher information (QFI) for the entire probe $Q_\mathrm{whole}$, the field subsystem $Q_\mathrm{field}$, and qubit subsystem $Q_\mathrm{qubit}$ as a function of $\mathcal{E}$ for some choices of $N$. (d)–(f) Wigner function of the field subsystem as a function of the phase-space variables $x$ and $p$ for $N=56$ and for several values of $\mathcal{E}/g$. (g)–(i) Expectation values of the qubit components $\langle\sigma_\alpha\rangle$ ($\alpha{=}x{,}y{,}z$) as a function of $\mathcal{E}/g$ and various values of the ratio $N$.}
    \label{fig_on_resonance}
\end{figure}

\textit{On resonance quantum-enhanced sensor---} We first focus on the sensing capabilities of the dissipative probe when it operates on resonance, $\Delta{=}0$ in Eq.~\eqref{eq_Hint}. In Figs.~\ref{fig_on_resonance} (a)-(c), we plot the QFI for the entire probe $Q_\mathrm{whole}$ as well as for the field (qubit) subsystem $Q_\mathrm{field}$ ($Q_\mathrm{qubit}$) obtained from the reduced steady state $\rho_\mathrm{SS}^{\text{field}}{=}\mathrm{Tr_\mathrm{qubit}[\rho_\mathrm{SS}]}$ ($\rho_\mathrm{SS}^{\text{qubit}}{=}\mathrm{Tr_\mathrm{field}[\rho_\mathrm{SS}]}$) as a function of $\mathcal{E}$ for selected values of $N$. Several qualitatively insights can be drawn from Figs.~\ref{fig_on_resonance} (a)-(c). First, all QFI functions increase as the sensing resource ratio $N$ becomes larger. This supports the choice of $N$ as sensing resource. Note that $g$ is taken as the reference frequency. Hence, we include a factor of $g^{2}$ so that the product $g^{2}\mathrm{QFI}$ remains invariant under changes in $N = \left( \frac{g}{2\kappa} \right)^{2}$---we set $\kappa{=}10^{-6}$ in our simulations. Second, the QFI of the field subsystem closely follows the behavior and values of the entire system, whereas the qubit subsystem does not. Hence, the field encodes significantly more information about $\mathcal{E}$ than the qubit which is also evident in the magnitude of the QFIs. Third, there is an abrupt change across all QFI functions near $2\mathcal{E}{\sim}g$. Fourth, the QFI exhibits a plateau with a moderate drop of it maximum value for parameters $2\mathcal{E}\gtrsim g$.
The above observed behavior can be explained by the fact that the system undergoes a well-established dissipative quantum phase transition at the critical point $2\mathcal{E}{=}g$, in the \textit{thermodynamic limit} where $N{\gg}1$~\cite{carmichael2015breakdown}. Specifically, for $2\mathcal{E}{<}g$, the quasienergy spectrum of $H_{\text{int}}$ is discrete with eigenvalues $E_0{=}0$ and doublets given by $E_{n,\pm}{=}\pm\sqrt{n}\hbar g \left[1{-}\left(\frac{2\mathcal{E}}{g}\right)^2 \right]^{3/4}$, with $n{\in}\mathbb{Z}_+$. Remarkably, at the critical point $2\mathcal{E}{=}g$, all quasienergies collapse to zero. For $2\mathcal{E}{>}g$, the discrete spectrum disappears and becomes continuous. Thus, the point $2\mathcal{E}{=}g$ marks and organizes a dissipative quantum phase transition~\cite{carmichael2015breakdown}. Note Figs.~\ref{fig_on_resonance} (a)-(c) address the key challenge of complex initial state preparation.

To gain deeper insight into the probe's dynamics, in Figs.~\ref{fig_on_resonance}(d)–(f) we plot the Wigner quasiprobability function of the field subsystem, defined as $W(x, p){=}\frac{1}{\pi\hbar}\int_{-\infty}^{\infty}\langle x{+}y| \rho_{\text{field}}|x{-}y\rangle e^{-2ipy/\hbar}dy$, as a function of the phase-space variables $x$ and $p$ for $N=56$ and various $\mathcal{E}/g$. As shown in Figs.~\ref{fig_on_resonance}(d)–(f), $W(x, p)$ remains strictly positive for all values of $\mathcal{E}/g$, implying the absence of nonclassical behavior. More importantly, the field subsystem undergoes an abrupt transition from a vacuum-like state to a quasiprobability distribution characterized by two well-separated peaks. This bistable behavior emerges around $2\mathcal{E}{\sim}g$, with the critical point $2\mathcal{E}{=}g$ in the thermodynamic limit $N{\to}\infty$ marking the sudden departure from the vacuum-like to the bistable state~\cite{carmichael2015breakdown}. In Figs.~\ref{fig_on_resonance}(g)–(i), we plot the expectation values of the qubit components, denoted as $\langle\sigma_\alpha\rangle$ for $\alpha{=}x{,}y{,}z$, as a function of $\mathcal{E}/g$ and several values of the ratio $N$. As shown in Figs.~\ref{fig_on_resonance}(g)–(i), $\langle\sigma_y\rangle$ is always zero---independent of the ratio $\mathcal{E}/g$, whereas the expected value along the $z$-axis transitions from non-zero when $2\mathcal{E}{<}g$ to zero when $2\mathcal{E}{>}g$. Along the $x$-axis, the qubit subsystem acquires a non-zero expected value across all values of $\mathcal{E}/g$. Notably, it approaches $\langle\sigma_x\rangle{\to}{-}1$ as $N{\gg}1$, signaling near-complete polarization in the $|-\rangle$ state. At the critical point $2\mathcal{E}=g$, in the thermodynamic limit, the dressed state becomes weakly entangled, that is the qubit subsystem spontaneously polarizes into the pure state $|+\rangle$, while the field subsystem becomes bistable---see Supplemental Material (SM)~\cite{SM} for entanglement and purity analysis.

\textit{On resonance scaling analysis---}
\begin{figure}
    \centering
    \includegraphics[width=\linewidth]{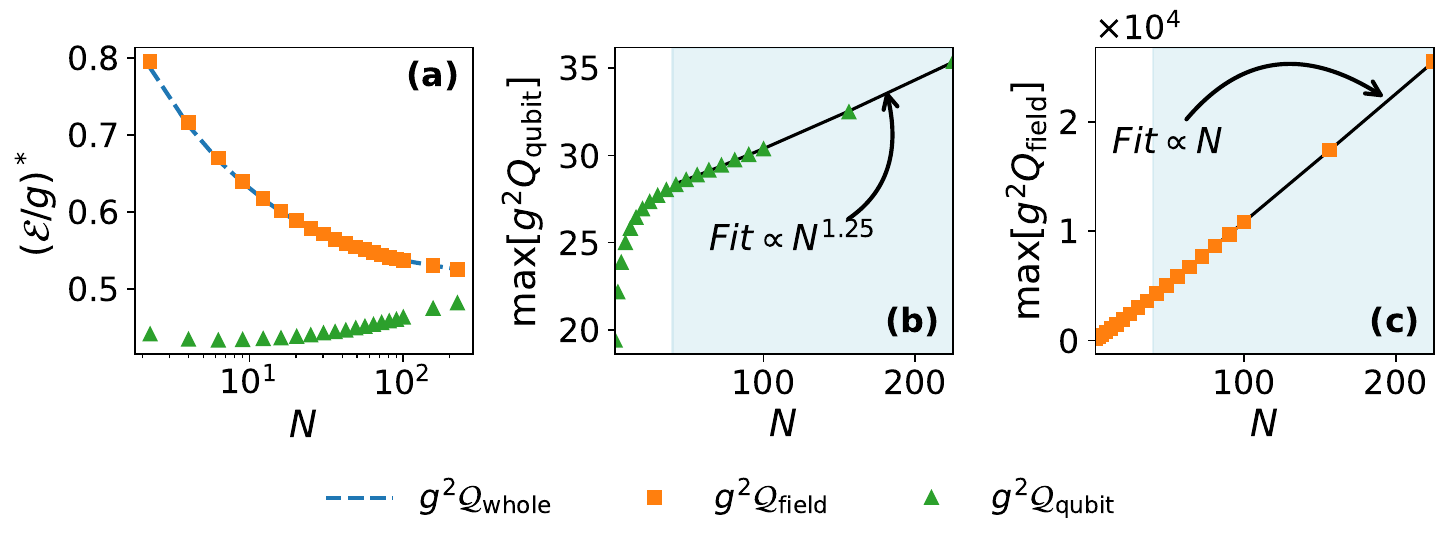}
    \caption{(a) Optimal driving amplitude $(\mathcal{E}{/}g)^*$ as a function of $N$. (b) Maximum QFI for the qubit subsystem as a function of $N$. (c) Maximum QFI as a function of the sensing resource $N$ for both the entire system and the field subsystem. In (b) and (c), the exponent $B$ obtained from fitting the data is explicitly shown in each plot.}
    \label{fig_scalings}
\end{figure}
We now analyze how the precision scales with the sensing resource $N$, considering both the entire system and its subsystems. To this end, we define the driving strength that yields the maximum QFI as
\begin{equation}
    \left(\frac{\mathcal{E}}{g}\right)^*_j=\arg\max_{\left\{\left(\frac{\mathcal{E}}{g}\right)\right\}}Q_j(\mathcal{E})\Big|_{\Delta = 0},
\end{equation}
where $j$ accounts for the whole, qubit, and field parties. In Fig.~\ref{fig_scalings}(a), we show how the optimal driving amplitude $(\mathcal{E}{/}g)^*$ varies with the sensing resource $N$ for the entire system and its subsystems. The optimal values $(\mathcal{E}/g)^*$ for the entire system and the field subsystem coincide across the full range of $N$, whereas the qubit subsystem exhibits a different trend. Notably, as $N$ increases, the optimal drive amplitude $(\mathcal{E}/g)^*$ for all three cases converges toward the value $1/2$, consistent with the critical point expected at resonance. In Fig.~\ref{fig_scalings}(b), we plot the maximum QFI for the qubit subsystem as a function of $N$. To capture this behavior, we fit the data using a function of the form $AN^B{+}C$. To exclude finite-size effects, we fit the curve using only data points with $N{>}20$, highlighted by the light blue shaded region. Beyond this point, the data follow a polynomial trend with a super-linear $B{\approx}1.25$ exponent, a clear evidence for quantum-enhanced sensitivity. In Fig.~\ref{fig_scalings}(c), we plot the maximum QFI as a function of the sensing resource $N$ for both the entire system and the field subsystem. The plot reveals a similar trend in both cases, characterized by clear polynomial growth. A fitting analysis yields a linear $B{\approx}1$ exponent, as shown in the main plot. 

\textit{Off resonance quantum-enhanced sensor---} Although the dissipative probe at resonance is robust and highly precise, it is natural to ask whether it is indeed the optimal probe. To address this, we now consider $\Delta{\neq}0$ in Eq.~\eqref{eq_Hint}. Specifically, we analyze the sensing precision, for estimating $\mathcal{E}$, at the optimal detuning $\left(\frac{\Delta}{g}\right)_j^*$ and optimal driving amplitude $\left(\frac{\mathcal{E}}{g}\right)_j^*$, which maximizes the QFI $Q_j$ for each $j{=}\text{whole{,}field{,}qubit}$ system. These optimal values are defined as:
\begin{equation}
\left[\left(\frac{\Delta}{g}\right)_j^*, \left(\frac{\mathcal{E}}{g}\right)_j^* \right]= \arg\max_{\left\{\frac{\Delta}{g},\frac{\mathcal{E}}{g}\right\}}Q_j(\mathcal{E},\Delta).\label{eq_opt_Deltas}
\end{equation}
For the purpose of this section, we focus on the detuning that maximizes $Q_\text{field}$ and refer to it simply as $\Delta{\neq}0$. All explicit values of $(\Delta/g)_j^*$ and $(\mathcal{E}/g)_j^*$ as shown in Eq.~\eqref{eq_opt_Deltas}, will be examined in later sections.

\begin{figure}[t]
    \centering
    \includegraphics[width=\linewidth]{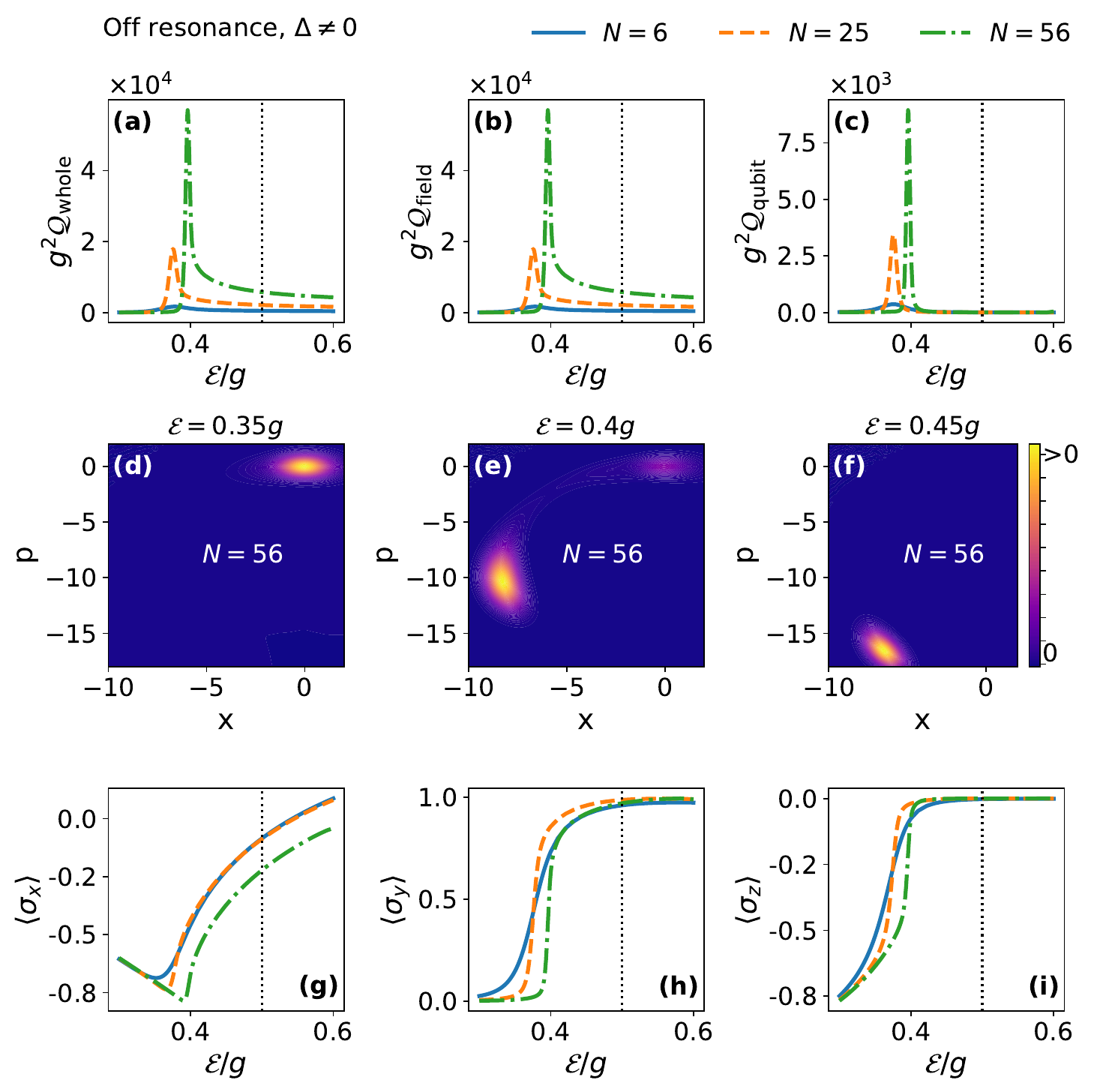}
    \caption{Off resonance case, $\Delta\neq 0$: QFI for the entire probe $Q_\mathrm{whole}$, the field subsystem $Q_\mathrm{field}$, and the qubit subsystem $Q_\mathrm{qubit}$ as a function of $\mathcal{E}$ for some choices of $N$. (d)-(f) Wigner quasiprobability function of the field subsystem in phase space for $N{=}56$ for several values of $\mathcal{E}{/}g$. (g)-(i) Expectation values of the qubit components $\langle \sigma_\alpha \rangle$ ($\alpha{=}x,y,z$) as a function of $\mathcal{E}/g$ for several choices of $N$.}
    \label{fig_off_resonance}
\end{figure}

In Figs.~\ref{fig_off_resonance}(a)-(c), we plot the QFI for the entire probe $Q_\mathrm{whole}$, the field subsystem $Q_\mathrm{field}$, and the qubit subsystem $Q_\mathrm{qubit}$ as a function of $\mathcal{E}{/}g$ for some choices of $N$. As shown in Figs.~\ref{fig_off_resonance}(a)–(c), the dissipative probe achieves higher sensitivity as $N$ increases, similar to the behavior observed in the on-resonance case. However, in contrast to the on-resonance results shown in Figs.~\ref{fig_on_resonance}(a)–(c), Figs.~\ref{fig_off_resonance}(a)–(c) reveal two stark differences: (i) a very sharp peak for all QFI functions emerges around $\mathcal{E}{/}g \sim 0.4$ and (ii) the QFI reaches significantly higher values---indicating improved precision---compared to the on-resonance case as $N{\gg}1$.

To better understand these differences, in Figs.~\ref{fig_off_resonance}(d)–(f) we show the Wigner quasiprobability distribution of the field subsystem in phase space for $N{=}56$ for driving values around $\mathcal{E}{/}g{\sim}0.4$. As seen from panels (d)–(f), the field transitions from a vacuum-like state for $\mathcal{E}{/}g{<}0.4$ to a strongly displaced state for $\mathcal{E}{/}g{>}0.4$. No signs of bistability, as seen in the on-resonance scenario, are observed. This indicates that, in the off-resonance case, the field exhibits a preferred localization in phase space. Note that choosing $\Delta{<}0$ preferentially selects the other bistable peak. To further analyze the system, Figs.~\ref{fig_off_resonance}(g)–(i) show the expectation values of the qubit components $\langle \sigma_\alpha \rangle$ ($\alpha{=}x,y,z$) as a function of $\mathcal{E}/g$ for several values of the sensing resource $N$. As shown in Figs.~\ref{fig_off_resonance}(g)–(i), increasing the sensing resource $N$ causes the qubit subsystem to become abruptly polarized in the XY-plane, along a superposition of the $|-\rangle$ and $|{+}i\rangle$ directions---$|\pm i\rangle=(|e\rangle\pm i|g\rangle)/\sqrt{2}$. The emergence of a nonzero component along the $y$-axis, due to $\Delta{\neq}0$, breaks the bistability and leads to the preferred localization of the field in phase space---choosing $\Delta{<}0$ (i.e., the right bistable peak) causes the qubit's $y$-component to acquire a phase, resulting in the state $|{-}i\rangle$. See SM~\cite{SM} for entanglement and purity details for this case.

\textit{Off resonance scaling analysis---}
\begin{figure}[t]
\centering\includegraphics[width=\linewidth]{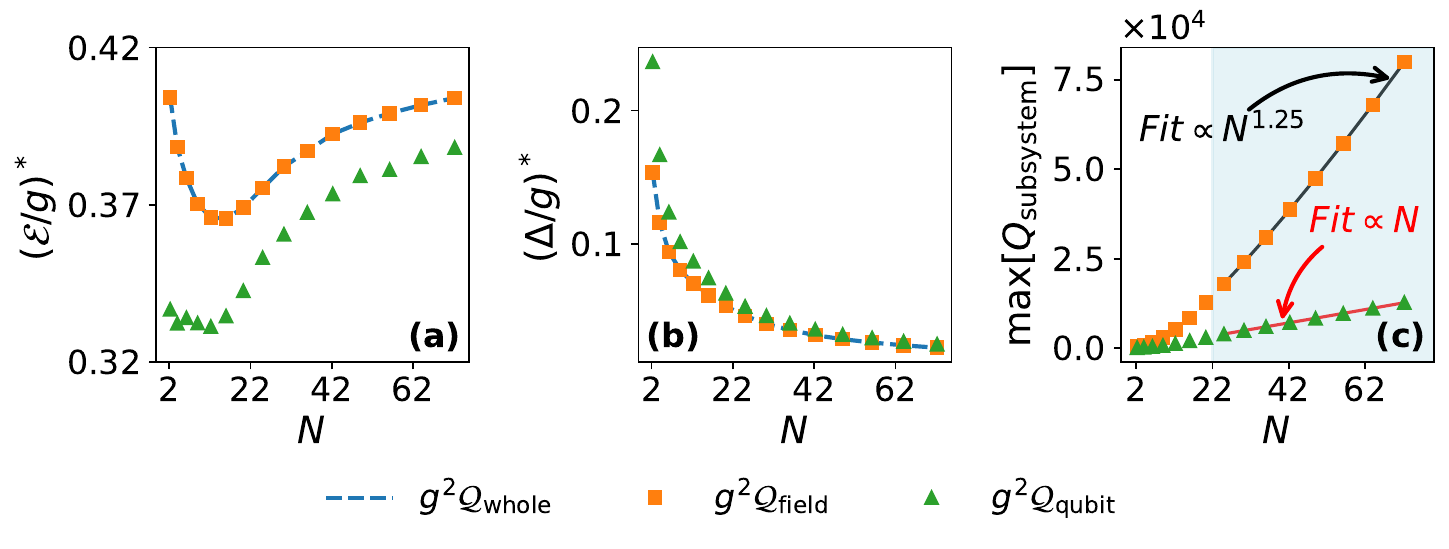}
    \caption{(a) Optimal driving amplitude $(\mathcal{E}/g)^*$ as a function of $N$. (b) Optimal detuning $(\Delta/g)^*$ as a function of the sensing resource $N$. (c) Fit of the optimized maximum QFI as a function of $N$.}
    \label{fig_scalings_off}
\end{figure}
In Fig.~\ref{fig_scalings_off}(a), we show how the optimal driving amplitude $(\mathcal{E}/g)^*$ varies with $N$. We consider $N{\gtrsim}20$ to exclude finite-size effects. As seen in the figure, a clear convergence toward $\mathcal{E}/g \sim 0.4$ is observed across subsystems. In Fig.~\ref{fig_scalings_off}(b), we plot the optimal detuning $(\Delta/g)^*$ as a function of the sensing resource $N$ for each subsystem. As $N$ increases, the differences between optimal detunings across subsystems become smaller. In Fig.~\ref{fig_scalings_off}(c), we show a fit of the maximum QFI---optimized over $(\Delta/g)^*$ and $(\mathcal{E}/g)^*$---using a polynomial of the form $AN^B{+}C$ as a function of $N$. We avoid finite-size effects by fitting the curve from $N>20$. As seen in the figure, the exponent $B$ can be either linear or super-linear, depending on the specific subsystem. Importantly, the off-resonance sensing performance significantly exceeds that of the on-resonance case, making it the optimal choice for quantum-enhanced sensing. 

\textit{Actual estimation---} The precision bounds discussed above demonstrate that the majority of the information about $\mathcal{E}$ is encoded in the field's degree of freedom. Thus, we focus on extracting the information about $\mathcal{E}$ from the field subsystem. To this end, we propose a feasible homodyne detection scheme combined with a Bayesian estimator. The full technical details of the homodyne (and heterodyne) measurement and estimation protocol are provided in the SM~\cite{SM}.
\begin{figure}
    \centering
    \includegraphics[width=\linewidth]{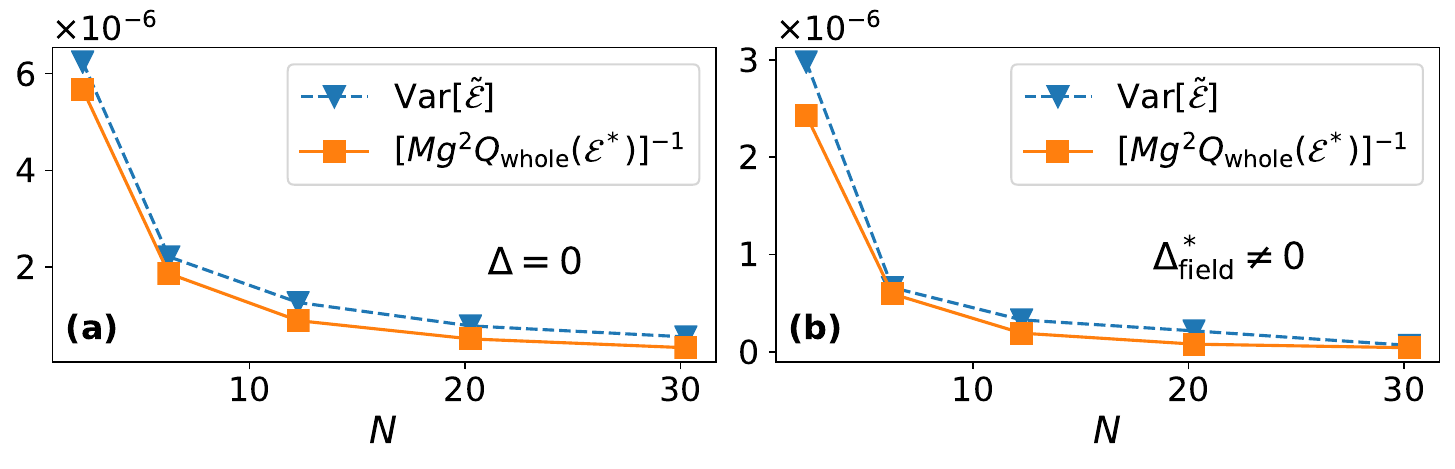}
    \caption{Estimator variance $\mathrm{Var}[\tilde{\mathcal{E}}]$ and inverse of the QFI $[Mg^2Q_\mathrm{whole}(\mathcal{E}^*)]^{-1}$ as a function of $N$. (a) on-resonance case. (b) off-resonance case.}
    \label{fig_bayesian}
\end{figure}

In Fig.~\ref{fig_bayesian}, we compare the estimator variance $\mathrm{Var}[\tilde{\mathcal{E}}]$ with the inverse of the global scaled QFI $(Mg^2Q_\mathrm{whole})^{-1}$ at the critical point $(\mathcal{E}{/}g)^*$ as a function of the ratio $N$. Panels (a) and (b) show the results for the on-resonance and off-resonance cases, respectively. Notably, as clearly seen from Figs.~\ref{fig_bayesian}(a)-(b), homodyne detection combined with Bayesian estimation nearly saturates the ultimate limit in precision quantified by the global quantum Cram\'{e}r-Rao bound, $\mathrm{Var}[\tilde{\mathcal{E}}]{\geq}[Mg^2Q_\mathrm{whole}\left( \mathcal{E}\right)]^{-1}$, when only partial access to the system is available. This near-ultimate performance can be attributed to two key factors: (i) the non-trivial qubit-field disentangling mechanisms at criticality, which effectively transfer most of the information to the field degree of freedom, and (ii) the fact that the field's phase-space quasiprobability distribution (see middle panels of Figs.~\ref{fig_on_resonance} and~\ref{fig_off_resonance}) remains positive and approximately Gaussian-shaped, making homodyne detection particularly effective in this scenario. 

\textit{Conclusions---} In this Letter, we address key challenges in quantum sensing, specifically: complex state initialization, demanding measurement requirements, and achieving quantum-enhanced sensitivity using only local access. To demonstrate our main findings, we investigate the precision limits of the driven-dissipative Jaynes–Cummings model undergoing a dissipative quantum phase transition. Defining the sensing resource as the interplay between coherent and incoherent processes, we report three main findings: First, detuning the dissipative probe off resonance significantly enhances sensing performance compared to the on-resonance case. Achieving quantum-enhanced sensitivity for the whole probe and the field subsystem. Our probe circumvents demanding state preparation requirements---such as those needed in ground-state metrology---by harnessing dissipative mechanisms. Second, analyzing each subsystem individually reveals that the field subsystem encodes nearly all the information about the unknown parameter. This overcomes the challenging requirement of accessing the entire system and instead allows us to probe only a part of the system. Finally, we show that homodyne detection combined with Bayesian estimation nearly saturates the ultimate precision bound set by the quantum Fisher information of the full system, despite accessing only part of it. This overcomes the difficulty of performing challenging measurements by relying instead on a straightforward measurement.

\textit{Acknowledgments---}  We acknowledge support from the National Natural Science Foundation of China Grants No. 12050410253, No. 92065115, No. 12274059, No. W2432005 and  No. 12374482.

\bibliography{Dissipative_v2}

\begin{thebibliography}{123}%
\makeatletter
\providecommand \@ifxundefined [1]{%
 \@ifx{#1\undefined}
}%
\providecommand \@ifnum [1]{%
 \ifnum #1\expandafter \@firstoftwo
 \else \expandafter \@secondoftwo
 \fi
}%
\providecommand \@ifx [1]{%
 \ifx #1\expandafter \@firstoftwo
 \else \expandafter \@secondoftwo
 \fi
}%
\providecommand \natexlab [1]{#1}%
\providecommand \enquote  [1]{``#1''}%
\providecommand \bibnamefont  [1]{#1}%
\providecommand \bibfnamefont [1]{#1}%
\providecommand \citenamefont [1]{#1}%
\providecommand \href@noop [0]{\@secondoftwo}%
\providecommand \href [0]{\begingroup \@sanitize@url \@href}%
\providecommand \@href[1]{\@@startlink{#1}\@@href}%
\providecommand \@@href[1]{\endgroup#1\@@endlink}%
\providecommand \@sanitize@url [0]{\catcode `\\12\catcode `\$12\catcode
  `\&12\catcode `\#12\catcode `\^12\catcode `\_12\catcode `\%12\relax}%
\providecommand \@@startlink[1]{}%
\providecommand \@@endlink[0]{}%
\providecommand \url  [0]{\begingroup\@sanitize@url \@url }%
\providecommand \@url [1]{\endgroup\@href {#1}{\urlprefix }}%
\providecommand \urlprefix  [0]{URL }%
\providecommand \Eprint [0]{\href }%
\providecommand \doibase [0]{http://dx.doi.org/}%
\providecommand \selectlanguage [0]{\@gobble}%
\providecommand \bibinfo  [0]{\@secondoftwo}%
\providecommand \bibfield  [0]{\@secondoftwo}%
\providecommand \translation [1]{[#1]}%
\providecommand \BibitemOpen [0]{}%
\providecommand \bibitemStop [0]{}%
\providecommand \bibitemNoStop [0]{.\EOS\space}%
\providecommand \EOS [0]{\spacefactor3000\relax}%
\providecommand \BibitemShut  [1]{\csname bibitem#1\endcsname}%
\let\auto@bib@innerbib\@empty
%</preamble>
\bibitem [{\citenamefont {Degen}\ \emph {et~al.}(2017)\citenamefont {Degen},
  \citenamefont {Reinhard},\ and\ \citenamefont
  {Cappellaro}}]{degen2017quantum}%
  \BibitemOpen
  \bibfield  {author} {\bibinfo {author} {\bibfnamefont {Christian~L}\
  \bibnamefont {Degen}}, \bibinfo {author} {\bibfnamefont {Friedemann}\
  \bibnamefont {Reinhard}}, \ and\ \bibinfo {author} {\bibfnamefont {Paola}\
  \bibnamefont {Cappellaro}},\ }\bibfield  {title} {\enquote {\bibinfo {title}
  {Quantum sensing},}\ }\href@noop {} {\bibfield  {journal} {\bibinfo
  {journal} {Reviews of Modern Physics}\ }\textbf {\bibinfo {volume} {89}},\
  \bibinfo {pages} {035002} (\bibinfo {year} {2017})}\BibitemShut {NoStop}%
\bibitem [{\citenamefont {Giovannetti}\ \emph {et~al.}(2006)\citenamefont
  {Giovannetti}, \citenamefont {Lloyd},\ and\ \citenamefont
  {Maccone}}]{giovannetti2006quantummetrology}%
  \BibitemOpen
  \bibfield  {author} {\bibinfo {author} {\bibfnamefont {Vittorio}\
  \bibnamefont {Giovannetti}}, \bibinfo {author} {\bibfnamefont {Seth}\
  \bibnamefont {Lloyd}}, \ and\ \bibinfo {author} {\bibfnamefont {Lorenzo}\
  \bibnamefont {Maccone}},\ }\bibfield  {title} {\enquote {\bibinfo {title}
  {Quantum metrology},}\ }\href {\doibase 10.1103/physrevlett.96.010401}
  {\bibfield  {journal} {\bibinfo  {journal} {Phys. Rev. Lett.}\ }\textbf
  {\bibinfo {volume} {96}},\ \bibinfo {pages} {010401} (\bibinfo {year}
  {2006})}\BibitemShut {NoStop}%
\bibitem [{\citenamefont {Giovannetti}\ \emph {et~al.}(2011)\citenamefont
  {Giovannetti}, \citenamefont {Lloyd},\ and\ \citenamefont
  {Maccone}}]{giovannetti2011advances}%
  \BibitemOpen
  \bibfield  {author} {\bibinfo {author} {\bibfnamefont {Vittorio}\
  \bibnamefont {Giovannetti}}, \bibinfo {author} {\bibfnamefont {Seth}\
  \bibnamefont {Lloyd}}, \ and\ \bibinfo {author} {\bibfnamefont {Lorenzo}\
  \bibnamefont {Maccone}},\ }\bibfield  {title} {\enquote {\bibinfo {title}
  {Advances in quantum metrology},}\ }\href@noop {} {\bibfield  {journal}
  {\bibinfo  {journal} {Nature photonics}\ }\textbf {\bibinfo {volume} {5}},\
  \bibinfo {pages} {222--229} (\bibinfo {year} {2011})}\BibitemShut {NoStop}%
\bibitem [{\citenamefont {Giovannetti}\ \emph {et~al.}(2004)\citenamefont
  {Giovannetti}, \citenamefont {Lloyd},\ and\ \citenamefont
  {Maccone}}]{giovannetti2004quantum}%
  \BibitemOpen
  \bibfield  {author} {\bibinfo {author} {\bibfnamefont {Vittorio}\
  \bibnamefont {Giovannetti}}, \bibinfo {author} {\bibfnamefont {Seth}\
  \bibnamefont {Lloyd}}, \ and\ \bibinfo {author} {\bibfnamefont {Lorenzo}\
  \bibnamefont {Maccone}},\ }\bibfield  {title} {\enquote {\bibinfo {title}
  {Quantum-enhanced measurements: beating the standard quantum limit},}\ }\href
  {https://www.science.org/doi/10.1126/science.1104149} {\bibfield  {journal}
  {\bibinfo  {journal} {Science}\ }\textbf {\bibinfo {volume} {306}},\ \bibinfo
  {pages} {1330--1336} (\bibinfo {year} {2004})}\BibitemShut {NoStop}%
\bibitem [{\citenamefont {Montenegro}\ \emph {et~al.}(2025)\citenamefont
  {Montenegro}, \citenamefont {Mukhopadhyay}, \citenamefont {Yousefjani},
  \citenamefont {Sarkar}, \citenamefont {Mishra}, \citenamefont {Paris},\ and\
  \citenamefont {Bayat}}]{montenegro2025review}%
  \BibitemOpen
  \bibfield  {author} {\bibinfo {author} {\bibfnamefont {Victor}\ \bibnamefont
  {Montenegro}}, \bibinfo {author} {\bibfnamefont {Chiranjib}\ \bibnamefont
  {Mukhopadhyay}}, \bibinfo {author} {\bibfnamefont {Rozhin}\ \bibnamefont
  {Yousefjani}}, \bibinfo {author} {\bibfnamefont {Saubhik}\ \bibnamefont
  {Sarkar}}, \bibinfo {author} {\bibfnamefont {Utkarsh}\ \bibnamefont
  {Mishra}}, \bibinfo {author} {\bibfnamefont {Matteo~G.A.}\ \bibnamefont
  {Paris}}, \ and\ \bibinfo {author} {\bibfnamefont {Abolfazl}\ \bibnamefont
  {Bayat}},\ }\bibfield  {title} {\enquote {\bibinfo {title} {Review: Quantum
  metrology and sensing with many-body systems},}\ }\href {\doibase
  https://doi.org/10.1016/j.physrep.2025.05.005} {\bibfield  {journal}
  {\bibinfo  {journal} {Physics Reports}\ }\textbf {\bibinfo {volume} {1134}},\
  \bibinfo {pages} {1--62} (\bibinfo {year} {2025})}\BibitemShut {NoStop}%
\bibitem [{\citenamefont {T{\'o}th}\ and\ \citenamefont
  {Apellaniz}(2014)}]{toth2014quantum}%
  \BibitemOpen
  \bibfield  {author} {\bibinfo {author} {\bibfnamefont {G{\'e}za}\
  \bibnamefont {T{\'o}th}}\ and\ \bibinfo {author} {\bibfnamefont {Iagoba}\
  \bibnamefont {Apellaniz}},\ }\bibfield  {title} {\enquote {\bibinfo {title}
  {Quantum metrology from a quantum information science perspective},}\
  }\href@noop {} {\bibfield  {journal} {\bibinfo  {journal} {Journal of Physics
  A: Mathematical and Theoretical}\ }\textbf {\bibinfo {volume} {47}},\
  \bibinfo {pages} {424006} (\bibinfo {year} {2014})}\BibitemShut {NoStop}%
\bibitem [{\citenamefont {Polino}\ \emph {et~al.}(2020)\citenamefont {Polino},
  \citenamefont {Valeri}, \citenamefont {Spagnolo},\ and\ \citenamefont
  {Sciarrino}}]{polino2020photonic}%
  \BibitemOpen
  \bibfield  {author} {\bibinfo {author} {\bibfnamefont {Emanuele}\
  \bibnamefont {Polino}}, \bibinfo {author} {\bibfnamefont {Mauro}\
  \bibnamefont {Valeri}}, \bibinfo {author} {\bibfnamefont {Nicol{\`o}}\
  \bibnamefont {Spagnolo}}, \ and\ \bibinfo {author} {\bibfnamefont {Fabio}\
  \bibnamefont {Sciarrino}},\ }\bibfield  {title} {\enquote {\bibinfo {title}
  {Photonic quantum metrology},}\ }\href@noop {} {\bibfield  {journal}
  {\bibinfo  {journal} {AVS Quantum Science}\ }\textbf {\bibinfo {volume} {2}}
  (\bibinfo {year} {2020})}\BibitemShut {NoStop}%
\bibitem [{\citenamefont {Mishra}\ and\ \citenamefont
  {Bayat}(2021{\natexlab{a}})}]{mishra2021driving}%
  \BibitemOpen
  \bibfield  {author} {\bibinfo {author} {\bibfnamefont {Utkarsh}\ \bibnamefont
  {Mishra}}\ and\ \bibinfo {author} {\bibfnamefont {Abolfazl}\ \bibnamefont
  {Bayat}},\ }\bibfield  {title} {\enquote {\bibinfo {title} {Driving enhanced
  quantum sensing in partially accessible many-body systems},}\ }\href@noop {}
  {\bibfield  {journal} {\bibinfo  {journal} {Physical Review Letters}\
  }\textbf {\bibinfo {volume} {127}},\ \bibinfo {pages} {080504} (\bibinfo
  {year} {2021}{\natexlab{a}})}\BibitemShut {NoStop}%
\bibitem [{\citenamefont {Mishra}\ and\ \citenamefont
  {Bayat}(2022)}]{mishra2022integrable}%
  \BibitemOpen
  \bibfield  {author} {\bibinfo {author} {\bibfnamefont {Utkarsh}\ \bibnamefont
  {Mishra}}\ and\ \bibinfo {author} {\bibfnamefont {Abolfazl}\ \bibnamefont
  {Bayat}},\ }\bibfield  {title} {\enquote {\bibinfo {title} {Integrable
  quantum many-body sensors for ac field sensing},}\ }\href@noop {} {\bibfield
  {journal} {\bibinfo  {journal} {Scientific Reports}\ }\textbf {\bibinfo
  {volume} {12}},\ \bibinfo {pages} {14760} (\bibinfo {year}
  {2022})}\BibitemShut {NoStop}%
\bibitem [{\citenamefont {Montenegro}\ \emph {et~al.}(2021)\citenamefont
  {Montenegro}, \citenamefont {Mishra},\ and\ \citenamefont
  {Bayat}}]{montenegro2021global}%
  \BibitemOpen
  \bibfield  {author} {\bibinfo {author} {\bibfnamefont {Victor}\ \bibnamefont
  {Montenegro}}, \bibinfo {author} {\bibfnamefont {Utkarsh}\ \bibnamefont
  {Mishra}}, \ and\ \bibinfo {author} {\bibfnamefont {Abolfazl}\ \bibnamefont
  {Bayat}},\ }\bibfield  {title} {\enquote {\bibinfo {title} {Global sensing
  and its impact for quantum many-body probes with criticality},}\ }\href@noop
  {} {\bibfield  {journal} {\bibinfo  {journal} {Physical Review Letters}\
  }\textbf {\bibinfo {volume} {126}},\ \bibinfo {pages} {200501} (\bibinfo
  {year} {2021})}\BibitemShut {NoStop}%
\bibitem [{\citenamefont {Marciniak}\ \emph {et~al.}(2022)\citenamefont
  {Marciniak}, \citenamefont {Feldker}, \citenamefont {Pogorelov},
  \citenamefont {Kaubruegger}, \citenamefont {Vasilyev}, \citenamefont {van
  Bijnen}, \citenamefont {Schindler}, \citenamefont {Zoller}, \citenamefont
  {Blatt},\ and\ \citenamefont {Monz}}]{marciniak2022optimal}%
  \BibitemOpen
  \bibfield  {author} {\bibinfo {author} {\bibfnamefont {Christian~D}\
  \bibnamefont {Marciniak}}, \bibinfo {author} {\bibfnamefont {Thomas}\
  \bibnamefont {Feldker}}, \bibinfo {author} {\bibfnamefont {Ivan}\
  \bibnamefont {Pogorelov}}, \bibinfo {author} {\bibfnamefont {Raphael}\
  \bibnamefont {Kaubruegger}}, \bibinfo {author} {\bibfnamefont {Denis~V}\
  \bibnamefont {Vasilyev}}, \bibinfo {author} {\bibfnamefont {Rick}\
  \bibnamefont {van Bijnen}}, \bibinfo {author} {\bibfnamefont {Philipp}\
  \bibnamefont {Schindler}}, \bibinfo {author} {\bibfnamefont {Peter}\
  \bibnamefont {Zoller}}, \bibinfo {author} {\bibfnamefont {Rainer}\
  \bibnamefont {Blatt}}, \ and\ \bibinfo {author} {\bibfnamefont {Thomas}\
  \bibnamefont {Monz}},\ }\bibfield  {title} {\enquote {\bibinfo {title}
  {Optimal metrology with programmable quantum sensors},}\ }\href {\doibase
  10.1038/s41586-022-04432-6} {\bibfield  {journal} {\bibinfo  {journal}
  {Nature}\ }\textbf {\bibinfo {volume} {603}},\ \bibinfo {pages} {604--609}
  (\bibinfo {year} {2022})}\BibitemShut {NoStop}%
\bibitem [{\citenamefont {He}\ \emph {et~al.}(2023)\citenamefont {He},
  \citenamefont {Yousefjani},\ and\ \citenamefont {Bayat}}]{he2023stark}%
  \BibitemOpen
  \bibfield  {author} {\bibinfo {author} {\bibfnamefont {Xingjian}\
  \bibnamefont {He}}, \bibinfo {author} {\bibfnamefont {Rozhin}\ \bibnamefont
  {Yousefjani}}, \ and\ \bibinfo {author} {\bibfnamefont {Abolfazl}\
  \bibnamefont {Bayat}},\ }\bibfield  {title} {\enquote {\bibinfo {title}
  {Stark localization as a resource for weak-field sensing with
  super-heisenberg precision},}\ }\href@noop {} {\bibfield  {journal} {\bibinfo
   {journal} {Physical Review Letters}\ }\textbf {\bibinfo {volume} {131}},\
  \bibinfo {pages} {010801} (\bibinfo {year} {2023})}\BibitemShut {NoStop}%
\bibitem [{\citenamefont {Mukhopadhyay}\ \emph {et~al.}(2025)\citenamefont
  {Mukhopadhyay}, \citenamefont {Montenegro},\ and\ \citenamefont
  {Bayat}}]{mukhopadhyay2025current}%
  \BibitemOpen
  \bibfield  {author} {\bibinfo {author} {\bibfnamefont {Chiranjib}\
  \bibnamefont {Mukhopadhyay}}, \bibinfo {author} {\bibfnamefont {Victor}\
  \bibnamefont {Montenegro}}, \ and\ \bibinfo {author} {\bibfnamefont
  {Abolfazl}\ \bibnamefont {Bayat}},\ }\bibfield  {title} {\enquote {\bibinfo
  {title} {Current trends in global quantum metrology},}\ }\href {\doibase
  10.1088/1751-8121/adb112} {\bibfield  {journal} {\bibinfo  {journal} {Journal
  of Physics A: Mathematical and Theoretical}\ }\textbf {\bibinfo {volume}
  {58}},\ \bibinfo {pages} {063001} (\bibinfo {year} {2025})}\BibitemShut
  {NoStop}%
\bibitem [{\citenamefont {Qvarfort}\ \emph {et~al.}(2018)\citenamefont
  {Qvarfort}, \citenamefont {Serafini}, \citenamefont {Barker},\ and\
  \citenamefont {Bose}}]{qvarfort2018gravimetry}%
  \BibitemOpen
  \bibfield  {author} {\bibinfo {author} {\bibfnamefont {Sofia}\ \bibnamefont
  {Qvarfort}}, \bibinfo {author} {\bibfnamefont {Alessio}\ \bibnamefont
  {Serafini}}, \bibinfo {author} {\bibfnamefont {P.~F.}\ \bibnamefont
  {Barker}}, \ and\ \bibinfo {author} {\bibfnamefont {Sougato}\ \bibnamefont
  {Bose}},\ }\bibfield  {title} {\enquote {\bibinfo {title} {Gravimetry through
  non-linear optomechanics},}\ }\href {\doibase 10.1038/s41467-018-06037-z}
  {\bibfield  {journal} {\bibinfo  {journal} {Nat. Commun.}\ }\textbf {\bibinfo
  {volume} {9}},\ \bibinfo {pages} {3690} (\bibinfo {year} {2018})}\BibitemShut
  {NoStop}%
\bibitem [{\citenamefont {Montenegro}(2025)}]{montenegro2025heisenberg}%
  \BibitemOpen
  \bibfield  {author} {\bibinfo {author} {\bibfnamefont {Victor}\ \bibnamefont
  {Montenegro}},\ }\bibfield  {title} {\enquote {\bibinfo {title}
  {Heisenberg-limited spin-mechanical gravimetry},}\ }\href {\doibase
  10.1103/PhysRevResearch.7.013016} {\bibfield  {journal} {\bibinfo  {journal}
  {Phys. Rev. Res.}\ }\textbf {\bibinfo {volume} {7}},\ \bibinfo {pages}
  {013016} (\bibinfo {year} {2025})}\BibitemShut {NoStop}%
\bibitem [{\citenamefont {Kwon}\ \emph {et~al.}(2020)\citenamefont {Kwon},
  \citenamefont {Yoon}, \citenamefont {Lee}, \citenamefont {Chen},
  \citenamefont {Liu}, \citenamefont {Schmid}, \citenamefont {Wu},
  \citenamefont {Choi},\ and\ \citenamefont {Won}}]{kwon2020magnetic}%
  \BibitemOpen
  \bibfield  {author} {\bibinfo {author} {\bibfnamefont {Hee~Young}\
  \bibnamefont {Kwon}}, \bibinfo {author} {\bibfnamefont {HG}~\bibnamefont
  {Yoon}}, \bibinfo {author} {\bibfnamefont {C}~\bibnamefont {Lee}}, \bibinfo
  {author} {\bibfnamefont {G}~\bibnamefont {Chen}}, \bibinfo {author}
  {\bibfnamefont {K}~\bibnamefont {Liu}}, \bibinfo {author} {\bibfnamefont
  {AK}~\bibnamefont {Schmid}}, \bibinfo {author} {\bibfnamefont
  {YZ}~\bibnamefont {Wu}}, \bibinfo {author} {\bibfnamefont {JW}~\bibnamefont
  {Choi}}, \ and\ \bibinfo {author} {\bibfnamefont {C}~\bibnamefont {Won}},\
  }\bibfield  {title} {\enquote {\bibinfo {title} {Magnetic hamiltonian
  parameter estimation using deep learning techniques},}\ }\href@noop {}
  {\bibfield  {journal} {\bibinfo  {journal} {Science advances}\ }\textbf
  {\bibinfo {volume} {6}},\ \bibinfo {pages} {eabb0872} (\bibinfo {year}
  {2020})}\BibitemShut {NoStop}%
\bibitem [{\citenamefont {Campos~Venuti}\ and\ \citenamefont
  {Zanardi}(2007)}]{campos2007quantum}%
  \BibitemOpen
  \bibfield  {author} {\bibinfo {author} {\bibfnamefont {Lorenzo}\ \bibnamefont
  {Campos~Venuti}}\ and\ \bibinfo {author} {\bibfnamefont {Paolo}\ \bibnamefont
  {Zanardi}},\ }\bibfield  {title} {\enquote {\bibinfo {title} {Quantum
  critical scaling of the geometric tensors},}\ }\href@noop {} {\bibfield
  {journal} {\bibinfo  {journal} {Physical Review Letters}\ }\textbf {\bibinfo
  {volume} {99}},\ \bibinfo {pages} {095701} (\bibinfo {year}
  {2007})}\BibitemShut {NoStop}%
\bibitem [{\citenamefont {Zanardi}\ \emph {et~al.}(2008)\citenamefont
  {Zanardi}, \citenamefont {Paris},\ and\ \citenamefont
  {Venuti}}]{zanardi2008quantum}%
  \BibitemOpen
  \bibfield  {author} {\bibinfo {author} {\bibfnamefont {Paolo}\ \bibnamefont
  {Zanardi}}, \bibinfo {author} {\bibfnamefont {Matteo~GA}\ \bibnamefont
  {Paris}}, \ and\ \bibinfo {author} {\bibfnamefont {Lorenzo~Campos}\
  \bibnamefont {Venuti}},\ }\bibfield  {title} {\enquote {\bibinfo {title}
  {Quantum criticality as a resource for quantum estimation},}\ }\href
  {https://link.aps.org/doi/10.1103/PhysRevA.78.042105} {\bibfield  {journal}
  {\bibinfo  {journal} {Phys. Rev. A}\ }\textbf {\bibinfo {volume} {78}},\
  \bibinfo {pages} {042105} (\bibinfo {year} {2008})}\BibitemShut {NoStop}%
\bibitem [{\citenamefont {Shabani}\ \emph {et~al.}(2011)\citenamefont
  {Shabani}, \citenamefont {Mohseni}, \citenamefont {Lloyd}, \citenamefont
  {Kosut},\ and\ \citenamefont {Rabitz}}]{shabani2011estimation}%
  \BibitemOpen
  \bibfield  {author} {\bibinfo {author} {\bibfnamefont {Alireza}\ \bibnamefont
  {Shabani}}, \bibinfo {author} {\bibfnamefont {Masoud}\ \bibnamefont
  {Mohseni}}, \bibinfo {author} {\bibfnamefont {Seth}\ \bibnamefont {Lloyd}},
  \bibinfo {author} {\bibfnamefont {Robert~L}\ \bibnamefont {Kosut}}, \ and\
  \bibinfo {author} {\bibfnamefont {Herschel}\ \bibnamefont {Rabitz}},\
  }\bibfield  {title} {\enquote {\bibinfo {title} {Estimation of many-body
  quantum hamiltonians via compressive sensing},}\ }\href@noop {} {\bibfield
  {journal} {\bibinfo  {journal} {Physical Review A—Atomic, Molecular, and
  Optical Physics}\ }\textbf {\bibinfo {volume} {84}},\ \bibinfo {pages}
  {012107} (\bibinfo {year} {2011})}\BibitemShut {NoStop}%
\bibitem [{\citenamefont {Dooley}(2021)}]{dooley2021robust}%
  \BibitemOpen
  \bibfield  {author} {\bibinfo {author} {\bibfnamefont {Shane}\ \bibnamefont
  {Dooley}},\ }\bibfield  {title} {\enquote {\bibinfo {title} {Robust quantum
  sensing in strongly interacting systems with many-body scars},}\ }\href@noop
  {} {\bibfield  {journal} {\bibinfo  {journal} {PRX Quantum}\ }\textbf
  {\bibinfo {volume} {2}},\ \bibinfo {pages} {020330} (\bibinfo {year}
  {2021})}\BibitemShut {NoStop}%
\bibitem [{\citenamefont {Sarkar}\ \emph {et~al.}(2022)\citenamefont {Sarkar},
  \citenamefont {Mukhopadhyay}, \citenamefont {Alase},\ and\ \citenamefont
  {Bayat}}]{sarkar2022free}%
  \BibitemOpen
  \bibfield  {author} {\bibinfo {author} {\bibfnamefont {Saubhik}\ \bibnamefont
  {Sarkar}}, \bibinfo {author} {\bibfnamefont {Chiranjib}\ \bibnamefont
  {Mukhopadhyay}}, \bibinfo {author} {\bibfnamefont {Abhijeet}\ \bibnamefont
  {Alase}}, \ and\ \bibinfo {author} {\bibfnamefont {Abolfazl}\ \bibnamefont
  {Bayat}},\ }\bibfield  {title} {\enquote {\bibinfo {title} {Free-fermionic
  topological quantum sensors},}\ }\href@noop {} {\bibfield  {journal}
  {\bibinfo  {journal} {Physical Review Letters}\ }\textbf {\bibinfo {volume}
  {129}},\ \bibinfo {pages} {090503} (\bibinfo {year} {2022})}\BibitemShut
  {NoStop}%
\bibitem [{\citenamefont {Montenegro}\ \emph
  {et~al.}(2022{\natexlab{a}})\citenamefont {Montenegro}, \citenamefont
  {Genoni}, \citenamefont {Bayat},\ and\ \citenamefont
  {Paris}}]{montenegro2022probing}%
  \BibitemOpen
  \bibfield  {author} {\bibinfo {author} {\bibfnamefont {V.}~\bibnamefont
  {Montenegro}}, \bibinfo {author} {\bibfnamefont {M.~G.}\ \bibnamefont
  {Genoni}}, \bibinfo {author} {\bibfnamefont {A.}~\bibnamefont {Bayat}}, \
  and\ \bibinfo {author} {\bibfnamefont {M.~G.~A.}\ \bibnamefont {Paris}},\
  }\bibfield  {title} {\enquote {\bibinfo {title} {Probing of nonlinear hybrid
  optomechanical systems via partial accessibility},}\ }\href {\doibase
  10.1103/PhysRevResearch.4.033036} {\bibfield  {journal} {\bibinfo  {journal}
  {Phys. Rev. Res.}\ }\textbf {\bibinfo {volume} {4}},\ \bibinfo {pages}
  {033036} (\bibinfo {year} {2022}{\natexlab{a}})}\BibitemShut {NoStop}%
\bibitem [{\citenamefont {Mukhopadhyay}\ and\ \citenamefont
  {Bayat}(2024{\natexlab{a}})}]{mukhopadhyay2024modular}%
  \BibitemOpen
  \bibfield  {author} {\bibinfo {author} {\bibfnamefont {Chiranjib}\
  \bibnamefont {Mukhopadhyay}}\ and\ \bibinfo {author} {\bibfnamefont
  {Abolfazl}\ \bibnamefont {Bayat}},\ }\bibfield  {title} {\enquote {\bibinfo
  {title} {Modular many-body quantum sensors},}\ }\href@noop {} {\bibfield
  {journal} {\bibinfo  {journal} {Physical Review Letters}\ }\textbf {\bibinfo
  {volume} {133}},\ \bibinfo {pages} {120601} (\bibinfo {year}
  {2024}{\natexlab{a}})}\BibitemShut {NoStop}%
\bibitem [{\citenamefont {Sahoo}\ \emph {et~al.}(2024)\citenamefont {Sahoo},
  \citenamefont {Mishra},\ and\ \citenamefont
  {Rakshit}}]{sahoo2024localization}%
  \BibitemOpen
  \bibfield  {author} {\bibinfo {author} {\bibfnamefont {Ayan}\ \bibnamefont
  {Sahoo}}, \bibinfo {author} {\bibfnamefont {Utkarsh}\ \bibnamefont {Mishra}},
  \ and\ \bibinfo {author} {\bibfnamefont {Debraj}\ \bibnamefont {Rakshit}},\
  }\bibfield  {title} {\enquote {\bibinfo {title} {Localization-driven quantum
  sensing},}\ }\href@noop {} {\bibfield  {journal} {\bibinfo  {journal}
  {Physical Review A}\ }\textbf {\bibinfo {volume} {109}},\ \bibinfo {pages}
  {L030601} (\bibinfo {year} {2024})}\BibitemShut {NoStop}%
\bibitem [{\citenamefont {Ye}\ and\ \citenamefont
  {Zoller}(2024)}]{ye2024essay}%
  \BibitemOpen
  \bibfield  {author} {\bibinfo {author} {\bibfnamefont {Jun}\ \bibnamefont
  {Ye}}\ and\ \bibinfo {author} {\bibfnamefont {Peter}\ \bibnamefont
  {Zoller}},\ }\bibfield  {title} {\enquote {\bibinfo {title} {Essay: Quantum
  sensing with atomic, molecular, and optical platforms for fundamental
  physics},}\ }\href@noop {} {\bibfield  {journal} {\bibinfo  {journal}
  {Physical Review Letters}\ }\textbf {\bibinfo {volume} {132}},\ \bibinfo
  {pages} {190001} (\bibinfo {year} {2024})}\BibitemShut {NoStop}%
\bibitem [{\citenamefont {Zhang}\ \emph {et~al.}(2025)\citenamefont {Zhang},
  \citenamefont {Xu}, \citenamefont {Hu},\ and\ \citenamefont
  {Qiu}}]{zhang2025quantum}%
  \BibitemOpen
  \bibfield  {author} {\bibinfo {author} {\bibfnamefont {Tao}\ \bibnamefont
  {Zhang}}, \bibinfo {author} {\bibfnamefont {Peng}\ \bibnamefont {Xu}},
  \bibinfo {author} {\bibfnamefont {Jiazhong}\ \bibnamefont {Hu}}, \ and\
  \bibinfo {author} {\bibfnamefont {Xingze}\ \bibnamefont {Qiu}},\ }\bibfield
  {title} {\enquote {\bibinfo {title} {Quantum sensing with topological-paired
  bound states},}\ }\href@noop {} {\bibfield  {journal} {\bibinfo  {journal}
  {New Journal of Physics}\ } (\bibinfo {year} {2025})}\BibitemShut {NoStop}%
\bibitem [{\citenamefont {Burgarth}\ and\ \citenamefont
  {Ajoy}(2017)}]{burgarth2017evolution}%
  \BibitemOpen
  \bibfield  {author} {\bibinfo {author} {\bibfnamefont {Daniel}\ \bibnamefont
  {Burgarth}}\ and\ \bibinfo {author} {\bibfnamefont {Ashok}\ \bibnamefont
  {Ajoy}},\ }\bibfield  {title} {\enquote {\bibinfo {title} {Evolution-free
  hamiltonian parameter estimation through zeeman markers},}\ }\href@noop {}
  {\bibfield  {journal} {\bibinfo  {journal} {Physical Review Letters}\
  }\textbf {\bibinfo {volume} {119}},\ \bibinfo {pages} {030402} (\bibinfo
  {year} {2017})}\BibitemShut {NoStop}%
\bibitem [{\citenamefont {Mehboudi}\ \emph {et~al.}(2019)\citenamefont
  {Mehboudi}, \citenamefont {Sanpera},\ and\ \citenamefont
  {Correa}}]{Mehboudi_2019}%
  \BibitemOpen
  \bibfield  {author} {\bibinfo {author} {\bibfnamefont {Mohammad}\
  \bibnamefont {Mehboudi}}, \bibinfo {author} {\bibfnamefont {Anna}\
  \bibnamefont {Sanpera}}, \ and\ \bibinfo {author} {\bibfnamefont {Luis~A}\
  \bibnamefont {Correa}},\ }\bibfield  {title} {\enquote {\bibinfo {title}
  {Thermometry in the quantum regime: recent theoretical progress},}\ }\href
  {\doibase 10.1088/1751-8121/ab2828} {\bibfield  {journal} {\bibinfo
  {journal} {Journal of Physics A: Mathematical and Theoretical}\ }\textbf
  {\bibinfo {volume} {52}},\ \bibinfo {pages} {303001} (\bibinfo {year}
  {2019})}\BibitemShut {NoStop}%
\bibitem [{\citenamefont {Razavian}\ \emph {et~al.}(2019)\citenamefont
  {Razavian}, \citenamefont {Benedetti}, \citenamefont {Bina}, \citenamefont
  {Akbari-Kourbolagh},\ and\ \citenamefont {Paris}}]{Razavian2019}%
  \BibitemOpen
  \bibfield  {author} {\bibinfo {author} {\bibfnamefont {Sholeh}\ \bibnamefont
  {Razavian}}, \bibinfo {author} {\bibfnamefont {Claudia}\ \bibnamefont
  {Benedetti}}, \bibinfo {author} {\bibfnamefont {Matteo}\ \bibnamefont
  {Bina}}, \bibinfo {author} {\bibfnamefont {Yahya}\ \bibnamefont
  {Akbari-Kourbolagh}}, \ and\ \bibinfo {author} {\bibfnamefont {Matteo G.~A.}\
  \bibnamefont {Paris}},\ }\bibfield  {title} {\enquote {\bibinfo {title}
  {Quantum thermometry by single-qubit dephasing},}\ }\href {\doibase
  10.1140/epjp/i2019-12708-9} {\bibfield  {journal} {\bibinfo  {journal} {The
  European Physical Journal Plus}\ }\textbf {\bibinfo {volume} {134}},\
  \bibinfo {pages} {284} (\bibinfo {year} {2019})}\BibitemShut {NoStop}%
\bibitem [{\citenamefont {Mihailescu}\ \emph {et~al.}(2023)\citenamefont
  {Mihailescu}, \citenamefont {Campbell},\ and\ \citenamefont
  {Mitchell}}]{mihailescu2023thermometry}%
  \BibitemOpen
  \bibfield  {author} {\bibinfo {author} {\bibfnamefont {George}\ \bibnamefont
  {Mihailescu}}, \bibinfo {author} {\bibfnamefont {Steve}\ \bibnamefont
  {Campbell}}, \ and\ \bibinfo {author} {\bibfnamefont {Andrew~K}\ \bibnamefont
  {Mitchell}},\ }\bibfield  {title} {\enquote {\bibinfo {title} {Thermometry of
  strongly correlated fermionic quantum systems using impurity probes},}\
  }\href@noop {} {\bibfield  {journal} {\bibinfo  {journal} {Physical Review
  A}\ }\textbf {\bibinfo {volume} {107}},\ \bibinfo {pages} {042614} (\bibinfo
  {year} {2023})}\BibitemShut {NoStop}%
\bibitem [{\citenamefont {Montenegro}\ \emph {et~al.}(2020)\citenamefont
  {Montenegro}, \citenamefont {Genoni}, \citenamefont {Bayat},\ and\
  \citenamefont {Paris}}]{montenegro2020mechanical}%
  \BibitemOpen
  \bibfield  {author} {\bibinfo {author} {\bibfnamefont {V.}~\bibnamefont
  {Montenegro}}, \bibinfo {author} {\bibfnamefont {M.~G.}\ \bibnamefont
  {Genoni}}, \bibinfo {author} {\bibfnamefont {A.}~\bibnamefont {Bayat}}, \
  and\ \bibinfo {author} {\bibfnamefont {M.~G.~A.}\ \bibnamefont {Paris}},\
  }\bibfield  {title} {\enquote {\bibinfo {title} {Mechanical oscillator
  thermometry in the nonlinear optomechanical regime},}\ }\href {\doibase
  10.1103/PhysRevResearch.2.043338} {\bibfield  {journal} {\bibinfo  {journal}
  {Phys. Rev. Res.}\ }\textbf {\bibinfo {volume} {2}},\ \bibinfo {pages}
  {043338} (\bibinfo {year} {2020})}\BibitemShut {NoStop}%
\bibitem [{\citenamefont {Srivastava}\ \emph {et~al.}(2023)\citenamefont
  {Srivastava}, \citenamefont {Bhattacharya}, \citenamefont {Lewenstein},\ and\
  \citenamefont {Płodzień}}]{srivastava2023topological}%
  \BibitemOpen
  \bibfield  {author} {\bibinfo {author} {\bibfnamefont {Anubhav~Kumar}\
  \bibnamefont {Srivastava}}, \bibinfo {author} {\bibfnamefont {Utso}\
  \bibnamefont {Bhattacharya}}, \bibinfo {author} {\bibfnamefont {Maciej}\
  \bibnamefont {Lewenstein}}, \ and\ \bibinfo {author} {\bibfnamefont {Marcin}\
  \bibnamefont {Płodzień}},\ }\href@noop {} {\enquote {\bibinfo {title}
  {Topological quantum thermometry},}\ } (\bibinfo {year} {2023}),\ \Eprint
  {http://arxiv.org/abs/2311.14524} {arXiv:2311.14524 [quant-ph]} \BibitemShut
  {NoStop}%
\bibitem [{\citenamefont {Sekatski}\ and\ \citenamefont
  {Perarnau-Llobet}(2022)}]{Sekatski_2022}%
  \BibitemOpen
  \bibfield  {author} {\bibinfo {author} {\bibfnamefont {Pavel}\ \bibnamefont
  {Sekatski}}\ and\ \bibinfo {author} {\bibfnamefont {Martí}\ \bibnamefont
  {Perarnau-Llobet}},\ }\bibfield  {title} {\enquote {\bibinfo {title} {Optimal
  nonequilibrium thermometry in markovian environments},}\ }\href {\doibase
  10.22331/q-2022-12-07-869} {\bibfield  {journal} {\bibinfo  {journal}
  {Quantum}\ }\textbf {\bibinfo {volume} {6}},\ \bibinfo {pages} {869}
  (\bibinfo {year} {2022})}\BibitemShut {NoStop}%
\bibitem [{\citenamefont {Glatthard}\ \emph {et~al.}(2022)\citenamefont
  {Glatthard}, \citenamefont {Rubio}, \citenamefont {Sawant}, \citenamefont
  {Hewitt}, \citenamefont {Barontini},\ and\ \citenamefont
  {Correa}}]{glatthard2022optimal}%
  \BibitemOpen
  \bibfield  {author} {\bibinfo {author} {\bibfnamefont {Jonas}\ \bibnamefont
  {Glatthard}}, \bibinfo {author} {\bibfnamefont {Jes\'us}\ \bibnamefont
  {Rubio}}, \bibinfo {author} {\bibfnamefont {Rahul}\ \bibnamefont {Sawant}},
  \bibinfo {author} {\bibfnamefont {Thomas}\ \bibnamefont {Hewitt}}, \bibinfo
  {author} {\bibfnamefont {Giovanni}\ \bibnamefont {Barontini}}, \ and\
  \bibinfo {author} {\bibfnamefont {Luis~A.}\ \bibnamefont {Correa}},\
  }\bibfield  {title} {\enquote {\bibinfo {title} {Optimal cold atom
  thermometry using adaptive bayesian strategies},}\ }\href {\doibase
  10.1103/PRXQuantum.3.040330} {\bibfield  {journal} {\bibinfo  {journal} {PRX
  Quantum}\ }\textbf {\bibinfo {volume} {3}},\ \bibinfo {pages} {040330}
  (\bibinfo {year} {2022})}\BibitemShut {NoStop}%
\bibitem [{\citenamefont {Mok}\ \emph {et~al.}(2021)\citenamefont {Mok},
  \citenamefont {Bharti}, \citenamefont {Kwek},\ and\ \citenamefont
  {Bayat}}]{mok2021optimal}%
  \BibitemOpen
  \bibfield  {author} {\bibinfo {author} {\bibfnamefont {Wai-Keong}\
  \bibnamefont {Mok}}, \bibinfo {author} {\bibfnamefont {Kishor}\ \bibnamefont
  {Bharti}}, \bibinfo {author} {\bibfnamefont {Leong-Chuan}\ \bibnamefont
  {Kwek}}, \ and\ \bibinfo {author} {\bibfnamefont {Abolfazl}\ \bibnamefont
  {Bayat}},\ }\bibfield  {title} {\enquote {\bibinfo {title} {Optimal probes
  for global quantum thermometry},}\ }\href
  {https://www.nature.com/articles/s42005-021-00572-w} {\bibfield  {journal}
  {\bibinfo  {journal} {Commun. Phys.}\ }\textbf {\bibinfo {volume} {4}},\
  \bibinfo {pages} {62} (\bibinfo {year} {2021})}\BibitemShut {NoStop}%
\bibitem [{\citenamefont {Rubio}\ \emph {et~al.}(2021)\citenamefont {Rubio},
  \citenamefont {Anders},\ and\ \citenamefont {Correa}}]{rubio2021global}%
  \BibitemOpen
  \bibfield  {author} {\bibinfo {author} {\bibfnamefont {Jes{\'u}s}\
  \bibnamefont {Rubio}}, \bibinfo {author} {\bibfnamefont {Janet}\ \bibnamefont
  {Anders}}, \ and\ \bibinfo {author} {\bibfnamefont {Luis~A}\ \bibnamefont
  {Correa}},\ }\bibfield  {title} {\enquote {\bibinfo {title} {Global quantum
  thermometry},}\ }\href@noop {} {\bibfield  {journal} {\bibinfo  {journal}
  {Physical Review Letters}\ }\textbf {\bibinfo {volume} {127}},\ \bibinfo
  {pages} {190402} (\bibinfo {year} {2021})}\BibitemShut {NoStop}%
\bibitem [{\citenamefont {Campbell}\ \emph {et~al.}(2017)\citenamefont
  {Campbell}, \citenamefont {Mehboudi}, \citenamefont {Chiara},\ and\
  \citenamefont {Paternostro}}]{campbell2017global}%
  \BibitemOpen
  \bibfield  {author} {\bibinfo {author} {\bibfnamefont {Steve}\ \bibnamefont
  {Campbell}}, \bibinfo {author} {\bibfnamefont {Mohammad}\ \bibnamefont
  {Mehboudi}}, \bibinfo {author} {\bibfnamefont {Gabriele~De}\ \bibnamefont
  {Chiara}}, \ and\ \bibinfo {author} {\bibfnamefont {Mauro}\ \bibnamefont
  {Paternostro}},\ }\bibfield  {title} {\enquote {\bibinfo {title} {Global and
  local thermometry schemes in coupled quantum systems},}\ }\href {\doibase
  10.1088/1367-2630/aa7fac} {\bibfield  {journal} {\bibinfo  {journal} {New
  Journal of Physics}\ }\textbf {\bibinfo {volume} {19}},\ \bibinfo {pages}
  {103003} (\bibinfo {year} {2017})}\BibitemShut {NoStop}%
\bibitem [{\citenamefont {Hou}\ \emph {et~al.}(2019)\citenamefont {Hou},
  \citenamefont {Wang}, \citenamefont {Tang}, \citenamefont {Yuan},
  \citenamefont {Xiang}, \citenamefont {Li},\ and\ \citenamefont
  {Guo}}]{hou2019control}%
  \BibitemOpen
  \bibfield  {author} {\bibinfo {author} {\bibfnamefont {Zhibo}\ \bibnamefont
  {Hou}}, \bibinfo {author} {\bibfnamefont {Rui-Jia}\ \bibnamefont {Wang}},
  \bibinfo {author} {\bibfnamefont {Jun-Feng}\ \bibnamefont {Tang}}, \bibinfo
  {author} {\bibfnamefont {Haidong}\ \bibnamefont {Yuan}}, \bibinfo {author}
  {\bibfnamefont {Guo-Yong}\ \bibnamefont {Xiang}}, \bibinfo {author}
  {\bibfnamefont {Chuan-Feng}\ \bibnamefont {Li}}, \ and\ \bibinfo {author}
  {\bibfnamefont {Guang-Can}\ \bibnamefont {Guo}},\ }\bibfield  {title}
  {\enquote {\bibinfo {title} {Control-enhanced sequential scheme for general
  quantum parameter estimation at the heisenberg limit},}\ }\href@noop {}
  {\bibfield  {journal} {\bibinfo  {journal} {Phys. Rev. Lett.}\ }\textbf
  {\bibinfo {volume} {123}},\ \bibinfo {pages} {040501} (\bibinfo {year}
  {2019})}\BibitemShut {NoStop}%
\bibitem [{\citenamefont {Liu}\ and\ \citenamefont
  {Yuan}(2017)}]{liu2017quantum}%
  \BibitemOpen
  \bibfield  {author} {\bibinfo {author} {\bibfnamefont {Jing}\ \bibnamefont
  {Liu}}\ and\ \bibinfo {author} {\bibfnamefont {Haidong}\ \bibnamefont
  {Yuan}},\ }\bibfield  {title} {\enquote {\bibinfo {title} {Quantum parameter
  estimation with optimal control},}\ }\href {\doibase
  10.1103/PhysRevA.96.012117} {\bibfield  {journal} {\bibinfo  {journal} {Phys.
  Rev. A}\ }\textbf {\bibinfo {volume} {96}},\ \bibinfo {pages} {012117}
  (\bibinfo {year} {2017})}\BibitemShut {NoStop}%
\bibitem [{\citenamefont {Hentschel}\ and\ \citenamefont
  {Sanders}(2011)}]{hentschel2011efficient}%
  \BibitemOpen
  \bibfield  {author} {\bibinfo {author} {\bibfnamefont {Alexander}\
  \bibnamefont {Hentschel}}\ and\ \bibinfo {author} {\bibfnamefont {Barry~C.}\
  \bibnamefont {Sanders}},\ }\bibfield  {title} {\enquote {\bibinfo {title}
  {Efficient algorithm for optimizing adaptive quantum metrology processes},}\
  }\href {\doibase 10.1103/PhysRevLett.107.233601} {\bibfield  {journal}
  {\bibinfo  {journal} {Phys. Rev. Lett.}\ }\textbf {\bibinfo {volume} {107}},\
  \bibinfo {pages} {233601} (\bibinfo {year} {2011})}\BibitemShut {NoStop}%
\bibitem [{\citenamefont {Cimini}\ \emph {et~al.}(2019)\citenamefont {Cimini},
  \citenamefont {Gianani}, \citenamefont {Spagnolo}, \citenamefont {Leccese},
  \citenamefont {Sciarrino},\ and\ \citenamefont
  {Barbieri}}]{cimini2019calibration}%
  \BibitemOpen
  \bibfield  {author} {\bibinfo {author} {\bibfnamefont {Valeria}\ \bibnamefont
  {Cimini}}, \bibinfo {author} {\bibfnamefont {Ilaria}\ \bibnamefont
  {Gianani}}, \bibinfo {author} {\bibfnamefont {Nicol\`o}\ \bibnamefont
  {Spagnolo}}, \bibinfo {author} {\bibfnamefont {Fabio}\ \bibnamefont
  {Leccese}}, \bibinfo {author} {\bibfnamefont {Fabio}\ \bibnamefont
  {Sciarrino}}, \ and\ \bibinfo {author} {\bibfnamefont {Marco}\ \bibnamefont
  {Barbieri}},\ }\bibfield  {title} {\enquote {\bibinfo {title} {Calibration of
  quantum sensors by neural networks},}\ }\href {\doibase
  10.1103/PhysRevLett.123.230502} {\bibfield  {journal} {\bibinfo  {journal}
  {Phys. Rev. Lett.}\ }\textbf {\bibinfo {volume} {123}},\ \bibinfo {pages}
  {230502} (\bibinfo {year} {2019})}\BibitemShut {NoStop}%
\bibitem [{\citenamefont {Yuan}\ and\ \citenamefont
  {Fung}(2015)}]{yuan2015optimal}%
  \BibitemOpen
  \bibfield  {author} {\bibinfo {author} {\bibfnamefont {Haidong}\ \bibnamefont
  {Yuan}}\ and\ \bibinfo {author} {\bibfnamefont {Chi-Hang~Fred}\ \bibnamefont
  {Fung}},\ }\bibfield  {title} {\enquote {\bibinfo {title} {Optimal feedback
  scheme and universal time scaling for hamiltonian parameter estimation},}\
  }\href {\doibase 10.1103/PhysRevLett.115.110401} {\bibfield  {journal}
  {\bibinfo  {journal} {Phys. Rev. Lett.}\ }\textbf {\bibinfo {volume} {115}},\
  \bibinfo {pages} {110401} (\bibinfo {year} {2015})}\BibitemShut {NoStop}%
\bibitem [{\citenamefont {Lang}\ \emph {et~al.}(2015)\citenamefont {Lang},
  \citenamefont {Liu},\ and\ \citenamefont {Monteiro}}]{lang2015dynamical}%
  \BibitemOpen
  \bibfield  {author} {\bibinfo {author} {\bibfnamefont {JE}~\bibnamefont
  {Lang}}, \bibinfo {author} {\bibfnamefont {Ren-Bao}\ \bibnamefont {Liu}}, \
  and\ \bibinfo {author} {\bibfnamefont {TS}~\bibnamefont {Monteiro}},\
  }\bibfield  {title} {\enquote {\bibinfo {title} {Dynamical-decoupling-based
  quantum sensing: Floquet spectroscopy},}\ }\href
  {https://link.aps.org/doi/10.1103/PhysRevX.5.041016} {\bibfield  {journal}
  {\bibinfo  {journal} {Phys. Rev. X}\ }\textbf {\bibinfo {volume} {5}},\
  \bibinfo {pages} {041016} (\bibinfo {year} {2015})}\BibitemShut {NoStop}%
\bibitem [{\citenamefont {Montenegro}\ \emph
  {et~al.}(2022{\natexlab{b}})\citenamefont {Montenegro}, \citenamefont
  {Jones}, \citenamefont {Bose},\ and\ \citenamefont
  {Bayat}}]{montenegro2022sequential}%
  \BibitemOpen
  \bibfield  {author} {\bibinfo {author} {\bibfnamefont {Victor}\ \bibnamefont
  {Montenegro}}, \bibinfo {author} {\bibfnamefont {Gareth~Si\^on}\ \bibnamefont
  {Jones}}, \bibinfo {author} {\bibfnamefont {Sougato}\ \bibnamefont {Bose}}, \
  and\ \bibinfo {author} {\bibfnamefont {Abolfazl}\ \bibnamefont {Bayat}},\
  }\bibfield  {title} {\enquote {\bibinfo {title} {Sequential measurements for
  quantum-enhanced magnetometry in spin chain probes},}\ }\href {\doibase
  10.1103/PhysRevLett.129.120503} {\bibfield  {journal} {\bibinfo  {journal}
  {Phys. Rev. Lett.}\ }\textbf {\bibinfo {volume} {129}},\ \bibinfo {pages}
  {120503} (\bibinfo {year} {2022}{\natexlab{b}})}\BibitemShut {NoStop}%
\bibitem [{\citenamefont {Burgarth}\ \emph {et~al.}(2015)\citenamefont
  {Burgarth}, \citenamefont {Giovannetti}, \citenamefont {Kato},\ and\
  \citenamefont {Yuasa}}]{burgarth2015quantum}%
  \BibitemOpen
  \bibfield  {author} {\bibinfo {author} {\bibfnamefont {Daniel}\ \bibnamefont
  {Burgarth}}, \bibinfo {author} {\bibfnamefont {Vittorio}\ \bibnamefont
  {Giovannetti}}, \bibinfo {author} {\bibfnamefont {Airi~N}\ \bibnamefont
  {Kato}}, \ and\ \bibinfo {author} {\bibfnamefont {Kazuya}\ \bibnamefont
  {Yuasa}},\ }\bibfield  {title} {\enquote {\bibinfo {title} {Quantum
  estimation via sequential measurements},}\ }\href@noop {} {\bibfield
  {journal} {\bibinfo  {journal} {New Journal of Physics}\ }\textbf {\bibinfo
  {volume} {17}},\ \bibinfo {pages} {113055} (\bibinfo {year}
  {2015})}\BibitemShut {NoStop}%
\bibitem [{\citenamefont {Yang}\ \emph {et~al.}(2023)\citenamefont {Yang},
  \citenamefont {Montenegro},\ and\ \citenamefont
  {Bayat}}]{yang2023extractable}%
  \BibitemOpen
  \bibfield  {author} {\bibinfo {author} {\bibfnamefont {Yaoling}\ \bibnamefont
  {Yang}}, \bibinfo {author} {\bibfnamefont {Victor}\ \bibnamefont
  {Montenegro}}, \ and\ \bibinfo {author} {\bibfnamefont {Abolfazl}\
  \bibnamefont {Bayat}},\ }\bibfield  {title} {\enquote {\bibinfo {title}
  {Extractable information capacity in sequential measurements metrology},}\
  }\href {\doibase 10.1103/PhysRevResearch.5.043273} {\bibfield  {journal}
  {\bibinfo  {journal} {Phys. Rev. Res.}\ }\textbf {\bibinfo {volume} {5}},\
  \bibinfo {pages} {043273} (\bibinfo {year} {2023})}\BibitemShut {NoStop}%
\bibitem [{\citenamefont {L\"u}\ \emph {et~al.}(2022)\citenamefont {L\"u},
  \citenamefont {Ning}, \citenamefont {Zhu}, \citenamefont {Wu}, \citenamefont
  {Shen}, \citenamefont {Yang},\ and\ \citenamefont {Zheng}}]{lu2022critical}%
  \BibitemOpen
  \bibfield  {author} {\bibinfo {author} {\bibfnamefont {Jia-Hao}\ \bibnamefont
  {L\"u}}, \bibinfo {author} {\bibfnamefont {Wen}\ \bibnamefont {Ning}},
  \bibinfo {author} {\bibfnamefont {Xin}\ \bibnamefont {Zhu}}, \bibinfo
  {author} {\bibfnamefont {Fan}\ \bibnamefont {Wu}}, \bibinfo {author}
  {\bibfnamefont {Li-Tuo}\ \bibnamefont {Shen}}, \bibinfo {author}
  {\bibfnamefont {Zhen-Biao}\ \bibnamefont {Yang}}, \ and\ \bibinfo {author}
  {\bibfnamefont {Shi-Biao}\ \bibnamefont {Zheng}},\ }\bibfield  {title}
  {\enquote {\bibinfo {title} {Critical quantum sensing based on the
  jaynes-cummings model with a squeezing drive},}\ }\href {\doibase
  10.1103/PhysRevA.106.062616} {\bibfield  {journal} {\bibinfo  {journal}
  {Phys. Rev. A}\ }\textbf {\bibinfo {volume} {106}},\ \bibinfo {pages}
  {062616} (\bibinfo {year} {2022})}\BibitemShut {NoStop}%
\bibitem [{\citenamefont {Salvatori}\ \emph {et~al.}(2014)\citenamefont
  {Salvatori}, \citenamefont {Mandarino},\ and\ \citenamefont
  {Paris}}]{salvatori2014quantum}%
  \BibitemOpen
  \bibfield  {author} {\bibinfo {author} {\bibfnamefont {Giulio}\ \bibnamefont
  {Salvatori}}, \bibinfo {author} {\bibfnamefont {Antonio}\ \bibnamefont
  {Mandarino}}, \ and\ \bibinfo {author} {\bibfnamefont {Matteo~GA}\
  \bibnamefont {Paris}},\ }\bibfield  {title} {\enquote {\bibinfo {title}
  {Quantum metrology in lipkin-meshkov-glick critical systems},}\ }\href
  {https://link.aps.org/doi/10.1103/PhysRevA.90.022111} {\bibfield  {journal}
  {\bibinfo  {journal} {Phys. Rev. A}\ }\textbf {\bibinfo {volume} {90}},\
  \bibinfo {pages} {022111} (\bibinfo {year} {2014})}\BibitemShut {NoStop}%
\bibitem [{\citenamefont {Zhu}\ \emph {et~al.}(2023)\citenamefont {Zhu},
  \citenamefont {Lü}, \citenamefont {Ning}, \citenamefont {Wu}, \citenamefont
  {Shen}, \citenamefont {Yang},\ and\ \citenamefont
  {Zheng}}]{Zhu2023criticality}%
  \BibitemOpen
  \bibfield  {author} {\bibinfo {author} {\bibfnamefont {Xin}\ \bibnamefont
  {Zhu}}, \bibinfo {author} {\bibfnamefont {Jia-Hao}\ \bibnamefont {Lü}},
  \bibinfo {author} {\bibfnamefont {Wen}\ \bibnamefont {Ning}}, \bibinfo
  {author} {\bibfnamefont {Fan}\ \bibnamefont {Wu}}, \bibinfo {author}
  {\bibfnamefont {Li-Tuo}\ \bibnamefont {Shen}}, \bibinfo {author}
  {\bibfnamefont {Zhen-Biao}\ \bibnamefont {Yang}}, \ and\ \bibinfo {author}
  {\bibfnamefont {Shi-Biao}\ \bibnamefont {Zheng}},\ }\bibfield  {title}
  {\enquote {\bibinfo {title} {Criticality-enhanced quantum sensing in the
  anisotropic quantum rabi model},}\ }\href {\doibase
  10.1007/s11433-022-2073-9} {\bibfield  {journal} {\bibinfo  {journal} {Sci.
  China Phys. Mech. Astron.}\ }\textbf {\bibinfo {volume} {66}} (\bibinfo
  {year} {2023}),\ 10.1007/s11433-022-2073-9}\BibitemShut {NoStop}%
\bibitem [{\citenamefont {Garbe}\ \emph {et~al.}(2020)\citenamefont {Garbe},
  \citenamefont {Bina}, \citenamefont {Keller}, \citenamefont {Paris},\ and\
  \citenamefont {Felicetti}}]{garbe2020critical}%
  \BibitemOpen
  \bibfield  {author} {\bibinfo {author} {\bibfnamefont {Louis}\ \bibnamefont
  {Garbe}}, \bibinfo {author} {\bibfnamefont {Matteo}\ \bibnamefont {Bina}},
  \bibinfo {author} {\bibfnamefont {Arne}\ \bibnamefont {Keller}}, \bibinfo
  {author} {\bibfnamefont {Matteo~GA}\ \bibnamefont {Paris}}, \ and\ \bibinfo
  {author} {\bibfnamefont {Simone}\ \bibnamefont {Felicetti}},\ }\bibfield
  {title} {\enquote {\bibinfo {title} {Critical quantum metrology with a
  finite-component quantum phase transition},}\ }\href
  {https://link.aps.org/doi/10.1103/PhysRevLett.124.120504} {\bibfield
  {journal} {\bibinfo  {journal} {Phys. Rev. Lett.}\ }\textbf {\bibinfo
  {volume} {124}},\ \bibinfo {pages} {120504} (\bibinfo {year}
  {2020})}\BibitemShut {NoStop}%
\bibitem [{\citenamefont {Salvia}\ \emph {et~al.}(2023)\citenamefont {Salvia},
  \citenamefont {Mehboudi},\ and\ \citenamefont
  {Perarnau-Llobet}}]{salvia2023critical}%
  \BibitemOpen
  \bibfield  {author} {\bibinfo {author} {\bibfnamefont {Raffaele}\
  \bibnamefont {Salvia}}, \bibinfo {author} {\bibfnamefont {Mohammad}\
  \bibnamefont {Mehboudi}}, \ and\ \bibinfo {author} {\bibfnamefont
  {Mart\'{\i}}\ \bibnamefont {Perarnau-Llobet}},\ }\bibfield  {title} {\enquote
  {\bibinfo {title} {Critical quantum metrology assisted by real-time feedback
  control},}\ }\href {\doibase 10.1103/PhysRevLett.130.240803} {\bibfield
  {journal} {\bibinfo  {journal} {Phys. Rev. Lett.}\ }\textbf {\bibinfo
  {volume} {130}},\ \bibinfo {pages} {240803} (\bibinfo {year}
  {2023})}\BibitemShut {NoStop}%
\bibitem [{\citenamefont {Hotter}\ \emph {et~al.}(2024)\citenamefont {Hotter},
  \citenamefont {Ritsch},\ and\ \citenamefont {Gietka}}]{hotter2024combining}%
  \BibitemOpen
  \bibfield  {author} {\bibinfo {author} {\bibfnamefont {Christoph}\
  \bibnamefont {Hotter}}, \bibinfo {author} {\bibfnamefont {Helmut}\
  \bibnamefont {Ritsch}}, \ and\ \bibinfo {author} {\bibfnamefont {Karol}\
  \bibnamefont {Gietka}},\ }\bibfield  {title} {\enquote {\bibinfo {title}
  {Combining critical and quantum metrology},}\ }\href {\doibase
  10.1103/PhysRevLett.132.060801} {\bibfield  {journal} {\bibinfo  {journal}
  {Phys. Rev. Lett.}\ }\textbf {\bibinfo {volume} {132}},\ \bibinfo {pages}
  {060801} (\bibinfo {year} {2024})}\BibitemShut {NoStop}%
\bibitem [{\citenamefont {Invernizzi}\ \emph {et~al.}(2008)\citenamefont
  {Invernizzi}, \citenamefont {Korbman}, \citenamefont {Venuti},\ and\
  \citenamefont {Paris}}]{invernizzi2008optimal}%
  \BibitemOpen
  \bibfield  {author} {\bibinfo {author} {\bibfnamefont {Carmen}\ \bibnamefont
  {Invernizzi}}, \bibinfo {author} {\bibfnamefont {Michael}\ \bibnamefont
  {Korbman}}, \bibinfo {author} {\bibfnamefont {Lorenzo~Campos}\ \bibnamefont
  {Venuti}}, \ and\ \bibinfo {author} {\bibfnamefont {Matteo~GA}\ \bibnamefont
  {Paris}},\ }\bibfield  {title} {\enquote {\bibinfo {title} {Optimal quantum
  estimation in spin systems at criticality},}\ }\href
  {https://doi.org/10.1103/PhysRevA.78.042106} {\bibfield  {journal} {\bibinfo
  {journal} {Phys. Rev. A}\ }\textbf {\bibinfo {volume} {78}},\ \bibinfo
  {pages} {042106} (\bibinfo {year} {2008})}\BibitemShut {NoStop}%
\bibitem [{\citenamefont {Zanardi}\ \emph {et~al.}(2007)\citenamefont
  {Zanardi}, \citenamefont {Quan}, \citenamefont {Wang},\ and\ \citenamefont
  {Sun}}]{zanardi2007mixed}%
  \BibitemOpen
  \bibfield  {author} {\bibinfo {author} {\bibfnamefont {Paolo}\ \bibnamefont
  {Zanardi}}, \bibinfo {author} {\bibfnamefont {HT}~\bibnamefont {Quan}},
  \bibinfo {author} {\bibfnamefont {Xiaoguang}\ \bibnamefont {Wang}}, \ and\
  \bibinfo {author} {\bibfnamefont {CP}~\bibnamefont {Sun}},\ }\bibfield
  {title} {\enquote {\bibinfo {title} {Mixed-state fidelity and quantum
  criticality at finite temperature},}\ }\href
  {https://link.aps.org/doi/10.1103/PhysRevA.75.032109} {\bibfield  {journal}
  {\bibinfo  {journal} {Phys. Rev. A}\ }\textbf {\bibinfo {volume} {75}},\
  \bibinfo {pages} {032109} (\bibinfo {year} {2007})}\BibitemShut {NoStop}%
\bibitem [{\citenamefont {Zanardi}\ and\ \citenamefont
  {Paunkovi{\'c}}(2006)}]{zanardi2006ground}%
  \BibitemOpen
  \bibfield  {author} {\bibinfo {author} {\bibfnamefont {Paolo}\ \bibnamefont
  {Zanardi}}\ and\ \bibinfo {author} {\bibfnamefont {Nikola}\ \bibnamefont
  {Paunkovi{\'c}}},\ }\bibfield  {title} {\enquote {\bibinfo {title} {Ground
  state overlap and quantum phase transitions},}\ }\href
  {https://link.aps.org/doi/10.1103/PhysRevE.74.031123} {\bibfield  {journal}
  {\bibinfo  {journal} {Phys. Rev. E}\ }\textbf {\bibinfo {volume} {74}},\
  \bibinfo {pages} {031123} (\bibinfo {year} {2006})}\BibitemShut {NoStop}%
\bibitem [{\citenamefont {Fr{\'e}rot}\ and\ \citenamefont
  {Roscilde}(2018)}]{frerot2018quantum}%
  \BibitemOpen
  \bibfield  {author} {\bibinfo {author} {\bibfnamefont {Ir{\'e}n{\'e}e}\
  \bibnamefont {Fr{\'e}rot}}\ and\ \bibinfo {author} {\bibfnamefont {Tommaso}\
  \bibnamefont {Roscilde}},\ }\bibfield  {title} {\enquote {\bibinfo {title}
  {Quantum critical metrology},}\ }\href
  {https://doi.org/10.1103/PhysRevLett.121.020402} {\bibfield  {journal}
  {\bibinfo  {journal} {Phys. Rev. Lett.}\ }\textbf {\bibinfo {volume} {121}},\
  \bibinfo {pages} {020402} (\bibinfo {year} {2018})}\BibitemShut {NoStop}%
\bibitem [{\citenamefont {Mukhopadhyay}\ and\ \citenamefont
  {Bayat}(2024{\natexlab{b}})}]{mukhopadhyay2023modular}%
  \BibitemOpen
  \bibfield  {author} {\bibinfo {author} {\bibfnamefont {Chiranjib}\
  \bibnamefont {Mukhopadhyay}}\ and\ \bibinfo {author} {\bibfnamefont
  {Abolfazl}\ \bibnamefont {Bayat}},\ }\bibfield  {title} {\enquote {\bibinfo
  {title} {Modular many-body quantum sensors},}\ }\href {\doibase
  10.1103/PhysRevLett.133.120601} {\bibfield  {journal} {\bibinfo  {journal}
  {Phys. Rev. Lett.}\ }\textbf {\bibinfo {volume} {133}},\ \bibinfo {pages}
  {120601} (\bibinfo {year} {2024}{\natexlab{b}})}\BibitemShut {NoStop}%
\bibitem [{\citenamefont {Mihailescu}\ \emph {et~al.}(2025)\citenamefont
  {Mihailescu}, \citenamefont {Campbell},\ and\ \citenamefont
  {Gietka}}]{mihailescu2025uncertain}%
  \BibitemOpen
  \bibfield  {author} {\bibinfo {author} {\bibfnamefont {George}\ \bibnamefont
  {Mihailescu}}, \bibinfo {author} {\bibfnamefont {Steve}\ \bibnamefont
  {Campbell}}, \ and\ \bibinfo {author} {\bibfnamefont {Karol}\ \bibnamefont
  {Gietka}},\ }\bibfield  {title} {\enquote {\bibinfo {title} {Uncertain
  quantum critical metrology: From single- to multiparameter sensing},}\ }\href
  {\doibase 10.1103/PhysRevA.111.052621} {\bibfield  {journal} {\bibinfo
  {journal} {Phys. Rev. A}\ }\textbf {\bibinfo {volume} {111}},\ \bibinfo
  {pages} {052621} (\bibinfo {year} {2025})}\BibitemShut {NoStop}%
\bibitem [{\citenamefont {Gietka}\ \emph {et~al.}(2022)\citenamefont {Gietka},
  \citenamefont {Ruks},\ and\ \citenamefont {Busch}}]{gietka2022understanding}%
  \BibitemOpen
  \bibfield  {author} {\bibinfo {author} {\bibfnamefont {Karol}\ \bibnamefont
  {Gietka}}, \bibinfo {author} {\bibfnamefont {Lewis}\ \bibnamefont {Ruks}}, \
  and\ \bibinfo {author} {\bibfnamefont {Thomas}\ \bibnamefont {Busch}},\
  }\bibfield  {title} {\enquote {\bibinfo {title} {Understanding and
  {I}mproving {C}ritical {M}etrology. {Q}uenching {S}uperradiant
  {L}ight-{M}atter {S}ystems {B}eyond the {C}ritical {P}oint},}\ }\href
  {\doibase 10.22331/q-2022-04-27-700} {\bibfield  {journal} {\bibinfo
  {journal} {{Quantum}}\ }\textbf {\bibinfo {volume} {6}},\ \bibinfo {pages}
  {700} (\bibinfo {year} {2022})}\BibitemShut {NoStop}%
\bibitem [{\citenamefont {Alushi}\ \emph {et~al.}(2025)\citenamefont {Alushi},
  \citenamefont {Coppo}, \citenamefont {Brosco}, \citenamefont {Di~Candia},\
  and\ \citenamefont {Felicetti}}]{alushi2025collective}%
  \BibitemOpen
  \bibfield  {author} {\bibinfo {author} {\bibfnamefont {Uesli}\ \bibnamefont
  {Alushi}}, \bibinfo {author} {\bibfnamefont {Alessandro}\ \bibnamefont
  {Coppo}}, \bibinfo {author} {\bibfnamefont {Valentina}\ \bibnamefont
  {Brosco}}, \bibinfo {author} {\bibfnamefont {Roberto}\ \bibnamefont
  {Di~Candia}}, \ and\ \bibinfo {author} {\bibfnamefont {Simone}\ \bibnamefont
  {Felicetti}},\ }\bibfield  {title} {\enquote {\bibinfo {title} {Collective
  quantum enhancement in critical quantum sensing},}\ }\href {\doibase
  10.1038/s42005-025-01975-9} {\bibfield  {journal} {\bibinfo  {journal}
  {Communications Physics}\ }\textbf {\bibinfo {volume} {8}},\ \bibinfo {pages}
  {74} (\bibinfo {year} {2025})}\BibitemShut {NoStop}%
\bibitem [{\citenamefont {Zhao}\ and\ \citenamefont
  {Zhou}(2009)}]{zhao2009singularities}%
  \BibitemOpen
  \bibfield  {author} {\bibinfo {author} {\bibfnamefont {Jian-Hui}\
  \bibnamefont {Zhao}}\ and\ \bibinfo {author} {\bibfnamefont {Huan-Qiang}\
  \bibnamefont {Zhou}},\ }\bibfield  {title} {\enquote {\bibinfo {title}
  {Singularities in ground-state fidelity and quantum phase transitions for the
  kitaev model},}\ }\href {https://link.aps.org/doi/10.1103/PhysRevB.80.014403}
  {\bibfield  {journal} {\bibinfo  {journal} {Phys. Rev. B}\ }\textbf {\bibinfo
  {volume} {80}},\ \bibinfo {pages} {014403} (\bibinfo {year}
  {2009})}\BibitemShut {NoStop}%
\bibitem [{\citenamefont {Venuti}\ and\ \citenamefont
  {Zanardi}(2007)}]{venuti2007quantum}%
  \BibitemOpen
  \bibfield  {author} {\bibinfo {author} {\bibfnamefont {Lorenzo~Campos}\
  \bibnamefont {Venuti}}\ and\ \bibinfo {author} {\bibfnamefont {Paolo}\
  \bibnamefont {Zanardi}},\ }\bibfield  {title} {\enquote {\bibinfo {title}
  {Quantum critical scaling of the geometric tensors},}\ }\href
  {https://link.aps.org/doi/10.1103/PhysRevLett.99.095701} {\bibfield
  {journal} {\bibinfo  {journal} {Phys. Rev. Lett.}\ }\textbf {\bibinfo
  {volume} {99}},\ \bibinfo {pages} {095701} (\bibinfo {year}
  {2007})}\BibitemShut {NoStop}%
\bibitem [{\citenamefont {Schwandt}\ \emph {et~al.}(2009)\citenamefont
  {Schwandt}, \citenamefont {Alet},\ and\ \citenamefont
  {Capponi}}]{schwandt2009quantum}%
  \BibitemOpen
  \bibfield  {author} {\bibinfo {author} {\bibfnamefont {David}\ \bibnamefont
  {Schwandt}}, \bibinfo {author} {\bibfnamefont {Fabien}\ \bibnamefont {Alet}},
  \ and\ \bibinfo {author} {\bibfnamefont {Sylvain}\ \bibnamefont {Capponi}},\
  }\bibfield  {title} {\enquote {\bibinfo {title} {Quantum monte carlo
  simulations of fidelity at magnetic quantum phase transitions},}\ }\href
  {https://doi.org/10.1103/PhysRevLett.103.170501} {\bibfield  {journal}
  {\bibinfo  {journal} {Phys. Rev. Lett.}\ }\textbf {\bibinfo {volume} {103}},\
  \bibinfo {pages} {170501} (\bibinfo {year} {2009})}\BibitemShut {NoStop}%
\bibitem [{\citenamefont {Albuquerque}\ \emph {et~al.}(2010)\citenamefont
  {Albuquerque}, \citenamefont {Alet}, \citenamefont {Sire},\ and\
  \citenamefont {Capponi}}]{albuquerque2010quantum}%
  \BibitemOpen
  \bibfield  {author} {\bibinfo {author} {\bibfnamefont {A~Fabricio}\
  \bibnamefont {Albuquerque}}, \bibinfo {author} {\bibfnamefont {Fabien}\
  \bibnamefont {Alet}}, \bibinfo {author} {\bibfnamefont {Cl{\'e}ment}\
  \bibnamefont {Sire}}, \ and\ \bibinfo {author} {\bibfnamefont {Sylvain}\
  \bibnamefont {Capponi}},\ }\bibfield  {title} {\enquote {\bibinfo {title}
  {Quantum critical scaling of fidelity susceptibility},}\ }\href
  {https://link.aps.org/doi/10.1103/PhysRevB.81.064418} {\bibfield  {journal}
  {\bibinfo  {journal} {Phys. Rev. B}\ }\textbf {\bibinfo {volume} {81}},\
  \bibinfo {pages} {064418} (\bibinfo {year} {2010})}\BibitemShut {NoStop}%
\bibitem [{\citenamefont {Gritsev}\ and\ \citenamefont
  {Polkovnikov}(2009)}]{gritsev2009universal}%
  \BibitemOpen
  \bibfield  {author} {\bibinfo {author} {\bibfnamefont {Vladimir}\
  \bibnamefont {Gritsev}}\ and\ \bibinfo {author} {\bibfnamefont {Anatoli}\
  \bibnamefont {Polkovnikov}},\ }\bibfield  {title} {\enquote {\bibinfo {title}
  {Universal dynamics near quantum critical points},}\ }\href
  {https://doi.org/10.48550/arXiv.0910.3692} {\bibfield  {journal} {\bibinfo
  {journal} {arXiv:0910.3692}\ } (\bibinfo {year} {2009})}\BibitemShut
  {NoStop}%
\bibitem [{\citenamefont {Gu}\ \emph {et~al.}(2008)\citenamefont {Gu},
  \citenamefont {Kwok}, \citenamefont {Ning}, \citenamefont {Lin} \emph
  {et~al.}}]{gu2008fidelity}%
  \BibitemOpen
  \bibfield  {author} {\bibinfo {author} {\bibfnamefont {Shi-Jian}\
  \bibnamefont {Gu}}, \bibinfo {author} {\bibfnamefont {Ho-Man}\ \bibnamefont
  {Kwok}}, \bibinfo {author} {\bibfnamefont {Wen-Qiang}\ \bibnamefont {Ning}},
  \bibinfo {author} {\bibfnamefont {Hai-Qing}\ \bibnamefont {Lin}},  \emph
  {et~al.},\ }\bibfield  {title} {\enquote {\bibinfo {title} {Fidelity
  susceptibility, scaling, and universality in quantum critical phenomena},}\
  }\href {https://link.aps.org/doi/10.1103/PhysRevB.77.245109} {\bibfield
  {journal} {\bibinfo  {journal} {Phys. Rev. B}\ }\textbf {\bibinfo {volume}
  {77}},\ \bibinfo {pages} {245109} (\bibinfo {year} {2008})}\BibitemShut
  {NoStop}%
\bibitem [{\citenamefont {Greschner}\ \emph {et~al.}(2013)\citenamefont
  {Greschner}, \citenamefont {Kolezhuk},\ and\ \citenamefont
  {Vekua}}]{greschner2013fidelity}%
  \BibitemOpen
  \bibfield  {author} {\bibinfo {author} {\bibfnamefont {Sebastian}\
  \bibnamefont {Greschner}}, \bibinfo {author} {\bibfnamefont {AK}~\bibnamefont
  {Kolezhuk}}, \ and\ \bibinfo {author} {\bibfnamefont {T}~\bibnamefont
  {Vekua}},\ }\bibfield  {title} {\enquote {\bibinfo {title} {Fidelity
  susceptibility and conductivity of the current in one-dimensional lattice
  models with open or periodic boundary conditions},}\ }\href
  {https://link.aps.org/doi/10.1103/PhysRevB.88.195101} {\bibfield  {journal}
  {\bibinfo  {journal} {Phys. Rev. B}\ }\textbf {\bibinfo {volume} {88}},\
  \bibinfo {pages} {195101} (\bibinfo {year} {2013})}\BibitemShut {NoStop}%
\bibitem [{\citenamefont {Mishra}\ and\ \citenamefont
  {Bayat}(2021{\natexlab{b}})}]{mishra2021integrable}%
  \BibitemOpen
  \bibfield  {author} {\bibinfo {author} {\bibfnamefont {Utkarsh}\ \bibnamefont
  {Mishra}}\ and\ \bibinfo {author} {\bibfnamefont {Abolfazl}\ \bibnamefont
  {Bayat}},\ }\href@noop {} {} (\bibinfo {year} {2021}{\natexlab{b}}),\ \Eprint
  {http://arxiv.org/abs/2105.13507} {arXiv:2105.13507 [quant-ph]} \BibitemShut
  {NoStop}%
\bibitem [{\citenamefont {Di~Candia}\ \emph {et~al.}(2023)\citenamefont
  {Di~Candia}, \citenamefont {Minganti}, \citenamefont {Petrovnin},
  \citenamefont {Paraoanu},\ and\ \citenamefont
  {Felicetti}}]{dicandia2023critical}%
  \BibitemOpen
  \bibfield  {author} {\bibinfo {author} {\bibfnamefont {R.}~\bibnamefont
  {Di~Candia}}, \bibinfo {author} {\bibfnamefont {F.}~\bibnamefont {Minganti}},
  \bibinfo {author} {\bibfnamefont {K.~V.}\ \bibnamefont {Petrovnin}}, \bibinfo
  {author} {\bibfnamefont {G.~S.}\ \bibnamefont {Paraoanu}}, \ and\ \bibinfo
  {author} {\bibfnamefont {S.}~\bibnamefont {Felicetti}},\ }\bibfield  {title}
  {\enquote {\bibinfo {title} {Critical parametric quantum sensing},}\ }\href
  {\doibase 10.1038/s41534-023-00690-z} {\bibfield  {journal} {\bibinfo
  {journal} {npj Quantum Information}\ }\textbf {\bibinfo {volume} {9}},\
  \bibinfo {pages} {23} (\bibinfo {year} {2023})}\BibitemShut {NoStop}%
\bibitem [{\citenamefont {Minganti}\ \emph {et~al.}(2018)\citenamefont
  {Minganti}, \citenamefont {Biella}, \citenamefont {Bartolo},\ and\
  \citenamefont {Ciuti}}]{minganti2018spectral}%
  \BibitemOpen
  \bibfield  {author} {\bibinfo {author} {\bibfnamefont {Fabrizio}\
  \bibnamefont {Minganti}}, \bibinfo {author} {\bibfnamefont {Alberto}\
  \bibnamefont {Biella}}, \bibinfo {author} {\bibfnamefont {Nicola}\
  \bibnamefont {Bartolo}}, \ and\ \bibinfo {author} {\bibfnamefont {Cristiano}\
  \bibnamefont {Ciuti}},\ }\bibfield  {title} {\enquote {\bibinfo {title}
  {Spectral theory of liouvillians for dissipative phase transitions},}\
  }\href@noop {} {\bibfield  {journal} {\bibinfo  {journal} {Physical Review
  A}\ }\textbf {\bibinfo {volume} {98}},\ \bibinfo {pages} {042118} (\bibinfo
  {year} {2018})}\BibitemShut {NoStop}%
\bibitem [{\citenamefont {Kessler}\ \emph {et~al.}(2012)\citenamefont
  {Kessler}, \citenamefont {Giedke}, \citenamefont {Imamoglu}, \citenamefont
  {Yelin}, \citenamefont {Lukin},\ and\ \citenamefont
  {Cirac}}]{kessler2012dissipative}%
  \BibitemOpen
  \bibfield  {author} {\bibinfo {author} {\bibfnamefont {Eric~M}\ \bibnamefont
  {Kessler}}, \bibinfo {author} {\bibfnamefont {Geza}\ \bibnamefont {Giedke}},
  \bibinfo {author} {\bibfnamefont {Atac}\ \bibnamefont {Imamoglu}}, \bibinfo
  {author} {\bibfnamefont {Susanne~F}\ \bibnamefont {Yelin}}, \bibinfo {author}
  {\bibfnamefont {Mikhail~D}\ \bibnamefont {Lukin}}, \ and\ \bibinfo {author}
  {\bibfnamefont {J~Ignacio}\ \bibnamefont {Cirac}},\ }\bibfield  {title}
  {\enquote {\bibinfo {title} {Dissipative phase transition in a central spin
  system},}\ }\href {https://link.aps.org/doi/10.1103/PhysRevA.86.012116}
  {\bibfield  {journal} {\bibinfo  {journal} {Phys. Rev. A}\ }\textbf {\bibinfo
  {volume} {86}},\ \bibinfo {pages} {012116} (\bibinfo {year}
  {2012})}\BibitemShut {NoStop}%
\bibitem [{\citenamefont {Dalla~Torre}\ \emph {et~al.}(2012)\citenamefont
  {Dalla~Torre}, \citenamefont {Demler}, \citenamefont {Giamarchi},\ and\
  \citenamefont {Altman}}]{dalla2012dynamics}%
  \BibitemOpen
  \bibfield  {author} {\bibinfo {author} {\bibfnamefont {Emanuele~G.}\
  \bibnamefont {Dalla~Torre}}, \bibinfo {author} {\bibfnamefont {Eugene}\
  \bibnamefont {Demler}}, \bibinfo {author} {\bibfnamefont {Thierry}\
  \bibnamefont {Giamarchi}}, \ and\ \bibinfo {author} {\bibfnamefont {Ehud}\
  \bibnamefont {Altman}},\ }\bibfield  {title} {\enquote {\bibinfo {title}
  {Dynamics and universality in noise-driven dissipative systems},}\ }\href
  {\doibase 10.1103/PhysRevB.85.184302} {\bibfield  {journal} {\bibinfo
  {journal} {Phys. Rev. B}\ }\textbf {\bibinfo {volume} {85}},\ \bibinfo
  {pages} {184302} (\bibinfo {year} {2012})}\BibitemShut {NoStop}%
\bibitem [{\citenamefont {Marino}\ and\ \citenamefont
  {Diehl}(2016)}]{marino2016driven}%
  \BibitemOpen
  \bibfield  {author} {\bibinfo {author} {\bibfnamefont {Jamir}\ \bibnamefont
  {Marino}}\ and\ \bibinfo {author} {\bibfnamefont {Sebastian}\ \bibnamefont
  {Diehl}},\ }\bibfield  {title} {\enquote {\bibinfo {title} {Driven markovian
  quantum criticality},}\ }\href
  {https://link.aps.org/doi/10.1103/PhysRevLett.116.070407} {\bibfield
  {journal} {\bibinfo  {journal} {Phys. Rev. Lett.}\ }\textbf {\bibinfo
  {volume} {116}},\ \bibinfo {pages} {070407} (\bibinfo {year}
  {2016})}\BibitemShut {NoStop}%
\bibitem [{\citenamefont {Fern{\'a}ndez-Lorenzo}\ and\ \citenamefont
  {Porras}(2017)}]{fernandez2017quantum}%
  \BibitemOpen
  \bibfield  {author} {\bibinfo {author} {\bibfnamefont {Samuel}\ \bibnamefont
  {Fern{\'a}ndez-Lorenzo}}\ and\ \bibinfo {author} {\bibfnamefont {Diego}\
  \bibnamefont {Porras}},\ }\bibfield  {title} {\enquote {\bibinfo {title}
  {Quantum sensing close to a dissipative phase transition: Symmetry breaking
  and criticality as metrological resources},}\ }\href
  {https://link.aps.org/doi/10.1103/PhysRevA.96.013817} {\bibfield  {journal}
  {\bibinfo  {journal} {Phys. Rev. A}\ }\textbf {\bibinfo {volume} {96}},\
  \bibinfo {pages} {013817} (\bibinfo {year} {2017})}\BibitemShut {NoStop}%
\bibitem [{\citenamefont {Lled{\'o}}\ and\ \citenamefont
  {Szyma{\'n}ska}(2020)}]{lledo2020dissipative}%
  \BibitemOpen
  \bibfield  {author} {\bibinfo {author} {\bibfnamefont {Crist{\'o}bal}\
  \bibnamefont {Lled{\'o}}}\ and\ \bibinfo {author} {\bibfnamefont {Marzena~H}\
  \bibnamefont {Szyma{\'n}ska}},\ }\bibfield  {title} {\enquote {\bibinfo
  {title} {A dissipative time crystal with or without z2 symmetry breaking},}\
  }\href {https://iopscience.iop.org/article/10.1088/1367-2630/ab9ae3}
  {\bibfield  {journal} {\bibinfo  {journal} {New J. Phys.}\ }\textbf {\bibinfo
  {volume} {22}},\ \bibinfo {pages} {075002} (\bibinfo {year}
  {2020})}\BibitemShut {NoStop}%
\bibitem [{\citenamefont {Bartolo}\ \emph {et~al.}(2016)\citenamefont
  {Bartolo}, \citenamefont {Minganti}, \citenamefont {Casteels},\ and\
  \citenamefont {Ciuti}}]{bartolo2016exact}%
  \BibitemOpen
  \bibfield  {author} {\bibinfo {author} {\bibfnamefont {Nicola}\ \bibnamefont
  {Bartolo}}, \bibinfo {author} {\bibfnamefont {Fabrizio}\ \bibnamefont
  {Minganti}}, \bibinfo {author} {\bibfnamefont {Wim}\ \bibnamefont
  {Casteels}}, \ and\ \bibinfo {author} {\bibfnamefont {Cristiano}\
  \bibnamefont {Ciuti}},\ }\bibfield  {title} {\enquote {\bibinfo {title}
  {Exact steady state of a kerr resonator with one- and two-photon driving and
  dissipation: Controllable wigner-function multimodality and dissipative phase
  transitions},}\ }\href {\doibase 10.1103/PhysRevA.94.033841} {\bibfield
  {journal} {\bibinfo  {journal} {Phys. Rev. A}\ }\textbf {\bibinfo {volume}
  {94}},\ \bibinfo {pages} {033841} (\bibinfo {year} {2016})}\BibitemShut
  {NoStop}%
\bibitem [{\citenamefont {Rota}\ \emph {et~al.}(2017)\citenamefont {Rota},
  \citenamefont {Storme}, \citenamefont {Bartolo}, \citenamefont {Fazio},\ and\
  \citenamefont {Ciuti}}]{rota2017critical}%
  \BibitemOpen
  \bibfield  {author} {\bibinfo {author} {\bibfnamefont {R.}~\bibnamefont
  {Rota}}, \bibinfo {author} {\bibfnamefont {F.}~\bibnamefont {Storme}},
  \bibinfo {author} {\bibfnamefont {N.}~\bibnamefont {Bartolo}}, \bibinfo
  {author} {\bibfnamefont {R.}~\bibnamefont {Fazio}}, \ and\ \bibinfo {author}
  {\bibfnamefont {C.}~\bibnamefont {Ciuti}},\ }\bibfield  {title} {\enquote
  {\bibinfo {title} {Critical behavior of dissipative two-dimensional spin
  lattices},}\ }\href {\doibase 10.1103/PhysRevB.95.134431} {\bibfield
  {journal} {\bibinfo  {journal} {Phys. Rev. B}\ }\textbf {\bibinfo {volume}
  {95}},\ \bibinfo {pages} {134431} (\bibinfo {year} {2017})}\BibitemShut
  {NoStop}%
\bibitem [{\citenamefont {Letscher}\ \emph {et~al.}(2017)\citenamefont
  {Letscher}, \citenamefont {Thomas}, \citenamefont {Niederpr\"um},
  \citenamefont {Fleischhauer},\ and\ \citenamefont
  {Ott}}]{letscher2016bistability}%
  \BibitemOpen
  \bibfield  {author} {\bibinfo {author} {\bibfnamefont {F.}~\bibnamefont
  {Letscher}}, \bibinfo {author} {\bibfnamefont {O.}~\bibnamefont {Thomas}},
  \bibinfo {author} {\bibfnamefont {T.}~\bibnamefont {Niederpr\"um}}, \bibinfo
  {author} {\bibfnamefont {M.}~\bibnamefont {Fleischhauer}}, \ and\ \bibinfo
  {author} {\bibfnamefont {H.}~\bibnamefont {Ott}},\ }\bibfield  {title}
  {\enquote {\bibinfo {title} {Bistability versus metastability in driven
  dissipative rydberg gases},}\ }\href {\doibase 10.1103/PhysRevX.7.021020}
  {\bibfield  {journal} {\bibinfo  {journal} {Phys. Rev. X}\ }\textbf {\bibinfo
  {volume} {7}},\ \bibinfo {pages} {021020} (\bibinfo {year}
  {2017})}\BibitemShut {NoStop}%
\bibitem [{\citenamefont {Overbeck}\ \emph {et~al.}(2017)\citenamefont
  {Overbeck}, \citenamefont {Maghrebi}, \citenamefont {Gorshkov},\ and\
  \citenamefont {Weimer}}]{overbeck2017multicritical}%
  \BibitemOpen
  \bibfield  {author} {\bibinfo {author} {\bibfnamefont {Vincent~R.}\
  \bibnamefont {Overbeck}}, \bibinfo {author} {\bibfnamefont {Mohammad~F.}\
  \bibnamefont {Maghrebi}}, \bibinfo {author} {\bibfnamefont {Alexey~V.}\
  \bibnamefont {Gorshkov}}, \ and\ \bibinfo {author} {\bibfnamefont {Hendrik}\
  \bibnamefont {Weimer}},\ }\bibfield  {title} {\enquote {\bibinfo {title}
  {Multicritical behavior in dissipative ising models},}\ }\href {\doibase
  10.1103/PhysRevA.95.042133} {\bibfield  {journal} {\bibinfo  {journal} {Phys.
  Rev. A}\ }\textbf {\bibinfo {volume} {95}},\ \bibinfo {pages} {042133}
  (\bibinfo {year} {2017})}\BibitemShut {NoStop}%
\bibitem [{\citenamefont {Iemini}\ \emph {et~al.}(2018)\citenamefont {Iemini},
  \citenamefont {Russomanno}, \citenamefont {Keeling}, \citenamefont
  {Schir{\`o}}, \citenamefont {Dalmonte},\ and\ \citenamefont
  {Fazio}}]{iemini2018boundary}%
  \BibitemOpen
  \bibfield  {author} {\bibinfo {author} {\bibfnamefont {F}~\bibnamefont
  {Iemini}}, \bibinfo {author} {\bibfnamefont {A}~\bibnamefont {Russomanno}},
  \bibinfo {author} {\bibfnamefont {J}~\bibnamefont {Keeling}}, \bibinfo
  {author} {\bibfnamefont {M}~\bibnamefont {Schir{\`o}}}, \bibinfo {author}
  {\bibfnamefont {M}~\bibnamefont {Dalmonte}}, \ and\ \bibinfo {author}
  {\bibfnamefont {R}~\bibnamefont {Fazio}},\ }\bibfield  {title} {\enquote
  {\bibinfo {title} {Boundary time crystals},}\ }\href
  {https://doi.org/10.1103/PhysRevLett.121.035301} {\bibfield  {journal}
  {\bibinfo  {journal} {Phys. Rev. Lett.}\ }\textbf {\bibinfo {volume} {121}},\
  \bibinfo {pages} {035301} (\bibinfo {year} {2018})}\BibitemShut {NoStop}%
\bibitem [{\citenamefont {Capriotti}\ \emph {et~al.}(2005)\citenamefont
  {Capriotti}, \citenamefont {Cuccoli}, \citenamefont {Fubini}, \citenamefont
  {Tognetti},\ and\ \citenamefont {Vaia}}]{capriotti2005dissipation}%
  \BibitemOpen
  \bibfield  {author} {\bibinfo {author} {\bibfnamefont {Luca}\ \bibnamefont
  {Capriotti}}, \bibinfo {author} {\bibfnamefont {Alessandro}\ \bibnamefont
  {Cuccoli}}, \bibinfo {author} {\bibfnamefont {Andrea}\ \bibnamefont
  {Fubini}}, \bibinfo {author} {\bibfnamefont {Valerio}\ \bibnamefont
  {Tognetti}}, \ and\ \bibinfo {author} {\bibfnamefont {Ruggero}\ \bibnamefont
  {Vaia}},\ }\bibfield  {title} {\enquote {\bibinfo {title} {Dissipation-driven
  phase transition in two-dimensional josephson arrays},}\ }\href {\doibase
  10.1103/PhysRevLett.94.157001} {\bibfield  {journal} {\bibinfo  {journal}
  {Phys. Rev. Lett.}\ }\textbf {\bibinfo {volume} {94}},\ \bibinfo {pages}
  {157001} (\bibinfo {year} {2005})}\BibitemShut {NoStop}%
\bibitem [{\citenamefont {Heugel}\ \emph {et~al.}(2019)\citenamefont {Heugel},
  \citenamefont {Biondi}, \citenamefont {Zilberberg},\ and\ \citenamefont
  {Chitra}}]{heugel2019quantum}%
  \BibitemOpen
  \bibfield  {author} {\bibinfo {author} {\bibfnamefont {Toni~L}\ \bibnamefont
  {Heugel}}, \bibinfo {author} {\bibfnamefont {Matteo}\ \bibnamefont {Biondi}},
  \bibinfo {author} {\bibfnamefont {Oded}\ \bibnamefont {Zilberberg}}, \ and\
  \bibinfo {author} {\bibfnamefont {Ramasubramanian}\ \bibnamefont {Chitra}},\
  }\bibfield  {title} {\enquote {\bibinfo {title} {Quantum transducer using a
  parametric driven-dissipative phase transition},}\ }\href
  {https://link.aps.org/doi/10.1103/PhysRevLett.123.173601} {\bibfield
  {journal} {\bibinfo  {journal} {Phys. Rev. Lett.}\ }\textbf {\bibinfo
  {volume} {123}},\ \bibinfo {pages} {173601} (\bibinfo {year}
  {2019})}\BibitemShut {NoStop}%
\bibitem [{\citenamefont {Sieberer}\ \emph {et~al.}(2014)\citenamefont
  {Sieberer}, \citenamefont {Huber}, \citenamefont {Altman},\ and\
  \citenamefont {Diehl}}]{sieberer2014nonequilibrium}%
  \BibitemOpen
  \bibfield  {author} {\bibinfo {author} {\bibfnamefont {L.~M.}\ \bibnamefont
  {Sieberer}}, \bibinfo {author} {\bibfnamefont {S.~D.}\ \bibnamefont {Huber}},
  \bibinfo {author} {\bibfnamefont {E.}~\bibnamefont {Altman}}, \ and\ \bibinfo
  {author} {\bibfnamefont {S.}~\bibnamefont {Diehl}},\ }\bibfield  {title}
  {\enquote {\bibinfo {title} {Nonequilibrium functional renormalization for
  driven-dissipative bose-einstein condensation},}\ }\href {\doibase
  10.1103/PhysRevB.89.134310} {\bibfield  {journal} {\bibinfo  {journal} {Phys.
  Rev. B}\ }\textbf {\bibinfo {volume} {89}},\ \bibinfo {pages} {134310}
  (\bibinfo {year} {2014})}\BibitemShut {NoStop}%
\bibitem [{\citenamefont {Carmichael}(2015)}]{carmichael2015breakdown}%
  \BibitemOpen
  \bibfield  {author} {\bibinfo {author} {\bibfnamefont {H.~J.}\ \bibnamefont
  {Carmichael}},\ }\bibfield  {title} {\enquote {\bibinfo {title} {Breakdown of
  photon blockade: A dissipative quantum phase transition in zero
  dimensions},}\ }\href {\doibase 10.1103/PhysRevX.5.031028} {\bibfield
  {journal} {\bibinfo  {journal} {Phys. Rev. X}\ }\textbf {\bibinfo {volume}
  {5}},\ \bibinfo {pages} {031028} (\bibinfo {year} {2015})}\BibitemShut
  {NoStop}%
\bibitem [{\citenamefont {Benito}\ \emph {et~al.}(2016)\citenamefont {Benito},
  \citenamefont {S\'anchez Mu\~noz},\ and\ \citenamefont
  {Navarrete-Benlloch}}]{benito2016degenerate}%
  \BibitemOpen
  \bibfield  {author} {\bibinfo {author} {\bibfnamefont {M\'onica}\
  \bibnamefont {Benito}}, \bibinfo {author} {\bibfnamefont {Carlos}\
  \bibnamefont {S\'anchez Mu\~noz}}, \ and\ \bibinfo {author} {\bibfnamefont
  {Carlos}\ \bibnamefont {Navarrete-Benlloch}},\ }\bibfield  {title} {\enquote
  {\bibinfo {title} {Degenerate parametric oscillation in quantum membrane
  optomechanics},}\ }\href {\doibase 10.1103/PhysRevA.93.023846} {\bibfield
  {journal} {\bibinfo  {journal} {Phys. Rev. A}\ }\textbf {\bibinfo {volume}
  {93}},\ \bibinfo {pages} {023846} (\bibinfo {year} {2016})}\BibitemShut
  {NoStop}%
\bibitem [{\citenamefont {Mendoza-Arenas}\ \emph {et~al.}(2016)\citenamefont
  {Mendoza-Arenas}, \citenamefont {Clark}, \citenamefont {Felicetti},
  \citenamefont {Romero}, \citenamefont {Solano}, \citenamefont {Angelakis},\
  and\ \citenamefont {Jaksch}}]{arenas2016beyond}%
  \BibitemOpen
  \bibfield  {author} {\bibinfo {author} {\bibfnamefont {J.~J.}\ \bibnamefont
  {Mendoza-Arenas}}, \bibinfo {author} {\bibfnamefont {S.~R.}\ \bibnamefont
  {Clark}}, \bibinfo {author} {\bibfnamefont {S.}~\bibnamefont {Felicetti}},
  \bibinfo {author} {\bibfnamefont {G.}~\bibnamefont {Romero}}, \bibinfo
  {author} {\bibfnamefont {E.}~\bibnamefont {Solano}}, \bibinfo {author}
  {\bibfnamefont {D.~G.}\ \bibnamefont {Angelakis}}, \ and\ \bibinfo {author}
  {\bibfnamefont {D.}~\bibnamefont {Jaksch}},\ }\bibfield  {title} {\enquote
  {\bibinfo {title} {Beyond mean-field bistability in driven-dissipative
  lattices: Bunching-antibunching transition and quantum simulation},}\ }\href
  {\doibase 10.1103/PhysRevA.93.023821} {\bibfield  {journal} {\bibinfo
  {journal} {Phys. Rev. A}\ }\textbf {\bibinfo {volume} {93}},\ \bibinfo
  {pages} {023821} (\bibinfo {year} {2016})}\BibitemShut {NoStop}%
\bibitem [{\citenamefont {Casteels}\ \emph {et~al.}(2017)\citenamefont
  {Casteels}, \citenamefont {Fazio},\ and\ \citenamefont
  {Ciuti}}]{casteels2017critical}%
  \BibitemOpen
  \bibfield  {author} {\bibinfo {author} {\bibfnamefont {Wim}\ \bibnamefont
  {Casteels}}, \bibinfo {author} {\bibfnamefont {Rosario}\ \bibnamefont
  {Fazio}}, \ and\ \bibinfo {author} {\bibfnamefont {Christiano}\ \bibnamefont
  {Ciuti}},\ }\bibfield  {title} {\enquote {\bibinfo {title} {Critical
  dynamical properties of a first-order dissipative phase transition},}\ }\href
  {https://link.aps.org/doi/10.1103/PhysRevA.95.012128} {\bibfield  {journal}
  {\bibinfo  {journal} {Phys. Rev. A}\ }\textbf {\bibinfo {volume} {95}},\
  \bibinfo {pages} {012128} (\bibinfo {year} {2017})}\BibitemShut {NoStop}%
\bibitem [{\citenamefont {Casteels}\ and\ \citenamefont
  {Ciuti}(2017)}]{casteels2017quantum}%
  \BibitemOpen
  \bibfield  {author} {\bibinfo {author} {\bibfnamefont {Wim}\ \bibnamefont
  {Casteels}}\ and\ \bibinfo {author} {\bibfnamefont {Cristiano}\ \bibnamefont
  {Ciuti}},\ }\bibfield  {title} {\enquote {\bibinfo {title} {Quantum
  entanglement in the spatial-symmetry-breaking phase transition of a
  driven-dissipative bose-hubbard dimer},}\ }\href {\doibase
  10.1103/PhysRevA.95.013812} {\bibfield  {journal} {\bibinfo  {journal} {Phys.
  Rev. A}\ }\textbf {\bibinfo {volume} {95}},\ \bibinfo {pages} {013812}
  (\bibinfo {year} {2017})}\BibitemShut {NoStop}%
\bibitem [{\citenamefont {Savona}(2017)}]{savona2017spontaneous}%
  \BibitemOpen
  \bibfield  {author} {\bibinfo {author} {\bibfnamefont {Vincenzo}\
  \bibnamefont {Savona}},\ }\bibfield  {title} {\enquote {\bibinfo {title}
  {Spontaneous symmetry breaking in a quadratically driven nonlinear photonic
  lattice},}\ }\href {\doibase 10.1103/PhysRevA.96.033826} {\bibfield
  {journal} {\bibinfo  {journal} {Phys. Rev. A}\ }\textbf {\bibinfo {volume}
  {96}},\ \bibinfo {pages} {033826} (\bibinfo {year} {2017})}\BibitemShut
  {NoStop}%
\bibitem [{\citenamefont {Sieberer}\ \emph {et~al.}(2013)\citenamefont
  {Sieberer}, \citenamefont {Huber}, \citenamefont {Altman},\ and\
  \citenamefont {Diehl}}]{sieberer2013dynamical}%
  \BibitemOpen
  \bibfield  {author} {\bibinfo {author} {\bibfnamefont {LM}~\bibnamefont
  {Sieberer}}, \bibinfo {author} {\bibfnamefont {Sebastian~D}\ \bibnamefont
  {Huber}}, \bibinfo {author} {\bibfnamefont {Ehud}\ \bibnamefont {Altman}}, \
  and\ \bibinfo {author} {\bibfnamefont {S}~\bibnamefont {Diehl}},\ }\bibfield
  {title} {\enquote {\bibinfo {title} {Dynamical critical phenomena in
  driven-dissipative systems},}\ }\href
  {https://link.aps.org/doi/10.1103/PhysRevLett.110.195301} {\bibfield
  {journal} {\bibinfo  {journal} {Phys. Rev. Lett.}\ }\textbf {\bibinfo
  {volume} {110}},\ \bibinfo {pages} {195301} (\bibinfo {year}
  {2013})}\BibitemShut {NoStop}%
\bibitem [{\citenamefont {Jin}\ \emph {et~al.}(2016)\citenamefont {Jin},
  \citenamefont {Biella}, \citenamefont {Viyuela}, \citenamefont {Mazza},
  \citenamefont {Keeling}, \citenamefont {Fazio},\ and\ \citenamefont
  {Rossini}}]{jin2016cluster}%
  \BibitemOpen
  \bibfield  {author} {\bibinfo {author} {\bibfnamefont {Jiasen}\ \bibnamefont
  {Jin}}, \bibinfo {author} {\bibfnamefont {Alberto}\ \bibnamefont {Biella}},
  \bibinfo {author} {\bibfnamefont {Oscar}\ \bibnamefont {Viyuela}}, \bibinfo
  {author} {\bibfnamefont {Leonardo}\ \bibnamefont {Mazza}}, \bibinfo {author}
  {\bibfnamefont {Jonathan}\ \bibnamefont {Keeling}}, \bibinfo {author}
  {\bibfnamefont {Rosario}\ \bibnamefont {Fazio}}, \ and\ \bibinfo {author}
  {\bibfnamefont {Davide}\ \bibnamefont {Rossini}},\ }\bibfield  {title}
  {\enquote {\bibinfo {title} {Cluster mean-field approach to the steady-state
  phase diagram of dissipative spin systems},}\ }\href {\doibase
  10.1103/PhysRevX.6.031011} {\bibfield  {journal} {\bibinfo  {journal} {Phys.
  Rev. X}\ }\textbf {\bibinfo {volume} {6}},\ \bibinfo {pages} {031011}
  (\bibinfo {year} {2016})}\BibitemShut {NoStop}%
\bibitem [{\citenamefont {Lee}\ \emph {et~al.}(2011)\citenamefont {Lee},
  \citenamefont {H\"affner},\ and\ \citenamefont
  {Cross}}]{lee2011antiferromagnetic}%
  \BibitemOpen
  \bibfield  {author} {\bibinfo {author} {\bibfnamefont {Tony~E.}\ \bibnamefont
  {Lee}}, \bibinfo {author} {\bibfnamefont {H.}~\bibnamefont {H\"affner}}, \
  and\ \bibinfo {author} {\bibfnamefont {M.~C.}\ \bibnamefont {Cross}},\
  }\bibfield  {title} {\enquote {\bibinfo {title} {Antiferromagnetic phase
  transition in a nonequilibrium lattice of rydberg atoms},}\ }\href {\doibase
  10.1103/PhysRevA.84.031402} {\bibfield  {journal} {\bibinfo  {journal} {Phys.
  Rev. A}\ }\textbf {\bibinfo {volume} {84}},\ \bibinfo {pages} {031402}
  (\bibinfo {year} {2011})}\BibitemShut {NoStop}%
\bibitem [{\citenamefont {Chan}\ \emph {et~al.}(2015)\citenamefont {Chan},
  \citenamefont {Lee},\ and\ \citenamefont {Gopalakrishnan}}]{chan2015limit}%
  \BibitemOpen
  \bibfield  {author} {\bibinfo {author} {\bibfnamefont {Ching-Kit}\
  \bibnamefont {Chan}}, \bibinfo {author} {\bibfnamefont {Tony~E.}\
  \bibnamefont {Lee}}, \ and\ \bibinfo {author} {\bibfnamefont {Sarang}\
  \bibnamefont {Gopalakrishnan}},\ }\bibfield  {title} {\enquote {\bibinfo
  {title} {Limit-cycle phase in driven-dissipative spin systems},}\ }\href
  {\doibase 10.1103/PhysRevA.91.051601} {\bibfield  {journal} {\bibinfo
  {journal} {Phys. Rev. A}\ }\textbf {\bibinfo {volume} {91}},\ \bibinfo
  {pages} {051601} (\bibinfo {year} {2015})}\BibitemShut {NoStop}%
\bibitem [{\citenamefont {Maghrebi}\ and\ \citenamefont
  {Gorshkov}(2016)}]{maghrebi2016nonequilibrium}%
  \BibitemOpen
  \bibfield  {author} {\bibinfo {author} {\bibfnamefont {Mohammad~F.}\
  \bibnamefont {Maghrebi}}\ and\ \bibinfo {author} {\bibfnamefont {Alexey~V.}\
  \bibnamefont {Gorshkov}},\ }\bibfield  {title} {\enquote {\bibinfo {title}
  {Nonequilibrium many-body steady states via keldysh formalism},}\ }\href
  {\doibase 10.1103/PhysRevB.93.014307} {\bibfield  {journal} {\bibinfo
  {journal} {Phys. Rev. B}\ }\textbf {\bibinfo {volume} {93}},\ \bibinfo
  {pages} {014307} (\bibinfo {year} {2016})}\BibitemShut {NoStop}%
\bibitem [{\citenamefont {Cai}\ \emph {et~al.}(2022)\citenamefont {Cai},
  \citenamefont {Liu}, \citenamefont {Jiang}, \citenamefont {Wu}, \citenamefont
  {Mei}, \citenamefont {Zhao}, \citenamefont {He}, \citenamefont {Zhang},
  \citenamefont {Zhou},\ and\ \citenamefont {Duan}}]{cai2022probing}%
  \BibitemOpen
  \bibfield  {author} {\bibinfo {author} {\bibfnamefont {M-L}\ \bibnamefont
  {Cai}}, \bibinfo {author} {\bibfnamefont {Z-D}\ \bibnamefont {Liu}}, \bibinfo
  {author} {\bibfnamefont {Y}~\bibnamefont {Jiang}}, \bibinfo {author}
  {\bibfnamefont {Y-K}\ \bibnamefont {Wu}}, \bibinfo {author} {\bibfnamefont
  {Q-X}\ \bibnamefont {Mei}}, \bibinfo {author} {\bibfnamefont {W-D}\
  \bibnamefont {Zhao}}, \bibinfo {author} {\bibfnamefont {L}~\bibnamefont
  {He}}, \bibinfo {author} {\bibfnamefont {X}~\bibnamefont {Zhang}}, \bibinfo
  {author} {\bibfnamefont {Z-C}\ \bibnamefont {Zhou}}, \ and\ \bibinfo {author}
  {\bibfnamefont {L-M}\ \bibnamefont {Duan}},\ }\bibfield  {title} {\enquote
  {\bibinfo {title} {Probing a dissipative phase transition with a trapped ion
  through reservoir engineering},}\ }\href
  {https://iopscience.iop.org/article/10.1088/0256-307X/39/2/020502} {\bibfield
   {journal} {\bibinfo  {journal} {Chinese Phys. Lett.}\ }\textbf {\bibinfo
  {volume} {39}},\ \bibinfo {pages} {020502} (\bibinfo {year}
  {2022})}\BibitemShut {NoStop}%
\bibitem [{\citenamefont {Ferioli}\ \emph {et~al.}(2023)\citenamefont
  {Ferioli}, \citenamefont {Glicenstein}, \citenamefont {Ferrier-Barbut},\ and\
  \citenamefont {Browaeys}}]{ferioli2022observation}%
  \BibitemOpen
  \bibfield  {author} {\bibinfo {author} {\bibfnamefont {Giovanni}\
  \bibnamefont {Ferioli}}, \bibinfo {author} {\bibfnamefont {Antoine}\
  \bibnamefont {Glicenstein}}, \bibinfo {author} {\bibfnamefont {Igor}\
  \bibnamefont {Ferrier-Barbut}}, \ and\ \bibinfo {author} {\bibfnamefont
  {Antoine}\ \bibnamefont {Browaeys}},\ }\bibfield  {title} {\enquote {\bibinfo
  {title} {A non-equilibrium superradiant phase transition in free space},}\
  }\href {https://doi.org/10.1038/s41567-023-02064-w} {\bibfield  {journal}
  {\bibinfo  {journal} {Nat. Phys.}\ }\textbf {\bibinfo {volume} {19}},\
  \bibinfo {pages} {1345--1349} (\bibinfo {year} {2023})}\BibitemShut {NoStop}%
\bibitem [{\citenamefont {Benary}\ \emph {et~al.}(2022)\citenamefont {Benary},
  \citenamefont {Baals}, \citenamefont {Bernhart}, \citenamefont {Jiang},
  \citenamefont {Röhrle},\ and\ \citenamefont {Ott}}]{benary2022experimental}%
  \BibitemOpen
  \bibfield  {author} {\bibinfo {author} {\bibfnamefont {J}~\bibnamefont
  {Benary}}, \bibinfo {author} {\bibfnamefont {C}~\bibnamefont {Baals}},
  \bibinfo {author} {\bibfnamefont {E}~\bibnamefont {Bernhart}}, \bibinfo
  {author} {\bibfnamefont {J}~\bibnamefont {Jiang}}, \bibinfo {author}
  {\bibfnamefont {M}~\bibnamefont {Röhrle}}, \ and\ \bibinfo {author}
  {\bibfnamefont {H}~\bibnamefont {Ott}},\ }\bibfield  {title} {\enquote
  {\bibinfo {title} {Experimental observation of a dissipative phase transition
  in a multi-mode many-body quantum system},}\ }\href {\doibase
  10.1088/1367-2630/ac97b6} {\bibfield  {journal} {\bibinfo  {journal} {New J.
  Phys.}\ }\textbf {\bibinfo {volume} {24}},\ \bibinfo {pages} {103034}
  (\bibinfo {year} {2022})}\BibitemShut {NoStop}%
\bibitem [{\citenamefont {Brennecke}\ \emph {et~al.}(2013)\citenamefont
  {Brennecke}, \citenamefont {Mottl}, \citenamefont {Baumann}, \citenamefont
  {Landig}, \citenamefont {Donner},\ and\ \citenamefont
  {Esslinger}}]{brennecke2013real}%
  \BibitemOpen
  \bibfield  {author} {\bibinfo {author} {\bibfnamefont {Ferdinand}\
  \bibnamefont {Brennecke}}, \bibinfo {author} {\bibfnamefont {Rafael}\
  \bibnamefont {Mottl}}, \bibinfo {author} {\bibfnamefont {Kristian}\
  \bibnamefont {Baumann}}, \bibinfo {author} {\bibfnamefont {Renate}\
  \bibnamefont {Landig}}, \bibinfo {author} {\bibfnamefont {Tobias}\
  \bibnamefont {Donner}}, \ and\ \bibinfo {author} {\bibfnamefont {Tilman}\
  \bibnamefont {Esslinger}},\ }\bibfield  {title} {\enquote {\bibinfo {title}
  {Real-time observation of fluctuations at the driven-dissipative dicke phase
  transition},}\ }\href {https://doi.org/10.1073/pnas.1306993110} {\bibfield
  {journal} {\bibinfo  {journal} {Proc. Natl. Acad. Sci.}\ }\textbf {\bibinfo
  {volume} {110}},\ \bibinfo {pages} {11763--11767} (\bibinfo {year}
  {2013})}\BibitemShut {NoStop}%
\bibitem [{\citenamefont {Ferri}\ \emph {et~al.}(2021)\citenamefont {Ferri},
  \citenamefont {Rosa-Medina}, \citenamefont {Finger}, \citenamefont {Dogra},
  \citenamefont {Soriente}, \citenamefont {Zilberberg}, \citenamefont
  {Donner},\ and\ \citenamefont {Esslinger}}]{ferri2021emerging}%
  \BibitemOpen
  \bibfield  {author} {\bibinfo {author} {\bibfnamefont {Francesco}\
  \bibnamefont {Ferri}}, \bibinfo {author} {\bibfnamefont {Rodrigo}\
  \bibnamefont {Rosa-Medina}}, \bibinfo {author} {\bibfnamefont {Fabian}\
  \bibnamefont {Finger}}, \bibinfo {author} {\bibfnamefont {Nishant}\
  \bibnamefont {Dogra}}, \bibinfo {author} {\bibfnamefont {Matteo}\
  \bibnamefont {Soriente}}, \bibinfo {author} {\bibfnamefont {Oded}\
  \bibnamefont {Zilberberg}}, \bibinfo {author} {\bibfnamefont {Tobias}\
  \bibnamefont {Donner}}, \ and\ \bibinfo {author} {\bibfnamefont {Tilman}\
  \bibnamefont {Esslinger}},\ }\bibfield  {title} {\enquote {\bibinfo {title}
  {Emerging dissipative phases in a superradiant quantum gas with tunable
  decay},}\ }\href {https://link.aps.org/doi/10.1103/PhysRevX.11.041046}
  {\bibfield  {journal} {\bibinfo  {journal} {Phys. Rev. X}\ }\textbf {\bibinfo
  {volume} {11}},\ \bibinfo {pages} {041046} (\bibinfo {year}
  {2021})}\BibitemShut {NoStop}%
\bibitem [{\citenamefont {Baumann}\ \emph {et~al.}(2010)\citenamefont
  {Baumann}, \citenamefont {Guerlin}, \citenamefont {Brennecke},\ and\
  \citenamefont {Esslinger}}]{baumann2010dicke}%
  \BibitemOpen
  \bibfield  {author} {\bibinfo {author} {\bibfnamefont {Kristian}\
  \bibnamefont {Baumann}}, \bibinfo {author} {\bibfnamefont {Christine}\
  \bibnamefont {Guerlin}}, \bibinfo {author} {\bibfnamefont {Ferdinand}\
  \bibnamefont {Brennecke}}, \ and\ \bibinfo {author} {\bibfnamefont {Tilman}\
  \bibnamefont {Esslinger}},\ }\bibfield  {title} {\enquote {\bibinfo {title}
  {Dicke quantum phase transition with a superfluid gas in an optical
  cavity},}\ }\href {https://www.nature.com/articles/nature09009} {\bibfield
  {journal} {\bibinfo  {journal} {Nature}\ }\textbf {\bibinfo {volume} {464}},\
  \bibinfo {pages} {1301--1306} (\bibinfo {year} {2010})}\BibitemShut {NoStop}%
\bibitem [{\citenamefont {Fink}\ \emph {et~al.}(2018)\citenamefont {Fink},
  \citenamefont {Schade}, \citenamefont {H{\"o}fling}, \citenamefont
  {Schneider},\ and\ \citenamefont {Imamoglu}}]{fink2018signatures}%
  \BibitemOpen
  \bibfield  {author} {\bibinfo {author} {\bibfnamefont {Thomas}\ \bibnamefont
  {Fink}}, \bibinfo {author} {\bibfnamefont {Anne}\ \bibnamefont {Schade}},
  \bibinfo {author} {\bibfnamefont {Sven}\ \bibnamefont {H{\"o}fling}},
  \bibinfo {author} {\bibfnamefont {Christian}\ \bibnamefont {Schneider}}, \
  and\ \bibinfo {author} {\bibfnamefont {Ata{\c{c}}}\ \bibnamefont
  {Imamoglu}},\ }\bibfield  {title} {\enquote {\bibinfo {title} {Signatures of
  a dissipative phase transition in photon correlation measurements},}\ }\href
  {https://www.nature.com/articles/s41567-017-0020-9} {\bibfield  {journal}
  {\bibinfo  {journal} {Nat. Phys.}\ }\textbf {\bibinfo {volume} {14}},\
  \bibinfo {pages} {365--369} (\bibinfo {year} {2018})}\BibitemShut {NoStop}%
\bibitem [{\citenamefont {Ohadi}\ \emph {et~al.}(2015)\citenamefont {Ohadi},
  \citenamefont {Dreismann}, \citenamefont {Rubo}, \citenamefont {Pinsker},
  \citenamefont {Redondo}, \citenamefont {Tsintzos}, \citenamefont
  {Hatzopoulos}, \citenamefont {Savvidis},\ and\ \citenamefont
  {Baumberg}}]{ohadi2015spontaneous}%
  \BibitemOpen
  \bibfield  {author} {\bibinfo {author} {\bibfnamefont {Hamid}\ \bibnamefont
  {Ohadi}}, \bibinfo {author} {\bibfnamefont {A}~\bibnamefont {Dreismann}},
  \bibinfo {author} {\bibfnamefont {YG}~\bibnamefont {Rubo}}, \bibinfo {author}
  {\bibfnamefont {F}~\bibnamefont {Pinsker}}, \bibinfo {author} {\bibfnamefont
  {Y~del Valle-Inclan}\ \bibnamefont {Redondo}}, \bibinfo {author}
  {\bibfnamefont {SI}~\bibnamefont {Tsintzos}}, \bibinfo {author}
  {\bibfnamefont {Z}~\bibnamefont {Hatzopoulos}}, \bibinfo {author}
  {\bibfnamefont {PG}~\bibnamefont {Savvidis}}, \ and\ \bibinfo {author}
  {\bibfnamefont {JJ}~\bibnamefont {Baumberg}},\ }\bibfield  {title} {\enquote
  {\bibinfo {title} {Spontaneous spin bifurcations and ferromagnetic phase
  transitions in a spinor exciton-polariton condensate},}\ }\href
  {https://link.aps.org/doi/10.1103/PhysRevX.5.031002} {\bibfield  {journal}
  {\bibinfo  {journal} {Phys. Rev. X}\ }\textbf {\bibinfo {volume} {5}},\
  \bibinfo {pages} {031002} (\bibinfo {year} {2015})}\BibitemShut {NoStop}%
\bibitem [{\citenamefont {Fitzpatrick}\ \emph {et~al.}(2017)\citenamefont
  {Fitzpatrick}, \citenamefont {Sundaresan}, \citenamefont {Li}, \citenamefont
  {Koch},\ and\ \citenamefont {Houck}}]{fitzpatrick2017observation}%
  \BibitemOpen
  \bibfield  {author} {\bibinfo {author} {\bibfnamefont {Mattias}\ \bibnamefont
  {Fitzpatrick}}, \bibinfo {author} {\bibfnamefont {Neereja~M}\ \bibnamefont
  {Sundaresan}}, \bibinfo {author} {\bibfnamefont {Andy~CY}\ \bibnamefont
  {Li}}, \bibinfo {author} {\bibfnamefont {Jens}\ \bibnamefont {Koch}}, \ and\
  \bibinfo {author} {\bibfnamefont {Andrew~A}\ \bibnamefont {Houck}},\
  }\bibfield  {title} {\enquote {\bibinfo {title} {Observation of a dissipative
  phase transition in a one-dimensional circuit {QED} lattice},}\ }\href
  {https://link.aps.org/doi/10.1103/PhysRevX.7.011016} {\bibfield  {journal}
  {\bibinfo  {journal} {Phys. Rev. X}\ }\textbf {\bibinfo {volume} {7}},\
  \bibinfo {pages} {011016} (\bibinfo {year} {2017})}\BibitemShut {NoStop}%
\bibitem [{\citenamefont {Collodo}\ \emph {et~al.}(2019)\citenamefont
  {Collodo}, \citenamefont {Poto{\v{c}}nik}, \citenamefont {Gasparinetti},
  \citenamefont {Besse}, \citenamefont {Pechal}, \citenamefont {Sameti},
  \citenamefont {Hartmann}, \citenamefont {Wallraff},\ and\ \citenamefont
  {Eichler}}]{collodo2019observation}%
  \BibitemOpen
  \bibfield  {author} {\bibinfo {author} {\bibfnamefont {Michele~C}\
  \bibnamefont {Collodo}}, \bibinfo {author} {\bibfnamefont {Anton}\
  \bibnamefont {Poto{\v{c}}nik}}, \bibinfo {author} {\bibfnamefont {Simone}\
  \bibnamefont {Gasparinetti}}, \bibinfo {author} {\bibfnamefont {Jean-Claude}\
  \bibnamefont {Besse}}, \bibinfo {author} {\bibfnamefont {Marek}\ \bibnamefont
  {Pechal}}, \bibinfo {author} {\bibfnamefont {Mahdi}\ \bibnamefont {Sameti}},
  \bibinfo {author} {\bibfnamefont {Michael~J}\ \bibnamefont {Hartmann}},
  \bibinfo {author} {\bibfnamefont {Andreas}\ \bibnamefont {Wallraff}}, \ and\
  \bibinfo {author} {\bibfnamefont {Christopher}\ \bibnamefont {Eichler}},\
  }\bibfield  {title} {\enquote {\bibinfo {title} {Observation of the crossover
  from photon ordering to delocalization in tunably coupled resonators},}\
  }\href {https://link.aps.org/doi/10.1103/PhysRevLett.122.183601} {\bibfield
  {journal} {\bibinfo  {journal} {Phys. Rev. Lett.}\ }\textbf {\bibinfo
  {volume} {122}},\ \bibinfo {pages} {183601} (\bibinfo {year}
  {2019})}\BibitemShut {NoStop}%
\bibitem [{\citenamefont {Ding}\ \emph {et~al.}(2020)\citenamefont {Ding},
  \citenamefont {Busche}, \citenamefont {Shi}, \citenamefont {Guo},\ and\
  \citenamefont {Adams}}]{ding2016phase}%
  \BibitemOpen
  \bibfield  {author} {\bibinfo {author} {\bibfnamefont {Dong-Sheng}\
  \bibnamefont {Ding}}, \bibinfo {author} {\bibfnamefont {Hannes}\ \bibnamefont
  {Busche}}, \bibinfo {author} {\bibfnamefont {Bao-Sen}\ \bibnamefont {Shi}},
  \bibinfo {author} {\bibfnamefont {Guang-Can}\ \bibnamefont {Guo}}, \ and\
  \bibinfo {author} {\bibfnamefont {Charles~S.}\ \bibnamefont {Adams}},\
  }\bibfield  {title} {\enquote {\bibinfo {title} {Phase diagram and
  self-organizing dynamics in a thermal ensemble of strongly interacting
  rydberg atoms},}\ }\href {\doibase 10.1103/PhysRevX.10.021023} {\bibfield
  {journal} {\bibinfo  {journal} {Phys. Rev. X}\ }\textbf {\bibinfo {volume}
  {10}},\ \bibinfo {pages} {021023} (\bibinfo {year} {2020})}\BibitemShut
  {NoStop}%
\bibitem [{\citenamefont {Beaulieu}\ \emph {et~al.}(2025)\citenamefont
  {Beaulieu}, \citenamefont {Minganti}, \citenamefont {Frasca}, \citenamefont
  {Savona}, \citenamefont {Felicetti}, \citenamefont {Di~Candia},\ and\
  \citenamefont
  {Scarlino}}]{beaulieu2023observationfirstsecondorderdissipative}%
  \BibitemOpen
  \bibfield  {author} {\bibinfo {author} {\bibfnamefont {Guillaume}\
  \bibnamefont {Beaulieu}}, \bibinfo {author} {\bibfnamefont {Fabrizio}\
  \bibnamefont {Minganti}}, \bibinfo {author} {\bibfnamefont {Simone}\
  \bibnamefont {Frasca}}, \bibinfo {author} {\bibfnamefont {Vincenzo}\
  \bibnamefont {Savona}}, \bibinfo {author} {\bibfnamefont {Simone}\
  \bibnamefont {Felicetti}}, \bibinfo {author} {\bibfnamefont {Roberto}\
  \bibnamefont {Di~Candia}}, \ and\ \bibinfo {author} {\bibfnamefont
  {Pasquale}\ \bibnamefont {Scarlino}},\ }\bibfield  {title} {\enquote
  {\bibinfo {title} {Observation of first- and second-order dissipative phase
  transitions in a two-photon driven kerr resonator},}\ }\href {\doibase
  10.1038/s41467-025-56830-w} {\bibfield  {journal} {\bibinfo  {journal}
  {Nature Communications}\ }\textbf {\bibinfo {volume} {16}},\ \bibinfo {pages}
  {1954} (\bibinfo {year} {2025})}\BibitemShut {NoStop}%
\bibitem [{\citenamefont {Xie}\ \emph {et~al.}(2020)\citenamefont {Xie},
  \citenamefont {Geng}, \citenamefont {Yu}, \citenamefont {Rong}, \citenamefont
  {Wang},\ and\ \citenamefont {Du}}]{xie2020dissipative}%
  \BibitemOpen
  \bibfield  {author} {\bibinfo {author} {\bibfnamefont {Yijin}\ \bibnamefont
  {Xie}}, \bibinfo {author} {\bibfnamefont {Jianpei}\ \bibnamefont {Geng}},
  \bibinfo {author} {\bibfnamefont {Huiyao}\ \bibnamefont {Yu}}, \bibinfo
  {author} {\bibfnamefont {Xing}\ \bibnamefont {Rong}}, \bibinfo {author}
  {\bibfnamefont {Ya}~\bibnamefont {Wang}}, \ and\ \bibinfo {author}
  {\bibfnamefont {Jiangfeng}\ \bibnamefont {Du}},\ }\bibfield  {title}
  {\enquote {\bibinfo {title} {Dissipative quantum sensing with a magnetometer
  based on nitrogen-vacancy centers in diamond},}\ }\href
  {https://doi.org/10.1103/PhysRevApplied.14.014013} {\bibfield  {journal}
  {\bibinfo  {journal} {Physical Review Applied}\ }\textbf {\bibinfo {volume}
  {14}},\ \bibinfo {pages} {014013} (\bibinfo {year} {2020})}\BibitemShut
  {NoStop}%
\bibitem [{\citenamefont {Raghunandan}\ \emph {et~al.}(2018)\citenamefont
  {Raghunandan}, \citenamefont {Wrachtrup},\ and\ \citenamefont
  {Weimer}}]{raghunandan2018highdensity}%
  \BibitemOpen
  \bibfield  {author} {\bibinfo {author} {\bibfnamefont {Meghana}\ \bibnamefont
  {Raghunandan}}, \bibinfo {author} {\bibfnamefont {Jörg}\ \bibnamefont
  {Wrachtrup}}, \ and\ \bibinfo {author} {\bibfnamefont {Hendrik}\ \bibnamefont
  {Weimer}},\ }\bibfield  {title} {\enquote {\bibinfo {title} {High-density
  quantum sensing with dissipative first order transitions},}\ }\href {\doibase
  10.1103/physrevlett.120.150501} {\bibfield  {journal} {\bibinfo  {journal}
  {Physical Review Letters}\ }\textbf {\bibinfo {volume} {120}} (\bibinfo
  {year} {2018}),\ 10.1103/physrevlett.120.150501}\BibitemShut {NoStop}%
\bibitem [{\citenamefont {Ivanov}(2020)}]{ivanov2020enhanced}%
  \BibitemOpen
  \bibfield  {author} {\bibinfo {author} {\bibfnamefont {Peter~A}\ \bibnamefont
  {Ivanov}},\ }\bibfield  {title} {\enquote {\bibinfo {title} {Enhanced
  two-parameter phase-space-displacement estimation close to a dissipative
  phase transition},}\ }\href@noop {} {\bibfield  {journal} {\bibinfo
  {journal} {Physical Review A}\ }\textbf {\bibinfo {volume} {102}},\ \bibinfo
  {pages} {052611} (\bibinfo {year} {2020})}\BibitemShut {NoStop}%
\bibitem [{\citenamefont {Arandes}\ and\ \citenamefont
  {Bergholtz}(2025)}]{arandes2025quantum}%
  \BibitemOpen
  \bibfield  {author} {\bibinfo {author} {\bibfnamefont {Oscar}\ \bibnamefont
  {Arandes}}\ and\ \bibinfo {author} {\bibfnamefont {Emil~J.}\ \bibnamefont
  {Bergholtz}},\ }\bibfield  {title} {\enquote {\bibinfo {title} {Quantum
  sensing with driven-dissipative su-schrieffer-heeger lattices},}\ }\href
  {\doibase 10.1103/PhysRevResearch.7.013309} {\bibfield  {journal} {\bibinfo
  {journal} {Phys. Rev. Res.}\ }\textbf {\bibinfo {volume} {7}},\ \bibinfo
  {pages} {013309} (\bibinfo {year} {2025})}\BibitemShut {NoStop}%
\bibitem [{\citenamefont {Montenegro}\ \emph {et~al.}(2023)\citenamefont
  {Montenegro}, \citenamefont {Genoni}, \citenamefont {Bayat},\ and\
  \citenamefont {Paris}}]{montenegro2023quantum}%
  \BibitemOpen
  \bibfield  {author} {\bibinfo {author} {\bibfnamefont {Victor}\ \bibnamefont
  {Montenegro}}, \bibinfo {author} {\bibfnamefont {Marco~G}\ \bibnamefont
  {Genoni}}, \bibinfo {author} {\bibfnamefont {Abolfazl}\ \bibnamefont
  {Bayat}}, \ and\ \bibinfo {author} {\bibfnamefont {Matteo~GA}\ \bibnamefont
  {Paris}},\ }\bibfield  {title} {\enquote {\bibinfo {title} {Quantum metrology
  with boundary time crystals},}\ }\href
  {https://www.nature.com/articles/s42005-023-01423-6} {\bibfield  {journal}
  {\bibinfo  {journal} {Commun. Phys.}\ }\textbf {\bibinfo {volume} {6}},\
  \bibinfo {pages} {304} (\bibinfo {year} {2023})}\BibitemShut {NoStop}%
\bibitem [{\citenamefont {Gribben}\ \emph {et~al.}(2025)\citenamefont
  {Gribben}, \citenamefont {Sanpera}, \citenamefont {Fazio}, \citenamefont
  {Marino},\ and\ \citenamefont {Iemini}}]{gribben2025boundary}%
  \BibitemOpen
  \bibfield  {author} {\bibinfo {author} {\bibfnamefont {Dominic}\ \bibnamefont
  {Gribben}}, \bibinfo {author} {\bibfnamefont {Anna}\ \bibnamefont {Sanpera}},
  \bibinfo {author} {\bibfnamefont {Rosario}\ \bibnamefont {Fazio}}, \bibinfo
  {author} {\bibfnamefont {Jamir}\ \bibnamefont {Marino}}, \ and\ \bibinfo
  {author} {\bibfnamefont {Fernando}\ \bibnamefont {Iemini}},\ }\bibfield
  {title} {\enquote {\bibinfo {title} {Boundary time crystals as ac sensors:
  Enhancements and constraints},}\ }\href@noop {} {\bibfield  {journal}
  {\bibinfo  {journal} {SciPost Physics}\ }\textbf {\bibinfo {volume} {18}},\
  \bibinfo {pages} {100} (\bibinfo {year} {2025})}\BibitemShut {NoStop}%
\bibitem [{\citenamefont {Pavlov}\ \emph {et~al.}(2023)\citenamefont {Pavlov},
  \citenamefont {Porras},\ and\ \citenamefont {Ivanov}}]{pavlov2023quantum}%
  \BibitemOpen
  \bibfield  {author} {\bibinfo {author} {\bibfnamefont {Venelin~P}\
  \bibnamefont {Pavlov}}, \bibinfo {author} {\bibfnamefont {Diego}\
  \bibnamefont {Porras}}, \ and\ \bibinfo {author} {\bibfnamefont {Peter~A}\
  \bibnamefont {Ivanov}},\ }\bibfield  {title} {\enquote {\bibinfo {title}
  {Quantum metrology with critical driven-dissipative collective spin
  system},}\ }\href {\doibase 10.1088/1402-4896/ace99f} {\bibfield  {journal}
  {\bibinfo  {journal} {Physica Scripta}\ }\textbf {\bibinfo {volume} {98}},\
  \bibinfo {pages} {095103} (\bibinfo {year} {2023})}\BibitemShut {NoStop}%
\bibitem [{\citenamefont {Cram{\'e}r}(1999)}]{cramer1999mathematical}%
  \BibitemOpen
  \bibfield  {author} {\bibinfo {author} {\bibfnamefont {Harald}\ \bibnamefont
  {Cram{\'e}r}},\ }\href@noop {} {\emph {\bibinfo {title} {Mathematical methods
  of statistics}}},\ Vol.~\bibinfo {volume} {26}\ (\bibinfo  {publisher}
  {Princeton university press},\ \bibinfo {year} {1999})\BibitemShut {NoStop}%
\bibitem [{\citenamefont {Le~Cam}(1986)}]{LeCam-1986}%
  \BibitemOpen
  \bibfield  {author} {\bibinfo {author} {\bibfnamefont {Lucien~M.}\
  \bibnamefont {Le~Cam}},\ }\href@noop {} {\emph {\bibinfo {title} {Asymptotic
  methods in statistical decision theory}}},\ Springer series in statistics\
  (\bibinfo  {publisher} {Springer-Verlag},\ \bibinfo {address} {New York},\
  \bibinfo {year} {1986})\BibitemShut {NoStop}%
\bibitem [{\citenamefont {Rao}(1992)}]{rao1992information}%
  \BibitemOpen
  \bibfield  {author} {\bibinfo {author} {\bibfnamefont {C~Radhakrishna}\
  \bibnamefont {Rao}},\ }\bibfield  {title} {\enquote {\bibinfo {title}
  {Information and the accuracy attainable in the estimation of statistical
  parameters},}\ }in\ \href@noop {} {\emph {\bibinfo {booktitle} {Breakthroughs
  in Statistics: Foundations and basic theory}}}\ (\bibinfo  {publisher}
  {Springer},\ \bibinfo {year} {1992})\ pp.\ \bibinfo {pages}
  {235--247}\BibitemShut {NoStop}%
\bibitem [{\citenamefont {Paris}\ and\ \citenamefont
  {Rehacek}(2004)}]{paris2004quantum}%
  \BibitemOpen
  \bibfield  {author} {\bibinfo {author} {\bibfnamefont {Matteo}\ \bibnamefont
  {Paris}}\ and\ \bibinfo {author} {\bibfnamefont {Jaroslav}\ \bibnamefont
  {Rehacek}},\ }\href@noop {} {\emph {\bibinfo {title} {Quantum state
  estimation}}},\ Vol.\ \bibinfo {volume} {649}\ (\bibinfo  {publisher}
  {Springer Science \& Business Media},\ \bibinfo {year} {2004})\BibitemShut
  {NoStop}%
\bibitem [{\citenamefont {Helstrom}(1976)}]{helstrom1976quantum}%
  \BibitemOpen
  \bibfield  {author} {\bibinfo {author} {\bibfnamefont {Carl~W}\ \bibnamefont
  {Helstrom}},\ }\href@noop {} {\emph {\bibinfo {title} {Quantum Detection and
  Estimation Theory}}}\ (\bibinfo  {publisher} {Academic Press},\ \bibinfo
  {year} {1976})\BibitemShut {NoStop}%
\bibitem [{\citenamefont {Van~Trees}(2004)}]{vantrees2004detection}%
  \BibitemOpen
  \bibfield  {author} {\bibinfo {author} {\bibfnamefont {Harry~L.}\
  \bibnamefont {Van~Trees}},\ }\href {\doibase 10.1002/0471221082} {\emph
  {\bibinfo {title} {Detection, Estimation, and Modulation Theory, Part I}}},\
  \bibinfo {edition} {2nd}\ ed.\ (\bibinfo  {publisher} {Wiley-Interscience},\
  \bibinfo {year} {2004})\BibitemShut {NoStop}%
\bibitem [{\citenamefont {Holevo}(2011)}]{holevo2011probabilistic}%
  \BibitemOpen
  \bibfield  {author} {\bibinfo {author} {\bibfnamefont {A.S.}\ \bibnamefont
  {Holevo}},\ }\href@noop {} {\emph {\bibinfo {title} {Probabilistic and
  Statistical Aspects of Quantum Theory}}}\ (\bibinfo  {publisher} {Edizioni
  della Normale, Pisa},\ \bibinfo {year} {2011})\BibitemShut {NoStop}%
\bibitem [{\citenamefont {Braunstein}\ and\ \citenamefont
  {Caves}(1994)}]{braunstein1994statistical}%
  \BibitemOpen
  \bibfield  {author} {\bibinfo {author} {\bibfnamefont {Samuel~L}\
  \bibnamefont {Braunstein}}\ and\ \bibinfo {author} {\bibfnamefont
  {Carlton~M}\ \bibnamefont {Caves}},\ }\bibfield  {title} {\enquote {\bibinfo
  {title} {Statistical distance and the geometry of quantum states},}\ }\href
  {https://link.aps.org/doi/10.1103/PhysRevLett.72.3439} {\bibfield  {journal}
  {\bibinfo  {journal} {Phys. Rev. Lett.}\ }\textbf {\bibinfo {volume} {72}},\
  \bibinfo {pages} {3439} (\bibinfo {year} {1994})}\BibitemShut {NoStop}%
\bibitem [{\citenamefont {Lambert}\ \emph {et~al.}(2024)\citenamefont
  {Lambert}, \citenamefont {Giguère}, \citenamefont {Menczel}, \citenamefont
  {Li}, \citenamefont {Hopf}, \citenamefont {Suárez}, \citenamefont {Gali},
  \citenamefont {Lishman}, \citenamefont {Gadhvi}, \citenamefont {Agarwal},
  \citenamefont {Galicia}, \citenamefont {Shammah}, \citenamefont {Nation},
  \citenamefont {Johansson}, \citenamefont {Ahmed}, \citenamefont {Cross},
  \citenamefont {Pitchford},\ and\ \citenamefont
  {Nori}}]{lambert2024qutip5quantumtoolbox}%
  \BibitemOpen
  \bibfield  {author} {\bibinfo {author} {\bibfnamefont {Neill}\ \bibnamefont
  {Lambert}}, \bibinfo {author} {\bibfnamefont {Eric}\ \bibnamefont
  {Giguère}}, \bibinfo {author} {\bibfnamefont {Paul}\ \bibnamefont
  {Menczel}}, \bibinfo {author} {\bibfnamefont {Boxi}\ \bibnamefont {Li}},
  \bibinfo {author} {\bibfnamefont {Patrick}\ \bibnamefont {Hopf}}, \bibinfo
  {author} {\bibfnamefont {Gerardo}\ \bibnamefont {Suárez}}, \bibinfo {author}
  {\bibfnamefont {Marc}\ \bibnamefont {Gali}}, \bibinfo {author} {\bibfnamefont
  {Jake}\ \bibnamefont {Lishman}}, \bibinfo {author} {\bibfnamefont {Rushiraj}\
  \bibnamefont {Gadhvi}}, \bibinfo {author} {\bibfnamefont {Rochisha}\
  \bibnamefont {Agarwal}}, \bibinfo {author} {\bibfnamefont {Asier}\
  \bibnamefont {Galicia}}, \bibinfo {author} {\bibfnamefont {Nathan}\
  \bibnamefont {Shammah}}, \bibinfo {author} {\bibfnamefont {Paul}\
  \bibnamefont {Nation}}, \bibinfo {author} {\bibfnamefont {J.~R.}\
  \bibnamefont {Johansson}}, \bibinfo {author} {\bibfnamefont {Shahnawaz}\
  \bibnamefont {Ahmed}}, \bibinfo {author} {\bibfnamefont {Simon}\ \bibnamefont
  {Cross}}, \bibinfo {author} {\bibfnamefont {Alexander}\ \bibnamefont
  {Pitchford}}, \ and\ \bibinfo {author} {\bibfnamefont {Franco}\ \bibnamefont
  {Nori}},\ }\bibfield  {title} {\enquote {\bibinfo {title} {Qutip 5: The
  quantum toolbox in python},}\ }\href {https://arxiv.org/abs/2412.04705} {\
  (\bibinfo {year} {2024})},\ \Eprint {http://arxiv.org/abs/2412.04705}
  {arXiv:2412.04705 [quant-ph]} \BibitemShut {NoStop}%
\bibitem [{SM()}]{SM}%
  \BibitemOpen
  \href@noop {} {\bibinfo  {journal} {See Supplemental Material for details}\
  }\BibitemShut {NoStop}%
\end{thebibliography}%

\clearpage
\onecolumngrid
\pagebreak
\widetext

\begin{center}
\textbf{\large Supplemental Material: Near-Ultimate Quantum-Enhanced Sensitivity in Dissipative Critical Sensing with Partial Access}
\end{center}

\date{\today}
\setcounter{equation}{0}
\setcounter{figure}{0}
\setcounter{table}{0}
\setcounter{page}{1}
\makeatletter
\renewcommand{\theequation}{S\arabic{equation}}
\renewcommand{\thefigure}{S\arabic{figure}}

\section{Quantum Entanglement and Purity Analysis}

In the middle and bottom panels of Fig.~\eqref{fig_on_resonance} (on-resonance case), we observe that the qubit's steady state of the qubit–field system abruptly polarizes along a single axis, while the field subsystem bifurcates into two bistable locations in phase space. Similarly, for the off-resonance scenario (middle and bottom panels of Fig.~\eqref{fig_off_resonance}), the steady state exhibits analogous behavior. The key difference is that, by appropriately choosing a detuning $\Delta \neq 0$, one can select a preferred location in phase space.

This behavior indicates, as discussed in the main text, that the system becomes weakly entangled and, due to the spontaneous polarization of the qubit, the steady state also becomes purified. To quantify this, we evaluate the entanglement of the steady state using the logarithmic negativity---a measure of entanglement valid for mixed states---defined as
\begin{equation}
E_N(\rho_\mathrm{SS}) = \log_2 ||\rho_\mathrm{SS}^{T_A}||_1,
\end{equation}
where $\rho_\mathrm{SS}^{T_A}$ denotes the partial transpose of the steady-state density matrix with respect to one subsystem, and $||\cdot||_1$ is the trace norm.

In addition, we compute the purity of the steady state as
\begin{equation}
\mathcal{P}(\rho_\mathrm{SS}) = \mathrm{Tr}[\rho_\mathrm{SS}^2].
\end{equation}

\begin{figure}[b]
    \centering
    \includegraphics[width=0.925\linewidth]{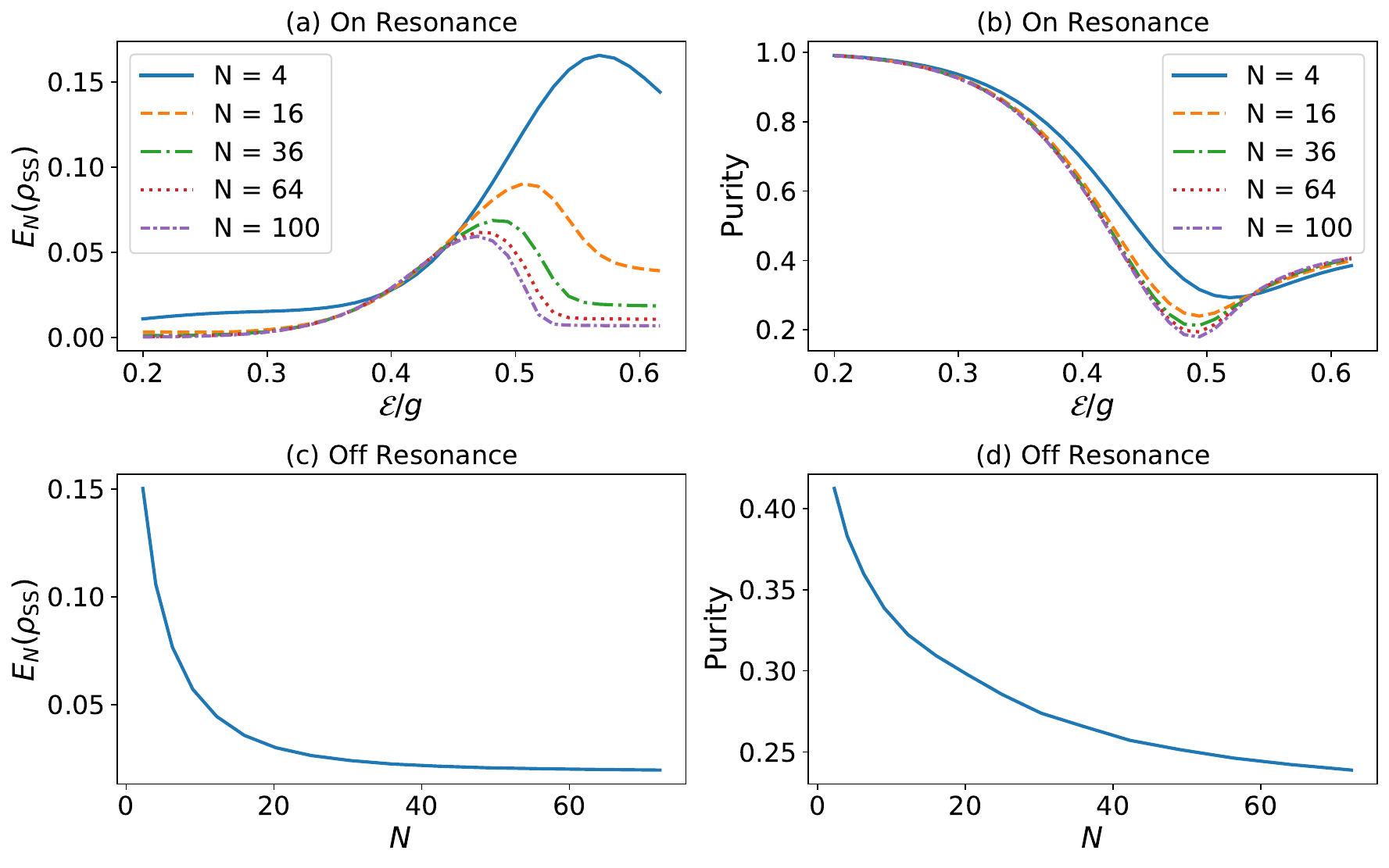}
    \caption{On resonant: (a) Logarithmic negativity $E_N(\rho_\mathrm{SS})$ plotted as a function of the driving amplitude $\mathcal{E}/g$ for different sensing resources $N$. (b) Purity as a function of the driving amplitude $\mathcal{E}/g$ for several values of $N$. Off resonant: (c) Logarithmic negativity as a function of $N$ at the optimal parameters $[(\Delta/g)^*, (\mathcal{E}/g)^*]$. (d) Purity as a function of $N$ at the optimal parameters $[(\Delta/g)^*, (\mathcal{E}/g)^*]$.}
    \label{fig_entanglement_purity}
\end{figure}

In the left column of Fig.~\ref{fig_entanglement_purity}, we show the logarithmic negativity for both the on-resonance and off-resonance cases. In Fig.~\ref{fig_entanglement_purity}(a), the logarithmic negativity is plotted as a function of the driving amplitude $\mathcal{E}/g$ for different sensing resources $N$. As the figure illustrates, the steady-state entanglement drops sharply both as $N$ increases and near the critical point, indicating that the qubit and field become only weakly entangled.

A similar trend is observed in the off-resonance case, shown in Fig.~\ref{fig_entanglement_purity}(c), where we plot the logarithmic negativity as a function of $N$ at the optimal parameters $[(\Delta/g)^*, (\mathcal{E}/g)^*]$. Here, the logarithmic negativity decreases as the sensing resource $N$ grows, signaling a partial disentanglement of the steady state in the thermodynamic limit.

In Fig.~\ref{fig_entanglement_purity}(b), we show the purity for the on-resonance case as a function of the driving amplitude $\mathcal{E}/g$ for several values of $N$. As the figure illustrates, consistent with the abrupt spontaneous polarization of the qubit, the purity exhibits a dip around the critical point. For larger driving amplitudes, the system becomes increasingly mixed.

For the off-resonance scenario, Fig.~\ref{fig_entanglement_purity}(d) shows the purity as a function of $N$ at the optimal parameters $[(\Delta/g)^*, (\mathcal{E}/g)^*]$. As the figure clearly demonstrates, the system becomes progressively more pure as $N$ increases. As stated in the main text, the combination of weak entanglement and increased purification allows the field subsystem, when measured locally via a homodyne scheme and analyzed with a Bayesian estimator, to nearly saturate the ultimate sensing precision of the entire probe.

\section{Continuous-Variable Bayesian Estimation}

This section clarifies the Bayesian estimation procedure illustrated in Fig.~\ref{fig_bayesian} of the main text.

In a nutshell, Bayesian analysis is a powerful parameter estimation method in which observed measurement outcomes are used to update a probability distribution for an unknown parameter, based on a given statistical model. This approach relies on three key ingredients: measurement outcomes, a statistical model (i.e., modeled probability distributions), and an update rule. The more likely a measurement outcome is under a specific model, the higher the likelihood assigned to the corresponding parameter value. Bayesian estimation also uses prior knowledge (or beliefs) about the parameter in question. This prior information is updated step-by-step using new data, yielding increasingly refined estimates of the parameter. At each step, the updated probability distribution (called the posterior) represents the best guess about the parameter given all the information available so far.

Mathematically, this update is governed by Bayes' rule:
\begin{equation}
P(\mathcal{E} | \mathrm{data}) = \frac{P(\mathrm{data} | \mathcal{E})  P(\mathcal{E})}{P(\mathrm{data})},
\end{equation}
where $\mathcal{E}$ is the unknown parameter we aim to estimate, and $\mathrm{\textit{data}}$ represents the observed outcomes. Here, $P(\mathcal{E})$ is the \textit{prior} distribution (the initial belief about $\mathcal{E}$), and $P(\mathrm{data} | \mathcal{E})$ is the \textit{likelihood} function---i.e., the probability of observing the data given the parameter $\mathcal{E}$; this constitutes the statistical model and can typically be computed or numerically simulated. The \textit{posterior} distribution $P(\mathcal{E} | \mathrm{data})$ reflects the updated belief about $\mathcal{E}$ after incorporating the observed data. Finally, $P(\mathrm{data})$, known as the \textit{evidence}, ensures normalization so that the posterior remains a valid probability distribution.

\subsection{Interlude I: Homodyne measurement}

As discussed above, three key ingredients are required for Bayesian estimation: a statistical model, measurement data, and an update rule. We have already introduced the update rule---Bayes' rule---and the measurement data will be (simulated) experimentally obtained. What remains is to specify the statistical model and its corresponding probability distributions. To construct this, we must define a measurement basis.

From the main text, we know that the subsystem offering the highest sensitivity across all values of $\mathcal{E}$ is the field degree of freedom. Therefore, we focus exclusively on measurements performed on the field state. Rather than pursuing optimal measurement bases---which are often experimentally infeasible---we restrict our attention to a readily available continuous-variable (CV) field measurements: homodyne detection. This is because, as shown in the main text, the bistable states and their preferred localization in phase space yield a positive Wigner quasiprobability distribution with an approximately Gaussian shape. This makes homodyne detection an excellent candidate for nearly saturating the ultimate precision bound in our sensing task.

Let us now construct the corresponding probability distributions from:
\begin{equation}
    p(X_\varphi|\mathcal{E})=\langle X_\varphi|\rho_\mathrm{SS}^\mathrm{field}|X_\varphi\rangle,\label{eq_SM_homodyne}
\end{equation}
where $\rho_\mathrm{SS}^\mathrm{field}=\mathrm{Tr}_\mathrm{qubit}[\rho_\mathrm{SS}]$ is the reduced density matrix of the field subsystem and $|X_\varphi\rangle$ is the eigenstate of the rotated field quadrature
\begin{equation}
    \hat{X}_\varphi = \frac{1}{\sqrt{2}} \left( \hat{a} e^{-i\varphi} + \hat{a}^\dagger e^{i\varphi} \right),
\end{equation}
where $\hat{a}$ and $\hat{a}^\dagger$ are the annihilation and creation operators of the field and $\varphi$ is the homodyne phase angle.

Hence, as a very first step in our estimation procedure, we optimize the measurement basis by identifying the value of $\varphi$ that yields the highest sensing performance within the chosen homodyne detection scheme. This optimal value of $\varphi$ is determined by maximizing the sensing performance for our specific homodyne scheme, namely by maximizing the classical Fisher information. As discussed in the main text, the CFI quantifies the sensitivity with respect to $\mathcal{E}$ for a given measurement. For clarity, we recall the expression for the CFI here as:
\begin{equation}
    \mathcal{F}(\varphi|\mathcal{E}) = \int_{-\infty}^\infty dX_\varphi \frac{1}{p(X_\varphi|\mathcal{E})} \left[ \frac{\partial p(X_\varphi|\mathcal{E})}{\partial \mathcal{E}} \right]^2, \label{eq_SM_cfi_homodyne}
\end{equation}
where $p(X_\varphi|\mathcal{E})$ accounts for the conditional probability of obtaining the value $X_\varphi$ assuming the value $\mathcal{E}$. To determine the optimal measurement basis, we maximize the homodyne angle $\varphi^*$ for a given set of parameters $\left(\frac{\mathcal{E}}{g}\right)^*_j$ and $\left(\frac{\Delta}{g}\right)^*_j$ and a given sensing resource $N$:
\begin{equation}
    \varphi^*=\arg\max_\varphi \mathcal{F}(\varphi|\mathcal{E}).
\end{equation}

\begin{figure}
    \centering
    \includegraphics[width=0.75\linewidth]{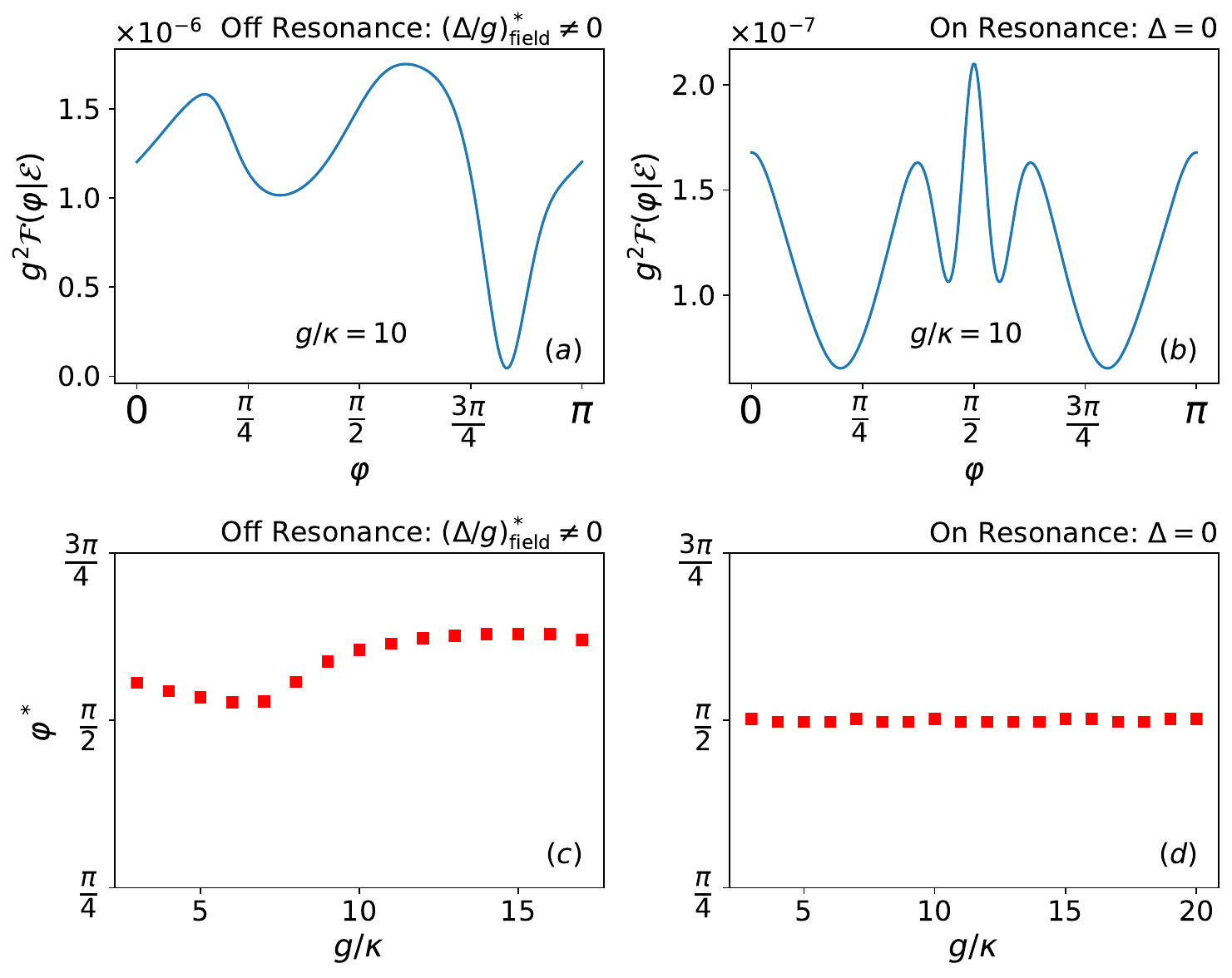}
    \caption{(a)-(b) CFI as a function of the homodyne angle $\varphi$ for a fixed value of $g/\kappa = 10$ for the off-resonance and on-resonance cases, respectively. (c)-(d) Optimized angle $\varphi^*$ as a function of $g/\kappa$ for the off-resonance and on-resonance scenarios, respectively.}
    \label{fig_homodyne_phase}
\end{figure}

In Figs.~\ref{fig_homodyne_phase}(a)–(b), we plot the CFI as a function of the homodyne angle $\varphi$ for a fixed value of $g/\kappa = 10$ for the off-resonance and on-resonance cases, respectively. From these plots, one can directly identify the optimal angle $\varphi^*$ that maximizes the CFI. In Figs.~\ref{fig_homodyne_phase}(c)–(d), we show how the optimized angle $\varphi^*$ varies with $g/\kappa$ for the off-resonance and on-resonance scenarios, respectively. As observed from the figures, in the off-resonance case, the optimal homodyne angle $\varphi^*$ oscillates within the range $\pi/2$ to $3\pi/4$. In contrast, for the on-resonance case, the optimal angle remains nearly constant around $\varphi^* \approx \pi/2$ across the entire range of $g/\kappa$ values considered.

\subsection{Interlude II: Heterodyne measurement}

Heterodyne measurement is another powerful experimental technique for probing the field state. In the main text, we argued that the homodyne scheme provides superior sensing performance. To quantify this claim, we define:
\begin{equation}
\text{Performance} = \frac{\mathcal{F}_\mathrm{field}(\mathcal{E})}{Q_\mathrm{whole}(\mathcal{E})},
\end{equation}
where $\mathcal{F}_\mathrm{field}(\mathcal{E})$ is the classical Fisher information, as described in the section ``Quantum Estimation Background" of the main text. It is given by Eq.~\eqref{eq_SM_cfi_homodyne} for the homodyne case, and by
\begin{equation}
\mathcal{F}_\mathrm{field}(\mathcal{E}) = \int d^2\Upsilon \frac{1}{p(\Upsilon|\mathcal{E})} \left( \frac{\partial p(\Upsilon|\mathcal{E})}{\partial \mathcal{E}} \right)^2
\end{equation}
for the heterodyne case, where the probability distribution is
\begin{equation}
p(\Upsilon|\mathcal{E}) = \frac{1}{\pi} \mathrm{Tr}\big[ |\Upsilon\rangle \langle \Upsilon| \rho \big],
\end{equation}
and $|\Upsilon\rangle$ is a coherent state.

\begin{figure}[t]
    \centering
    \includegraphics[width=0.45\linewidth]{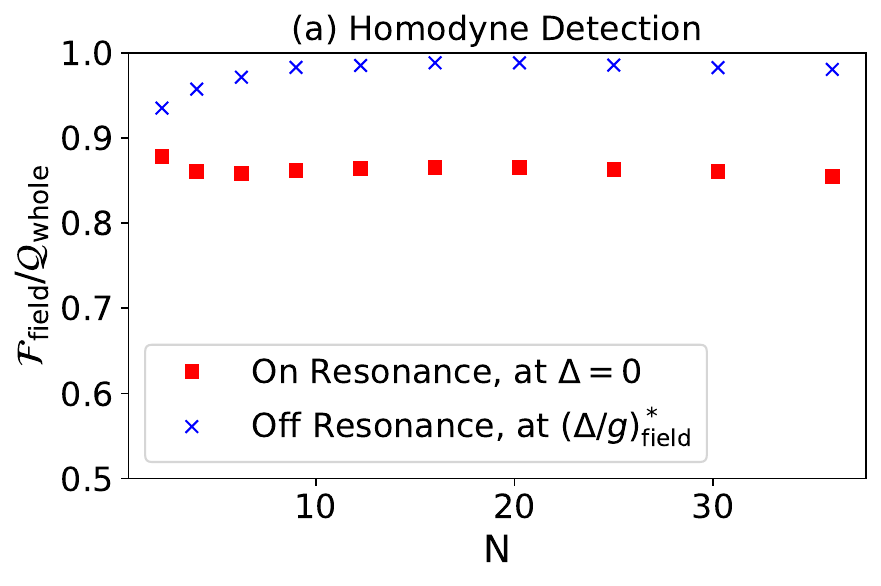}
    \includegraphics[width=0.45\linewidth]{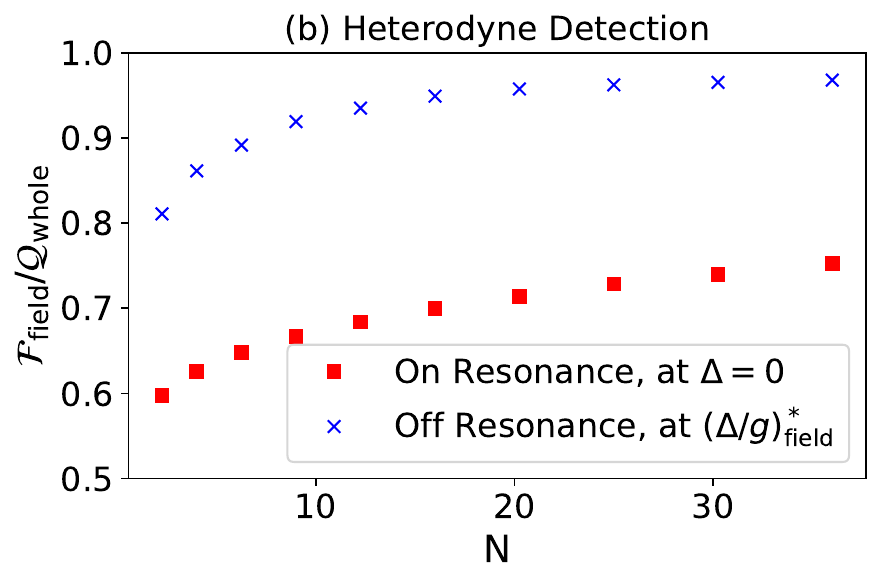}
    \caption{(a) Homodyne performance ratio $\frac{\mathcal{F}_\mathrm{field}(\mathcal{E})}{\mathcal{Q}_\mathrm{whole}(\mathcal{E})}$ for the probe on and off resonance as a function of $N$. (b) Heterodyne performance ratio $\frac{\mathcal{F}_\mathrm{field}(\mathcal{E})}{\mathcal{Q}_\mathrm{whole}(\mathcal{E})}$ for the probe on and off resonance as a function of $N$.}
    \label{fig_heterodyne_homodyne}
\end{figure}

In Fig.~\ref{fig_heterodyne_homodyne}(a), we show the performance ratio $\frac{\mathcal{F}_\mathrm{field}(\mathcal{E})}{\mathcal{Q}_\mathrm{whole}(\mathcal{E})}$ for the probe on and off resonance in the homodyne case as a function of $N$. As seen in the figure, off-resonance homodyne detection nearly saturates the ultimate sensing capability of the probe. This observation supports our Bayesian analysis, in which homodyne detection combined with Bayesian estimation approaches the ultimate sensitivity of the full probe.

In Fig.~\ref{fig_heterodyne_homodyne}(b), we show the same performance ratio for heterodyne detection. As the figure demonstrates, homodyne detection outperforms heterodyne detection across all values of $N$. Interestingly, as $N$ increases, heterodyne detection becomes comparable to homodyne detection. This is noteworthy because increasing $N$ makes numerical simulations for homodyne detection challenging due to truncation issues, whereas the heterodyne scheme remains more reliable and feasible.

\subsection{Likelihood Function}

Armed with the optimized homodyne angle $\varphi^*$, we now construct the likelihood function $P(\mathrm{data}|\mathcal{E})$ for the continuous-variable case. To this end, we discretize the continuous probability density function $p(X_{\varphi^*}|\mathcal{E})$ into a set of $m$ probabilities:
\begin{equation}
\mathcal{P}_m = \int_{y_m}^{y_{m+1}} p(X_{\varphi^*}|\mathcal{E}) dX,\label{eq_discretize}
\end{equation}
where the integration limits satisfy
\begin{equation}
y_{m+1} = y_m + W,
\end{equation}
with $W$ being a fixed bin width along the quadrature axis $X_{\varphi^*}$. $W$ is to be determined by the resolution of measurement device. The resulting likelihood function, assuming independent and identically distributed measurements, is then given by:
\begin{equation}
P(\mathrm{data}|\mathcal{E}) \sim  \prod_{m} \mathcal{P}_m^{C_m},  
\end{equation}
where the exponent $C_m$ denotes the number of measurement outcomes falling within the $m$-th bin. Recall that the probabilities $\mathcal{P}_m$ correspond to the theoretical outcome probabilities predicted by the model. In our case, they are fully determined by the steady state of the system and can be computed directly from it. In contrast, the counts $C_m$ represent the observed measurement data. These are simulated by drawing $M = \sum_m C_m$ independent random samples according to the distribution defined by the theoretical probabilities $\mathcal{P}_m$.

\begin{figure}
    \centering
    \includegraphics[width=\linewidth]{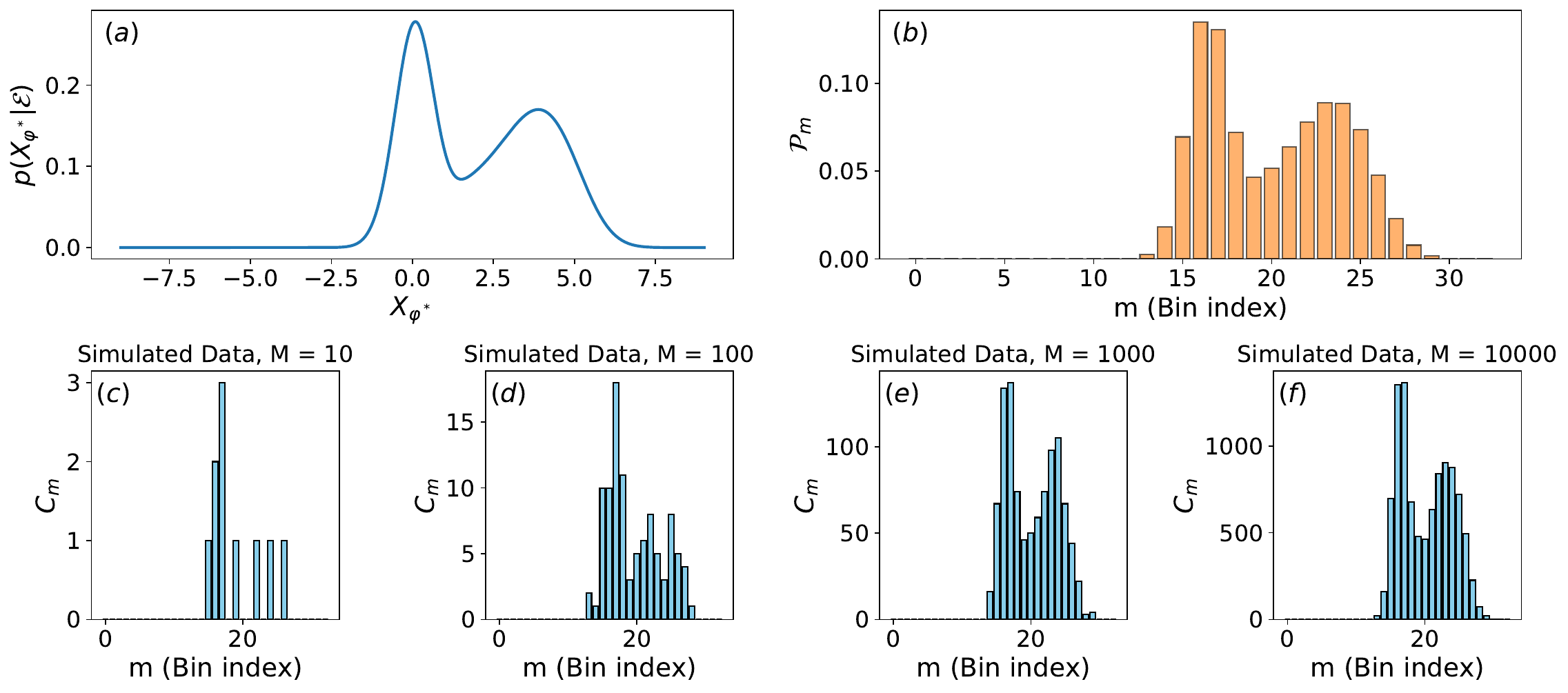}
    \caption{(a) Probability distribution $p(X_{\varphi}|\mathcal{E})$ for a fixed value of $\mathcal{E}$ and sensing resource $N$ as a function of the eigenvalue of the optimized quadrature $X_{\varphi^*}$. (b) Discretization of $p(X_{\varphi}|\mathcal{E})$. (c)–(f) Random counts $C_m$ for increasing values of $M$.}
    \label{fig_bins}
\end{figure}

In Fig.~\ref{fig_bins}, we illustrate the full procedure used to construct the observed counts $C_m$ and, consequently, the likelihood function. In Fig.~\ref{fig_bins}(a), we plot the modeled probability distribution $p(X_{\varphi}|\mathcal{E})$ for a fixed value of $\mathcal{E}$ and sensing resource $N$ as a function of the eigenvalue of the optimized quadrature $X_{\varphi^*}$. Fig.~\ref{fig_bins}(b) shows the discretization of this continuous distribution, where the probabilities for each bin are computed using Eq.~\eqref{eq_discretize}. As seen in the figure, the chosen number of bins fully captures the values of $X_{\varphi^*}$. Since any continuous probability density function can be discretized, we can use this to simulate observed data. Specifically, we generate random samples by drawing $M$ independent measurements according to the discretized distribution.
Figs.~\ref{fig_bins}(c)–(f) show examples of the resulting random counts $C_m$ for increasing values of $M$. As $M$ increases, the sampled counts more accurately reflect the discretized distribution from Fig.~\ref{fig_bins}(b), making the evaluation of the likelihood function straightforward.

\subsection{Data Simulation}

The simulated statistics is as follows: We perform 100 independent experiments. In each experiment, we simulate $M{=}1000$ measurements [we obtain a curve from simulated data as shown in Fig.~\ref{fig_bins}(e)], which are used to compute a single estimate of $\tilde{\mathcal{E}}$ by finding the value of $\mathcal{E}$ that maximizes the posterior function. This results in 100 independent estimates of $\tilde{\mathcal{E}}$. We then calculate the variance across these 100 values to obtain the estimator variance $\mathrm{Var}[\tilde{\mathcal{E}}]$. This procedure allows us to generate the Bayesian curves in Fig.~\ref{fig_bayesian}.

\end{document}